\documentclass[%
reprint,
amsmath,amssymb,
aps,
floatfix,
]{revtex4-2}

\usepackage{graphicx}
\usepackage{dcolumn}
\usepackage{bm}
\usepackage{hyperref}
\usepackage{xcolor}

\def\mchi{{m_{\chi}}}
\def\mphi{{m_{\phi}}}
\def\alphachi{{\alpha_{\chi}}}

\def\aap{AA}
\def\prl{Phys. Rev. Lett.}

\def\apjl{ApJL}

\def\mnras{MNRAS}
\def\apj{ApJ}
\def\apjs{ApJS}

\def\nat{Nat}

\def\prd{PhysRevD}
\def\jcap{JCAP}
\def\physrep{Phys. Rept.}

\begin{document}
	
	\preprint{APS/123-QED}
	
	\title[strong lensing constraints on SIDM]{Constraining resonant dark matter self-interactions with strong gravitational lenses}
	
	\author{Daniel Gilman$^{1}$}
	\thanks{daniel.gilman@utoronto.ca}
	\author{Yi-Ming Zhong$^{2}$}
	\author{Jo Bovy$^{1}$}
	\affiliation{%
		$^{1}$Department of Astronomy and Astrophysics, University of Toronto, Toronto, ON, M5S 3H4, Canada\\
		$^{2}$Kavli Institute for Cosmological Physics, University of Chicago, Chicago, IL 60637, USA}
	
	
	
	\date{\today}
	
	\begin{abstract}
		We devise a method to constrain self-interacting dark matter (SIDM) from observations of quadruply-imaged quasars, and apply it to five self-interaction potentials with a long-range dark force. We consider several SIDM models with an attractive potential that allows for the formation of quasi-bound states, giving rise to resonant features in the cross section localized at particular velocities below $50 \ \rm{km} \ \rm{s^{-1}}$. We propose these resonances, which amplify or suppress the cross section amplitude by over an order of magnitude, accelerate or delay the onset of core collapse in low-mass dark matter halos, and derive constraints on the timescale for core collapse for the five interaction potentials we consider. Our data strongly disfavors scenarios in which a majority of halos core collapse, with the strongest constraints obtained for cross section strengths exceeding $100 \ \rm{cm^2} \rm{g}^{-1}$ at relative velocities below $30 \ \rm{km} \ \rm{s^{-1}}$. This work opens a new avenue to explore the vast landscape of possible SIDM theories. 
	\end{abstract}
	
	\maketitle
	
	\section{Introduction} \label{sec:intro}
	Self-interacting dark matter (SIDM) has gained traction as a viable alternative to the concordance theory of cold dark matter (CDM). Support for SIDM comes primarily from observations of galaxies and dwarf galaxies, where dark self-interactions give rise to two physical processes relevant for dark matter halos. First, self-interactions transfer heat into the center of halos, producing a central core \citep{Spergel++00,AhnShapiro05,Rocha++13,Vogelsberger++12,Elbert++15}. Eventually, the halo undergoes a runaway contraction known as gravothermal catastrophe, or core collapse \citep{LyndenBellWood,LyndenBellEggleton80,Balberg++02,Elbert++15,Nishikawa++20,Sameie++20,Turner++21,Correa++20,Correa++22}. For self-interaction cross sections larger than $\mathcal{O}(1-10 \ \rm{cm}^2\, \rm{g}^{-1})$, the processes of core formation and collapse increase diversity between dwarf galaxy rotation curves, a consequence of SIDM that many authors argue provides a compelling explanation for the properties of low-mass galaxies \citep{Oman++15,Kaplinghat++16,Kamada++17,Tulin++18,Kahlhoefer++19,Zavala++19,Correa++20,Silverman++22,Adhikari++22,Slone++23}. 
	
	Analyses of galaxy clusters place stringent upper limits on elastic scattering with cross section strengths $\lesssim 0.1 - 1.0 \ \rm{cm^2} \, \rm{g^{-1}}$ \citep{Randall++08,Peter++13,Harvey++19,Sagunski++21,Andrade++22}, creating tension with models invoking SIDM to explain galactic rotation curves. The resolution to this apparent inconsistency involves adding a velocity dependence to the self-interaction cross section, suppressing it at scales $\sim 1,000 \ \rm{km} \ \rm{s^{-1}}$ relevant for galaxy clusters \citep{Yoshida2000,Kaplinghat++16,Correa++20}. Velocity-dependent cross sections arise naturally in many particle physics models with dark force mediators \citep{ArkaniHamed++09,BuckleyFox10,Foot++15,Boddy++16,Kahlhoefer++17,Cyr-Racine++16,Ryan++21}.
	
\begin{figure}
		\centering
		\includegraphics[trim=3.5cm 0cm 3.5cm
		0.cm,width=0.45\textwidth]{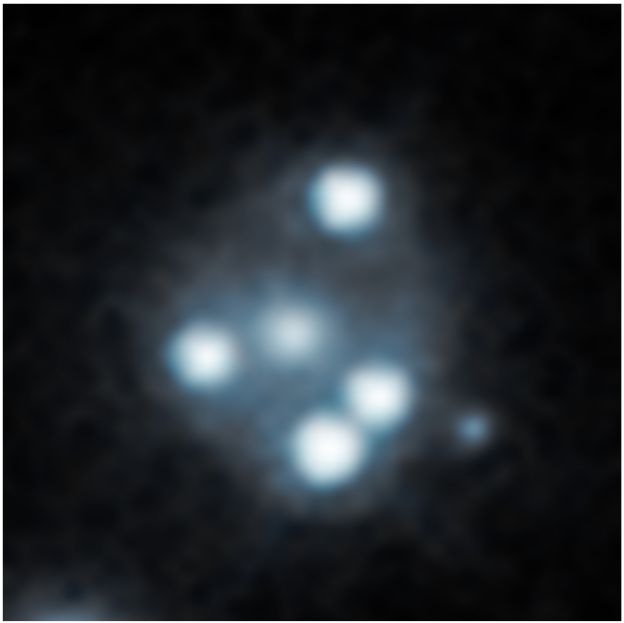}
		\caption{\label{fig:quadimg} An image from the Hubble Space Telescope of the strong lens system WFI 2033-4723, one of the eleven systems analyzed in this work. Deflection of light by a massive foreground galaxy ($z=0.66$, center) produces four highly-magnified images of a background ($z=1.66$) quasar, with a maximum image separation of approximately 2 arcseconds, or 14 kpc at the lens redshift. A small satellite galaxy of the main deflector is visible to the right of the merging image pair. Image courtesy of NASA, ESA, A. Nierenberg (JPL) and T. Treu (UCLA).}
	\end{figure}

	Due to the velocity dependence of the SIDM cross section, its strength can exceed $100 \  \rm{cm^2} \ 
	\rm{g^{-1}}$  low velocities, below $30 \ \rm{km} \ \rm{s^{-1}}$, causing significant fractions of low-mass halos to core collapse \citep{Turner++21,Gilman++21,Correa++22}. The high central density of collapsed objects transforms them into extremely efficient gravitational lenses \citep{Gilman++21,Yang++21}. Gravitational lensing refers to the deflection of light by gravitational fields, and strong lensing refers to a particular case where a massive foreground mass produces multiple images of a background source. We focus on a particular kind of lens system in which a quasar becomes quadruply imaged by a foreground galaxy, as depicted in Figure \ref{fig:quadimg}. The relative image magnifications (flux ratios) in quadruple-image systems provide a sensitive probe of dark halos along the line of sight between the observer and source, leading to constraints on dark matter theories that alter halo abundance and concentration \citep{Dalal++02,Nierenberg++14,Nierenberg++17,Gilman++20,Gilman++20b,Hsueh++20,Gilman++21,Gilman++22,Laroche++22}. As a direct gravitational probe of halo density profiles, extending down to masses of at least $ 10^7 M_{\odot}$ with existing flux ratio measurements, lensing avoids systematic uncertainties associated with studying SIDM with dwarf galaxies \citep[e.g][]{Pontzen++12,Chan++15,Sawala++16,Robles++19,Fitts++19,Kaplinghat++20,Sameie++21,Zentner++22,Roper++22}, because stellar feedback and baryonic contraction become inefficient in halos with viral masses below $10^{9} M_{\odot}$ \citep{Robles++17,Lazar++20,Rose++22}.
	
	In this work, we develop a method to constrain SIDM with quadruply-imaged quasars. We apply the method to analyze five velocity-dependent cross sections with resonances at low velocity. The inference method we use is an extension of methods developed and tested by \citet{Gilman++18,Gilman++19,Gilman++20,Gilman++21} applied to data collected over the last decade \citep{Chiba++05,Sugai++07,Stacey++18,Nierenberg++14,Nierenberg++17,Nierenberg++20}. We begin in Section \ref{sec:models} by describing the five benchmark cross sections we analyze. Section \ref{sec:ccmodel} presents the model for structure formation that we use to predict the fraction of core-collapsed halos as a function of halo mass for each cross section. Section \ref{sec:inference} presents our constraints on the cross sections from applying our method to analyze eleven lenses, and we conclude in Section \ref{sec:discussion}.
	
	\section{Benchmark models for the self-interaction cross section}
	\label{sec:models}
	We consider a long-range force between dark matter particles of mass $m_{\chi}$ described by a Yukawa potential
	\begin{equation}
		\label{eqn:pot}
		V\left(r\right) = \pm \alphachi\frac{e^{-\mphi r}}{r},
	\end{equation}
	where $\alpha_{\chi}$ is the strength of the potential, $m_{\phi}$ is the mediator mass, and $+/-$ represents a repulsive/attractive interaction. Given the differential scattering cross section $d\sigma / d \Omega$ corresponding to $V\left(r\right)$, we compute the viscosity transfer cross section \citep{Tulin++13,Colquhoun++21,Yang++22}
	\begin{equation}
		\label{eqn:viscositycross}
		\sigma_{V} = \int \frac{d\sigma}{d\Omega} \sin^2 \theta d \Omega,
	\end{equation}
	where $\theta$ is the scattering angle.
	
	We compute $\sigma_V$ using partial-wave analysis, where we expand the wave function of dark matter in spherical harmonics, compute the phase shift, $\delta_\ell$, of each scattered partial wave labelled by its angular momentum quantum number $\ell$, and sum over the phase shifts ($\hbar=c=1$) \citep{Colquhoun++21}
	\begin{equation}
		\label{eqn:sigmavsum}
		\sigma_{V} = \frac{4\pi}{k^2} \sum_{\ell=0}^{\ell_{\rm{max}}} \frac{\left(\ell+1\right)\left(\ell+2\right)}{2\ell +3}\sin^2\left(\delta_{\ell+2} - \delta_\ell\right).
	\end{equation}
	Here, $k = m_{\chi} v / 2$ is the momentum of the dark matter particle, and $v$ is the magnitude of the relative velocity. For the models we consider, we need only retain contributions up to $\ell_{\rm{max}} = 50$. 
	
	To compute the phase shifts, we recast the Schr\"{o}dinger equation as a first-order differential equation using the auxiliary function $\delta_\ell (r)$ suggested by \citet{Chu++20}
	\begin{eqnarray}
		\label{eqn:phaseshifts}
		\frac{\partial \delta_\ell \left(r\right)}{\partial r} = - k \mchi r^2 V\left(r\right)&\\
		\nonumber \times \bigl(\sin\left[\delta_\ell\left(r\right)\right] & j_{\ell}\left(k r\right) - \cos\left[\delta_\ell\left(r\right)\right] n_{\ell}\left(k r\right)\bigr)^2 
	\end{eqnarray} 
	where $j_l$ and $n_l$ are spherical Bessel functions. We solve this equation numerically with the boundary condition $\delta_\ell \left(0\right)=0$, and obtain the phase shifts by taking the limit $\delta_\ell = \lim_{r\to \infty} \delta_\ell(r)$. We evaluate Equations \ref{eqn:sigmavsum} and \ref{eqn:phaseshifts} using {\tt{Mathematica}}.
	
	Figure \ref{fig:benchmarks} shows five benchmark cross sections that emerge from these calculations, with values of $\alpha_{\chi}$, $m_{\chi}$, and $m_{\phi}$ listed in Table \ref{tab:benchmarkparams}. Each model has a suppression of the cross section amplitude at high speeds, evading upper limits on the cross section strength from galaaxy clusters at $v \sim 1,000\rm{km}\  \rm{s^{-1}}$. Among the many possible cross sections that exhibit this high-speed suppression, we choose this particular subset of five models because their cross section amplitudes at low speeds span a representative range of structure formation outcomes in terms of the number of core-collapsed halos predicted to form for each cross section (see Section \ref{sec:ccmodel}). 
	
	Model 1 has a repulsive potential in the semi-classical regime \citep{Colquhoun++21}, while Models 2--5 have an attractive potential. The particular combinations of the dark matter particle mass to the mediator mass ($m_{\chi} / m_{\phi}$) and the potential strength ($\alpha_{\chi}$) for each of the five benchmark models place them in the non-perturbative scattering regime, where non-perturbative effects manifest as \textit{resonances}, which refer to a suppression or enhancement of the cross section amplitude at particular speeds. These quantum mechanical interference effects appear due to the formation of quasi-bound states with the attractive potential \citep{BuckleyFox10,Tulin++13,Tulin++13a,Chu++19,Chu++20}, and cause order-of-magnitude enhancement or suppression of the scattering cross section in the range of halo masses with central velocity dispersion aligned with the position of resonance in the cross section. In the next section, we explore the consequences of this phenomenon by implementing a model that relates SIDM cross sections to the process of core collapse.
	
	\section{A physical model for core collapse in dark field halos and subhalos}
	\label{sec:ccmodel}
	
\begin{figure}
		\centering
		\includegraphics[trim=0.cm 0.7cm 0.cm
		0.0cm,width=0.45\textwidth]{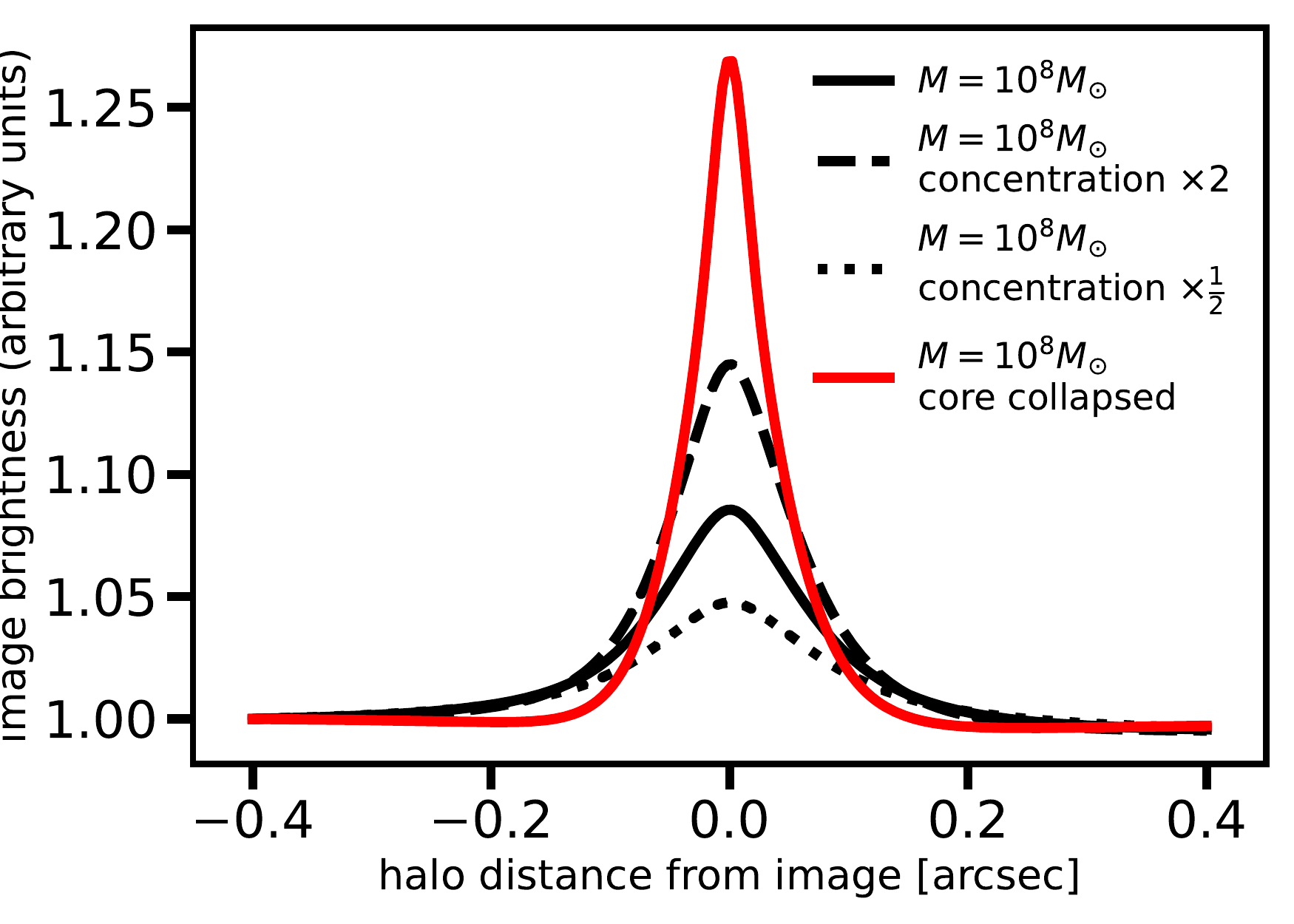}
		\caption{\label{fig:magcross} The magnification cross section for a single $10^8 M_{\odot}$ halo with varying concentration (black), and with a core-collapsed density profile given by Equation \ref{eqn:collapseprof} (see Section \ref{sec:ccmodel}). The high central density of collapsed halos expected to form in SIDM models with large cross sections become extremely efficient gravitational lenses. The magnification cross section also depends on source size; to create the figure we assume a source size of $\sim 50 \rm{pc}$, typical of the nuclear narrow-line emission region measured in our data.}
	\end{figure}

	The heat transfer through self-interacting dark matter halo density profiles drives a dynamic evolution of halo density profiles that culminates with a runaway contraction of the halo density profile typically referred to as core collapse, or the \textit{gravothermal catastrophe} \citep{LyndenBellWood}. This physical process is of particular interest for the lensing analysis presented in this work because more concentrated halos act as more efficient lenses. Figure \ref{fig:magcross} quantifies this intuition by showing the magnification cross section caused by a single $10^8 M_{\odot}$ halo with different density profile. The black, magenta, and green curves show the magnification cross section for a Navarro-Frenk-White (hereafter NFW) profile predicted by CDM \citep{Navarro++97} with a concentration varying around the CDM prediction. The red curve shows the magnification for a reasonable implementation of a collapsed halo profile (see Section \ref{sec:ccmodel}). 
	
	The factor of $\sim 2$ enhancement to the lensing efficiency of the collapsed objects suggests that, if large quantities of these objects exist, we should be able to statistically detect this signal by analyzing a sample of quadruply-imaged quasars, such as the lens system WFI 2033-4723 shown in Figure \ref{fig:quadimg}. Quadruple-image lens systems enable the statistical detection of dark matter halos across cosmological distance on mass scales below $10^8 M_{\odot}$, a regime where halos are not expected to host a luminous galaxy. To interpret a signal in the context of a particular SIDM cross section, however, we require a model that relates the SIDM cross section to the populations of collapsed field halos and subhalos  that perturb the lenses.
	
	\subsection{The self-similar time evolution of SIDM halos}
	\citet{ShengqiYang++22} and \citet{Yang++22} compare semi-analytical models of SIDM with N-body simulations, and show that halo evolution is nearly self-similar when expressed in terms of the characteristic timescale
	\begin{eqnarray}
		\nonumber t_0 \left(m, z, \sigma_V\right) = \left(\frac{1\, \rm{cm^2} \ \rm{g^{-1}}}{\langle \sigma_V v^5\rangle / \langle v^5\rangle }\right)   \left(\frac{100 \ \rm{km} \ \rm{s^{-1}}}{v_{\rm{max}}}\right) \\
		\times  
		\left(\frac{10^7 M_{\odot} \rm{kpc^{-3}}}{\rho_s}\right) \rm{Gyr}.
		\label{eqn:timescale}
	\end{eqnarray}
	This quantity depends on halo mass, $m$, through the maximum circular velocity $v_{\rm{max}} = 1.65 \sqrt{G \rho_s r_s^2}$ defined in terms of the scale radius ($r_s$) and the characteristic density ($\rho_s$) of a NFW profile. The equation also depends on the halo redshift, $z$, through the concentration-mass relation, which relates $\rho_s$ and $r_s$ to the halo mass as a function of redshift. 
	
	A careful derivation of the thermal conductivity \citep{LifshitzPitaevskii81,ShengqiYang++22,Yang++22} shows that the relevant timescale for heat transfer throughout the halo profile involves a velocity average performed with a $v^5$ kernel, giving the thermally-averaged cross section 
	\begin{equation}
		\label{eqn:thermalavg}
		\langle \sigma_V v^5\rangle = \frac{1}{2\sqrt{\pi}v_{0}^3} \int_{0}^{\infty} v^{\prime 5} \sigma_V \times v^{\prime 2} \exp\left(\frac{-v^{\prime2}}{4 v_{0}^2}\right) dv^{\prime},
	\end{equation}
	a quantity that appears in the denominator of Equation \ref{eqn:timescale}. The velocity scale $v_0$ varies proportionally with the maximum circular velocity $v_0 = 0.64 \ v_{\rm{max}}$ \citep{Outmezguine++22}. Figure \ref{fig:collapsetimes} shows $t_0$ as a function of halo mass  for the five benchmark models evaluated at $z = 0.5$ with the concentration-mass relation presented by \citet{Diemer++19}. As Figure \ref{fig:collapsetimes} clearly illustrates, resonances in the cross section at different speeds correspond to shorter evolution timescales in halos with particular masses.
	
	\begin{figure}
		\centering
		\includegraphics[trim=0.cm 0.7cm 0.cm
		0.0cm,width=0.48\textwidth]{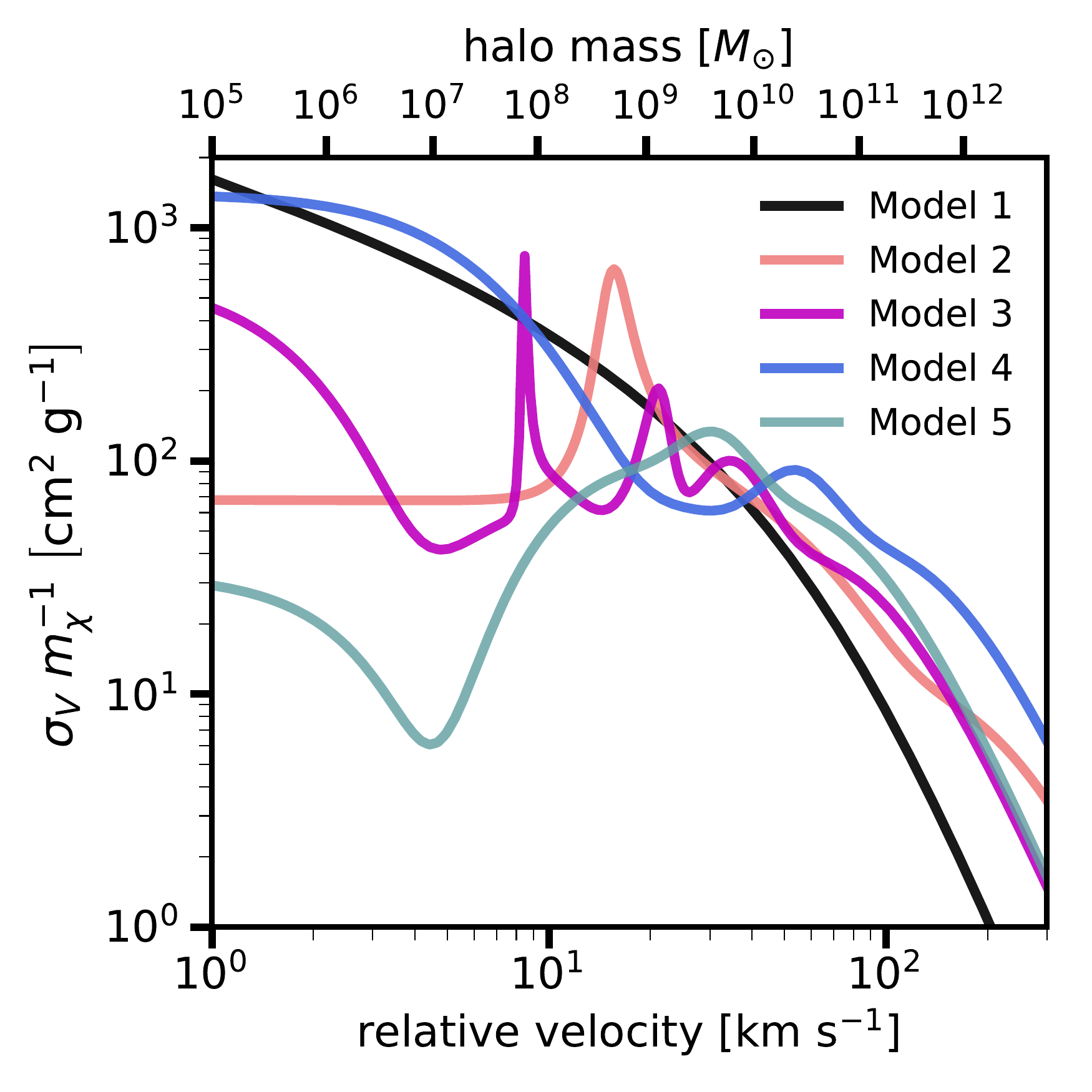}
		\caption{\label{fig:benchmarks} The viscosity transfer cross section divided by the dark matter particle mass $\sigma_V/m_{\chi}$ for the five benchmark models analyzed in this work, plotted as a function of the relative velocity $v$ (lower x-axis). The upper x-axis shows the halo mass with maximum circular velocity $v_{\rm{max}} = v$.}
	\end{figure} 
	\begin{figure}
		\centering
		\includegraphics[trim=0.cm 0.7cm 0.cm
		0.0cm,width=0.48\textwidth]{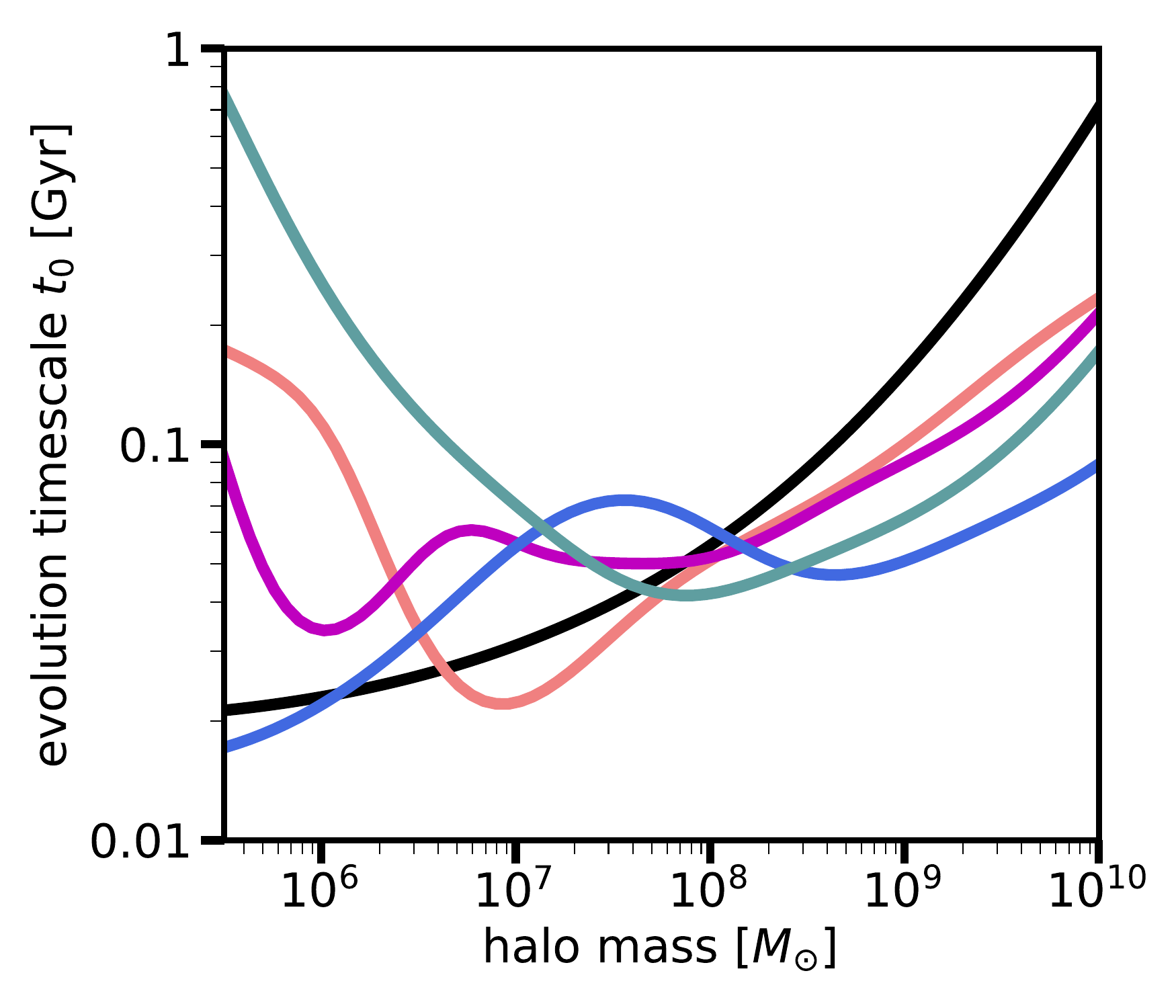}
		\caption{\label{fig:collapsetimes} The structural evolution timescale $t_0$ in Equation \ref{eqn:timescale} evaluated at $z=0.5$ for each of the five benchmark models The figure uses the same color scheme to identify the five benchmark models as Figure \ref{fig:benchmarks}. Short timescales correspond to higher fractions of collapsed objects (see Figure \ref{fig:collapsefrac}).}
	\end{figure} 
	\begin{figure}
		\centering
		\includegraphics[trim=0.cm 0.7cm 0.cm
		0.0cm,width=0.48\textwidth]{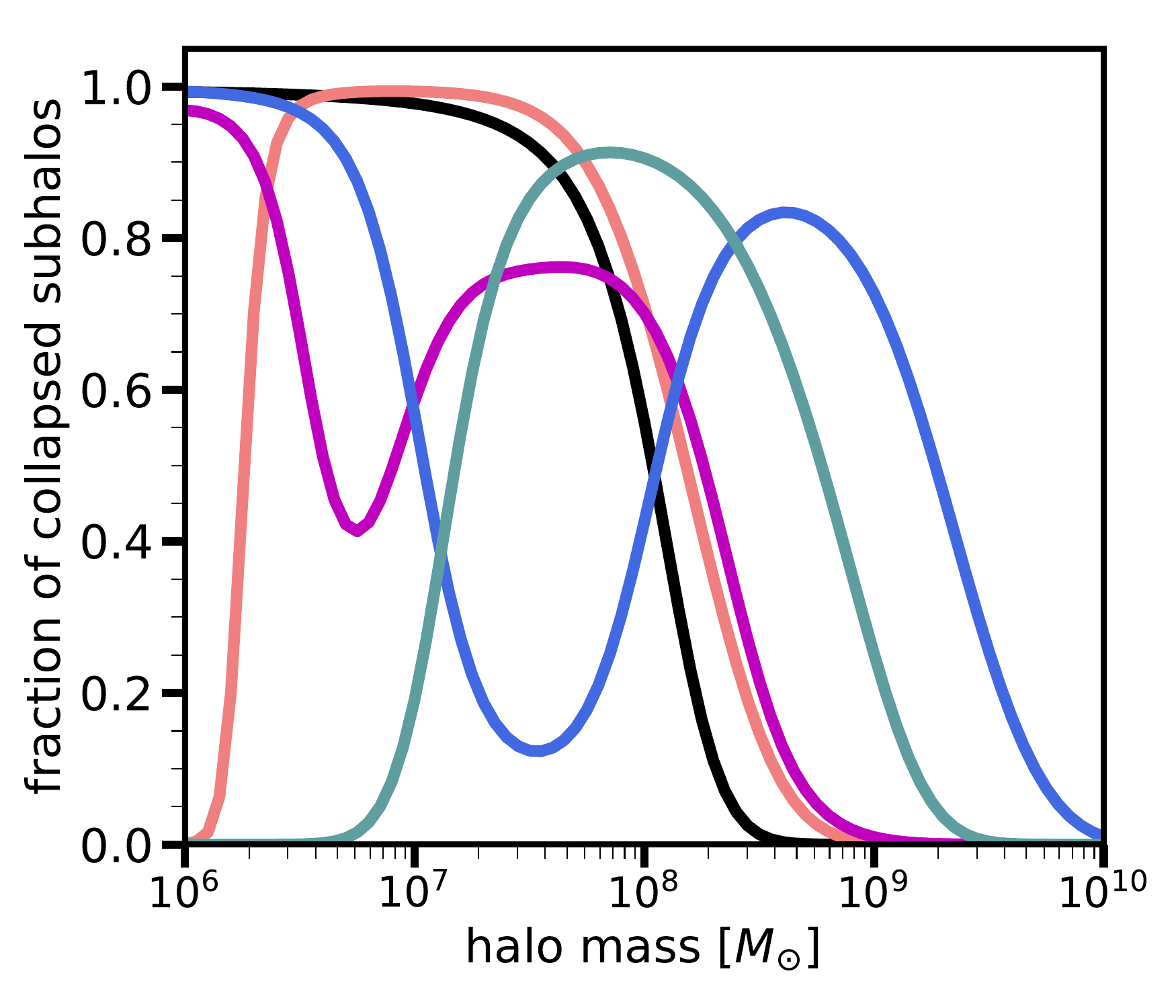}
		\caption{\label{fig:collapsefrac} The fraction of collapsed subhalos as a function of halo mass for each of the benchmark models, computed at $z=0.5$ using Equations \ref{eqn:timescale}, \ref{eqn:thermalavg}, and \ref{eqn:collapseprob}, with $s_{\rm{sub}} = 0.5 \ \rm{Gyr}$, $\lambda_{\rm{field}}=350$, $\lambda_{\rm{sub}} = 0.4$, such that $t_{\rm{sub}} = 140 \ t_0$. Very few field halos collapse for the chosen value of $\lambda_{\rm{field}}$.}
	\end{figure} 
	\begin{table}
		\caption{\label{tab:benchmarkparams} The dark matter particle mass $m_{\chi}$, mediator mass $m_{\phi}$, and potential strength $\alpha_{\chi}$ for each benchmark model.}
		\begin{ruledtabular}
			\begin{tabular}{cccc}
				& $m_{\chi} \left[\rm{GeV}\right]$& $m_{\phi} \left[\rm{MeV}\right]$ & $\alpha_{\chi}$ \\
				\hline 
				\\
				Model 1& 119 & 0.4 & $3.0\times10^{-3}$  \\
				Model 2& 20 & 3.5 & $1.6\times10^{-3}$  \\
				Model 3& 40 & 1.1 & $1.5\times10^{-3}$ \\
				Model 4& 18 & 1.7 & $1.3\times10^{-3}$ \\
				Model 5& 27 & 1.1 & $9.0\times10^{-4}$  \\
			\end{tabular}
		\end{ruledtabular}
	\end{table} 
	\begin{figure*}
		\centering
		\includegraphics[trim=0.75cm 0.5cm 0.75cm
		1.cm,width=0.95\textwidth]{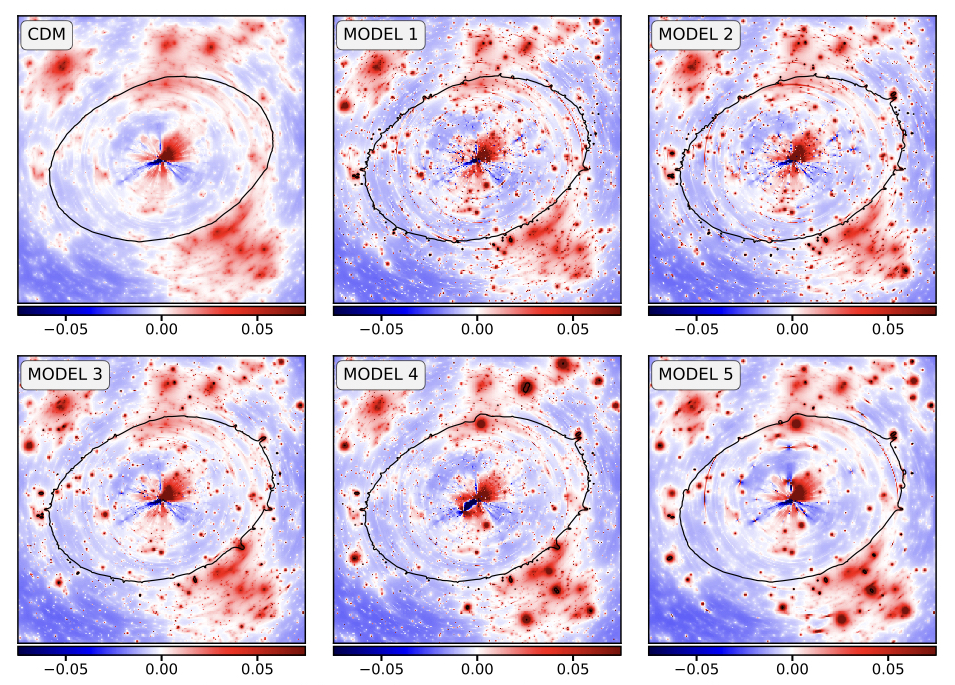}
		\caption{\label{fig:massdistributions} A possible realization of projected mass in dark matter field halos and subhalos in CDM (top left). Each other panel shows the same population of halos, but with a fraction of core-collapsed halos implemented according to the collapse probabilities shown in Figure \ref{fig:collapsefrac}. The color scale represents fluctuations of the projected mass around the average. Black lines show the critical curve, near where highly magnified lensed images appear. The semi-major axis is $\sim 1$ arcsecond. Halos along the line of sight appear warped and distorted in this representation. The visible deformations of the critical curve illustrate the efficient lensing properties of collapsed halos.}
	\end{figure*} 
	\subsection{Two distinct timescales for core collapse in dark subhalos and field halos}
	To predict the fraction of collapsed halos as a function of halo mass, we will introduce two dimensionless numbers, $\lambda_{\rm{field}}$ and $\lambda_{\rm{sub}}$, that determine the time of onset of core collapse in dark matter field halos and subhalos, respectively, relative to the physical timescale $t_0$ given by Equation \ref{eqn:timescale}
	\begin{eqnarray}
		t_{\rm{field}} &\equiv& \lambda_{\rm{field}} \ t_0 \\
		t_{\rm{sub}} &\equiv& \lambda_{\rm{sub}}\left(\sigma_V\right) t_{\rm{field}} = \lambda_{\rm{sub}}\left(\sigma_V\right) \lambda_{\rm{field}} t_0.
	\end{eqnarray}
	The concrete physical interpretation of $t_{\rm{field}}$ and $t_{\rm{sub}}$ is that half of all dark matter halos will collapse once their age exceeds $t_{\rm{field}}$, and half of all subhalos will have collapsed once their age exceeds $t_{\rm{sub}}$. Note that we have also allowed $t_{\rm{sub}}$ to depend explicitly on the cross section. This dependence exists, at least in principle, because dark matter particles bound to substructures can interact with the dark matter particles bound to the host dark matter halo (for example, through ram-pressure stripping \citep{Nadler++21} or evaporation \citep{Vogelsberger++19,Zeng++22}). The importance of this effect depends on the amplitude of $\sigma_V$ at a velocity scale comparable to the host halo central velocity dispersion, typically $\mathcal{O}\left(100\right) \rm{km} \ \rm{s^{-1}}$ for a galaxy. 
	
	The motivation behind introducing two distinct timescales between subhalos and field halos comes from considering the different environments relevant for each population of objects, and how these environments affect the process of core collapse. In contrast to field halos, galactic subhalos experience tidal disruption by a central galaxy, which can accelerate collapse (lower $\lambda_{\rm{sub}}$) in subhalos relative to field halos \citep{Kahlhoefer++19,Sameie++20,Nishikawa++20}. As mentioned in the previous paragraph, scattering between subhalo and host halo particles can also decelerate the onset of collapse in subhalos (higher $\lambda_{\rm{sub}}$) relative to field halos to a degree that that depends on $\sigma_{V}$ and its velocity dependence \cite{Vogelsberger++19,Nadler++20,Zeng++22}. Additional physical processes, such as the destruction of subhalos by ram-pressure stripping, can be implemented within this framework by enforcing covariance between the collapse timescales and the amplitude of the subhalo mass function through importance sampling. 
	
	We will now discuss the quantitative details of our model for core collapse in temrs of $\lambda_{\rm{sub}}$ and $\lambda_{\rm{field}}$. Given the age of a halo, $T\left(z\right)$, defined as the elapsed time between redshift $z=10$ (a typical formation time for low-mass halos) and the halo redshift $z$, we expect core collapse to occur once $T\left(z\right) \gg t_{\rm{sub}}$ or $T\left(z\right) \gg t_{\rm{field}}$. Different environments and evolutionary histories likely accelerate or delay collapse for individual halos. On a population level, we account for these effects by introducing scatter in the collapse times, represented by parameters $s_{\rm{sub}}$ and $s_{\rm{field}}$ for subhalos and field halos, respectively. A reasonable implementation of the collapse probability, $P_c$, should satisfy $P_{\rm{c}} \rightarrow 0$ ($P_{\rm{c}} \rightarrow 1$) when $t_{\rm{sub}} \gg T\left(z\right)$ ($t_{\rm{sub}} \ll T\left(z\right)$). The same trends should hold for field halos. We use a function for the collapse probability that meets these criteria 
	\begin{equation}
		\label{eqn:collapseprob}
		P_{\rm{c}}\left(m, z, \sigma_V\right) = \frac{1}{2} \left[1+ \tanh \left( \frac{T\left(z\right) - t_{\rm{sub}}\left(m, z, \sigma_V\right)}{2 s_{\rm{sub}}}\right)\right].
	\end{equation}
	The collapse probability for field halos follows a similar distribution, replacing $t_{\rm{sub}}$ and $s_{\rm{sub}}$ with $t_{\rm{field}}$ and $s_{\rm{field}}$. We predict the fraction of collapsed halos in a mass range between $m_a$ and $m_b$, $f_{a/b}\left(z,\sigma_V\right)$, by integrating over the halo mass function $dN/dm$ weighted by the probability of halos collapsing 
	\begin{equation}
		f_{a/b}\left(z,\sigma_V\right)=\frac{1}{\langle N\rangle}\int_{m_a}^{m_b} P_c\left(m,z,\sigma_V\right)\frac{dN}{dm} dm
	\end{equation}
	where $\langle N \rangle = \int_{m_a}^{m_b} \left(dN / dm\right) dm$. 
	
	Figure \ref{fig:collapsefrac} shows the fraction of core-collapsed subhalos as a function of halo mass, assuming $\lambda_{\rm{sub}} = 0.4$ and $\lambda_{\rm{field}}=350$, and $s_{\rm{sub}} = 0.5 \  \rm{Gyr}$. Our model predicts peaks in the fraction of collapsed halos associated with peaks in the scattering cross section. The $\sigma_{V} \propto v^{-1}$ behavior of the repulsive cross section (Model 1) causes most subhalos less massive than $10^8\, M_{\odot}$ to collapse, while the resonant enhancement of the cross section in Model 2 near $20 \ \rm{km} \ \rm{s^{-1}}$ causes nearly all of subhalos in the mass the range $5\times 10^6 M_{\odot} - 5 \times 10^7 M_{\odot}$ to core collapse. The multiple resonances in Models 3 and 4 produce bimodal distributions in the fraction of collapsed halos as a function of mass. For the value $\lambda_{\rm{field}} = 350$ used to create the figure, which roughly corresponds to the predicted timescale for core collapse in field halos from gravothermal fluid models with elastic scattering \cite{Outmezguine++22,ShengqiYang++22,Yang++22}, a negligible fraction of field halos collapse. Figures \ref{fig:ssubz} and \ref{fig:ssubscatter} in Appendix \ref{sec:supinference} show how the fraction of collapsed subhalos depends on redshift and $s_{\rm{sub}}$, respectively. 
	
	\subsection{The (sub)halo mass function and density profiles}
	We assume an SIDM model that does not alter the linear matter power spectrum \cite[e.g.][]{Vogelsberger++16,Cyr-Racine++16}, so we rely on halo density profiles to distinguish SIDM from CDM\footnote{We can easily extend our analysis to models with suppressed small-scale power, given a model for the halo mass function and concentration-mass relation in these scenarios.}. Prior to and during the early stages of collapse SIDM halos have cores, but the cores have a negligible impact on the magnification cross section \citep{Gilman++21}, so we model halos that have not collapsed with NFW profiles for a conservative estimation. If we determine, based on Equation \ref{eqn:collapseprob}, that a halo has collapsed, we model its density profile as 
	\begin{equation}
		\label{eqn:collapseprof}
		\rho\left(r,r_c,x_{\rm{match}}\right) = \rho_0\left(x_{\rm{match}}\right) \left(1 + r^2 / r_c^2\right)^{-\gamma/2}.
	\end{equation}
	The parameter $x_{\rm{match}}$, defined as the multiple of the halo scale radius where the collapsed profile encloses the same mass as an NFW profile, fixes the normalization $\rho_0$; simulations show that $x_{\rm{match}}$ lies in the range $2-3$ \citep{Zeng++22}. We include a core radius $r_c$ with size between $0.01 r_s -0.05 r_s$ to regularize the profile at the origin, and reproduce the small cores still present in collapsed halo profiles. Regarding the logarithmic slope $\gamma$, some N-body simulations and semi-analytical fluid models predict $\gamma \sim 2.2$ \citep{LyndenBellEggleton80,Jiang++22,Yang++22,Correa++22,Koda2011, Essig++19,Outmezguine++22}, while other simulations find $\gamma \sim 3$ \citep{Kahlhoefer++19,Turner++21,Zeng++22}. The differences may stem from the cosmological evolution of the halo, different implementations of the tidal evolution of subhalos, or the violation of the hydrostatic condition assumed in the fluid models. We explore how different models for the collapsed halo profile affect our results in Appendix \ref{sec:supcollapsedprofile}.
	
	Figure \ref{fig:massdistributions} illustrates how the existence of a population of collapsed objects dramatically enhances the magnification cross section, relative to a population of NFW profiles. The figure shows possible projected mass distributions of subhalos and field halos for a simulated lens system for each benchmark model. Each panel depicts the mass in substructure projected along the line of sight between observer and source. The realizations shown in each panel halos have the same coordinates, but we have forced different objects to core collapse based on the core collapse probabilities shown in Figure \ref{fig:collapsefrac}. Deformation of the critical curve (black lines) implies a large perturbation to the magnification of a nearby image, as shown explicitly in Figure \ref{fig:magcross}. By forward modeling image flux ratios with millions of realizations of dark matter structure, similar to those depicted in Figure \ref{fig:massdistributions}, we can statistically infer the fraction of collapsed halos perturbing strong lenses, and derive constraints on the five cross sections shown in Figure \ref{fig:benchmarks}.
	
	\section{Constraints on the benchmark models}
	\label{sec:inference}
	To constrain the five benchmark models, we use the inference framework developed and tested by \citet{Gilman++19,Gilman++20,Gilman++21}. The methodology involves forward modeling the lensed image positions and flux ratios in the presence of dark matter halos, where the abundance and density profiles of the halos are determined by a vector of hyper-parameters. The hyper-parameters can describe, for example, the amplitude and slope of the (sub)halo mass function, the concentration-mass relation, or the abundance of core-collapsed halos. By comparing simulated data with observed data using informative summary statistics, we compute the likelihood function of the data given the model parameters, from which we can derive relative likelihoods and posterior probability distributions. Appendix \ref{ssec:generalinf}, as well as previous analyses that use the method  \cite{Gilman++20,Gilman++21,Gilman++22,Laroche++22}, provide additional details regarding how we compute the likelihood function.
	
	\subsection{Dataset}
	\label{sec:supdata}
	We analyze a sample of eleven quadruply-imaged quasars with the sample selection subject to two criteria. First, the systems must have flux ratios measured from an extended region around the background quasar to eliminate contamination from microlensing by stars. The angular size of the background source determines the smallest deflection angle that can impact an image magnification, so measuring image brightness from emission lines that come from an extended region around the source eliminates contamination associated with micro-lensing by stars, while retaining sensitivity to milli-arcsecond scale deflections caused by dark matter halos with masses between $10^7 - 10^{10} M_{\odot}$. 
	
	Nine of the eleven systems have image magnifications measured from nuclear narrow-line emission: RXJ 1131+0231 \citep{Sugai++07}, B1422+231 \citep{Nierenberg++14}, HE0435-1223 \citep{Nierenberg++17}, WGD J0405-3308, RX J0911+0551, PS J1606-2333, WFI 2026-4536, WFI 2033-4723, WGD 2038-4008 \citep{Nierenberg++20}. One system (PG 1115+080) has measurements in the mid-IR \citep{Chiba++05}, and one system (MG0414+0534) has flux ratios measured from compact CO 11-10 emission \citep{Stacey++18}. 
	
	The second criteria that must be satisfied is that the lens system show no evidence for morphological complexity in the form of stellar disks, because these structures require explicit lens modeling \citep{Gilman++17,Hsueh++17}. Disks appear prominently in images of the deflector, allowing for the removal of problematic systems from the sample. The lens modeling applied to these data is discussed in detail in Appendix \ref{ssec:generalinf}. 
	
	As illustrated by Figures \ref{fig:magcross} and \ref{fig:massdistributions}, core-collapsed halos act as extremely efficient lenses, perturbing the relative image magnifications of the eleven systems we analyze more than we would expect in CDM. In the next section, we discuss how we compute the likelihood function using our data, and how we derive constraints on the core-collapse timescales $\lambda_{\rm{sub}}$ and $\lambda_{\rm{field}}$ from our data.
	
	\subsection{Derivation of the likelihood function}
	To constrain the five benchmark SIDM cross sections shown in Figure \ref{fig:benchmarks}, we use two vectors of hyper-parameters: First, we have a set of parameters ${\bf{q}}$ that specify the fraction of core-collapsed subhalos and field halos in three mass ranges logarithmically-spaced between $10^6 -10^{10}M_{\odot}$\footnote{This is the relevant mass range for substructure lensing because the size of the background source removes signal from less massive halos, while halos more massive than $10^{10} M_{\odot}$ are rare, and likely host a luminous galaxy, in which case we would explicitly model them.}. Appendix \ref{ssec:inf} gives additional details regrading how we define the vector of hyper-parameters ${\bf{q}}$ in the lensing analysis. Second, we have a set of hyper-parameters ${\bf{v}}$ that specify the amplitude and logarithmic slope of the (sub)halo mass function. For each lens, we compute $\mathcal{L}\left({\bf{d}_n} | {\bf{q}},{\bf{v}} \right)$, the likelihood of the data of the $n$th lens in terms of these parameters.  When computing $\mathcal{L}\left({\bf{d}_n} | {\bf{q}},{\bf{v}} \right)$, we marginalize over uniform priors on $\gamma$ between $2.7-3.3$, on $r_c$ between $0.01 r_s$ and $0.05 r_s$, on $x_{\rm{match}}$ between $2-3$, as well as the size of the lensed background source, and the mass profile of the main deflector. Appendix \ref{sec:supmodels} describes how we parameterize the subhalo and field halo mass functions, the background source, and the mass profile of the main deflector.
	\begin{figure*}
		\centering
		\includegraphics[trim=10.cm 2.cm 10.cm
		2.0cm,width=0.95\textwidth]{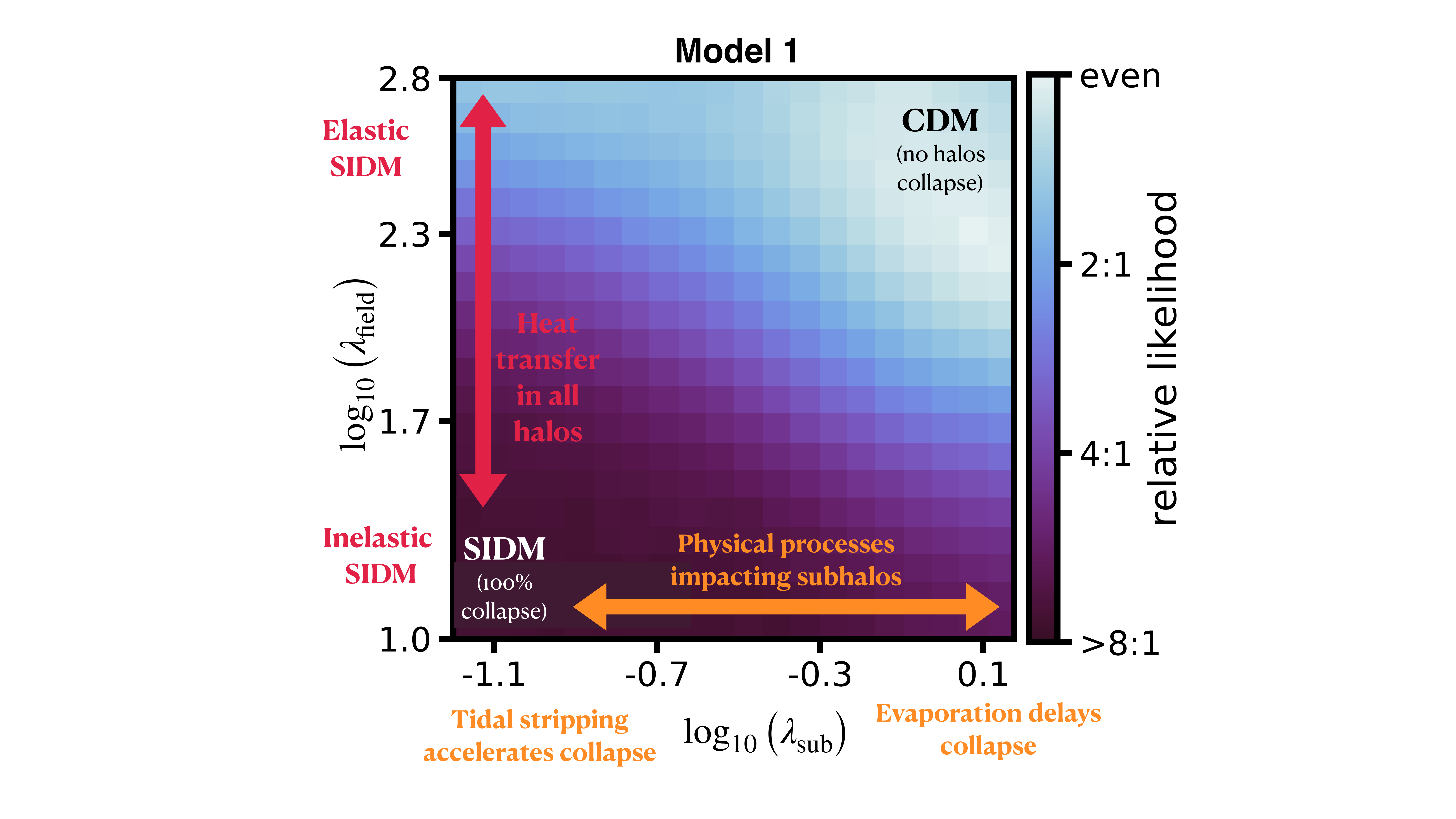}
		\caption{\label{fig:model1} The joint likelihood of the core-collapse timescale for all halos, $\lambda_{\rm{field}}$, and the collapse timescale for subhalos relative to field halos, $\lambda_{\rm{sub}}$. The color scale shows the relative likelihood of each combination of $\lambda_{\rm{field}}$ and $\lambda_{\rm{sub}}$ allowed by the data after marginalizing over uncertainties associated with the amplitude of the subhalo mass function. The top right (bottom left) regions of parameter space correspond to models in which almost zero (almost all) subhalos and field halos collapse. Moving along the y-axis changes the overall timescale for core collapse in all halos, which corresponds to changing the efficiency with which self-interactions can move heat through the halo profile through, for example, inelastic scattering. Moving along the x-axis explores different models for core collapse in subhalos. Small values of $\lambda_{\rm{sub}}$ correspond to accelerated collapse of subhalos relative to field halos by processes such as tidal stripping, while larger $\lambda_{\rm{sub}}$ correspond to situations in which processes such as host-subhalo scattering, also referred to as evaporation, delay collapse.}
	\end{figure*} 
	\begin{figure*}
		\centering
		\includegraphics[trim=0.0cm 0.cm 0.0cm
		0.0cm,width=0.45\textwidth]{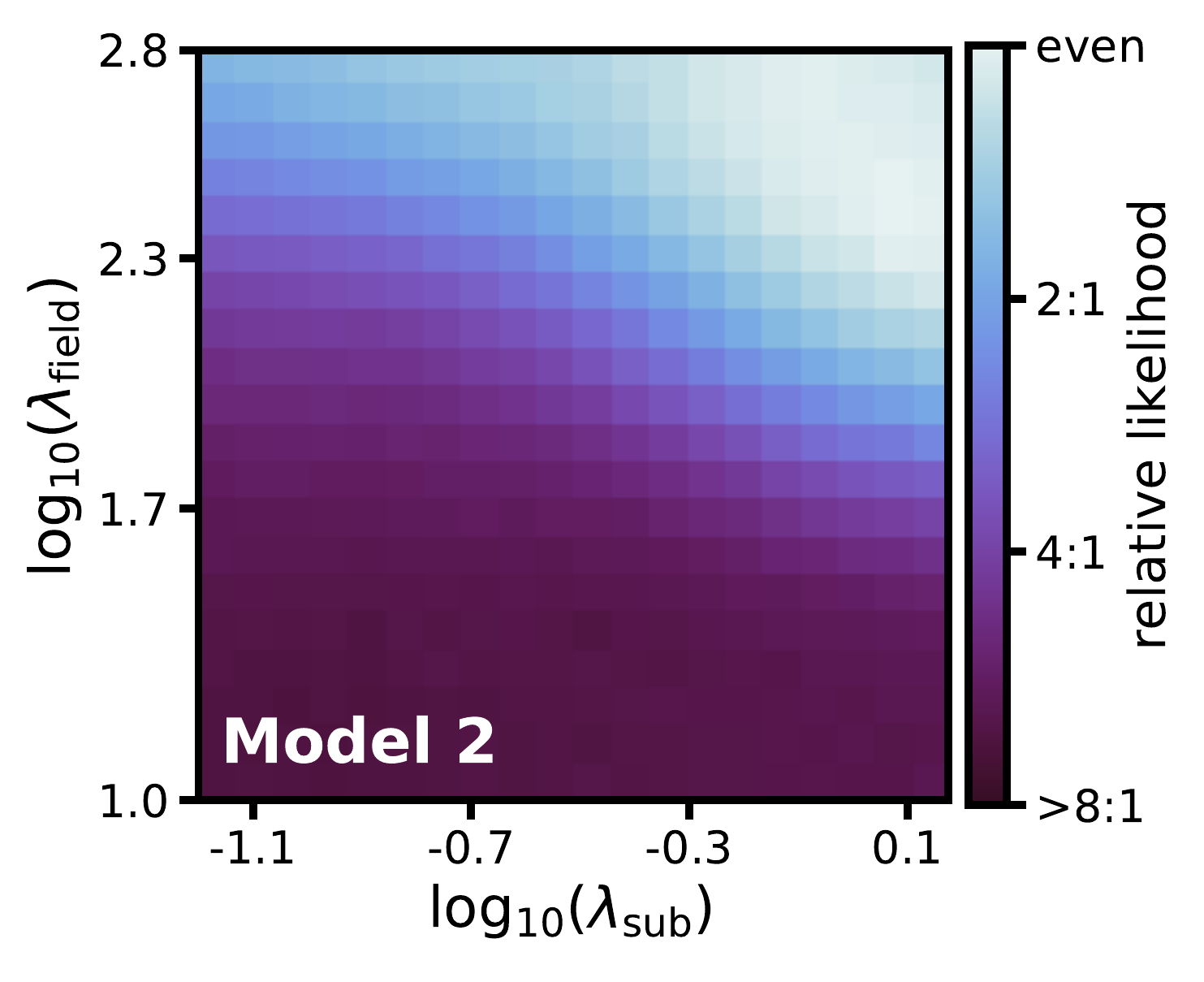}
		\includegraphics[trim=0.cm 0.cm 0.cm
		0.0cm,width=0.45\textwidth]{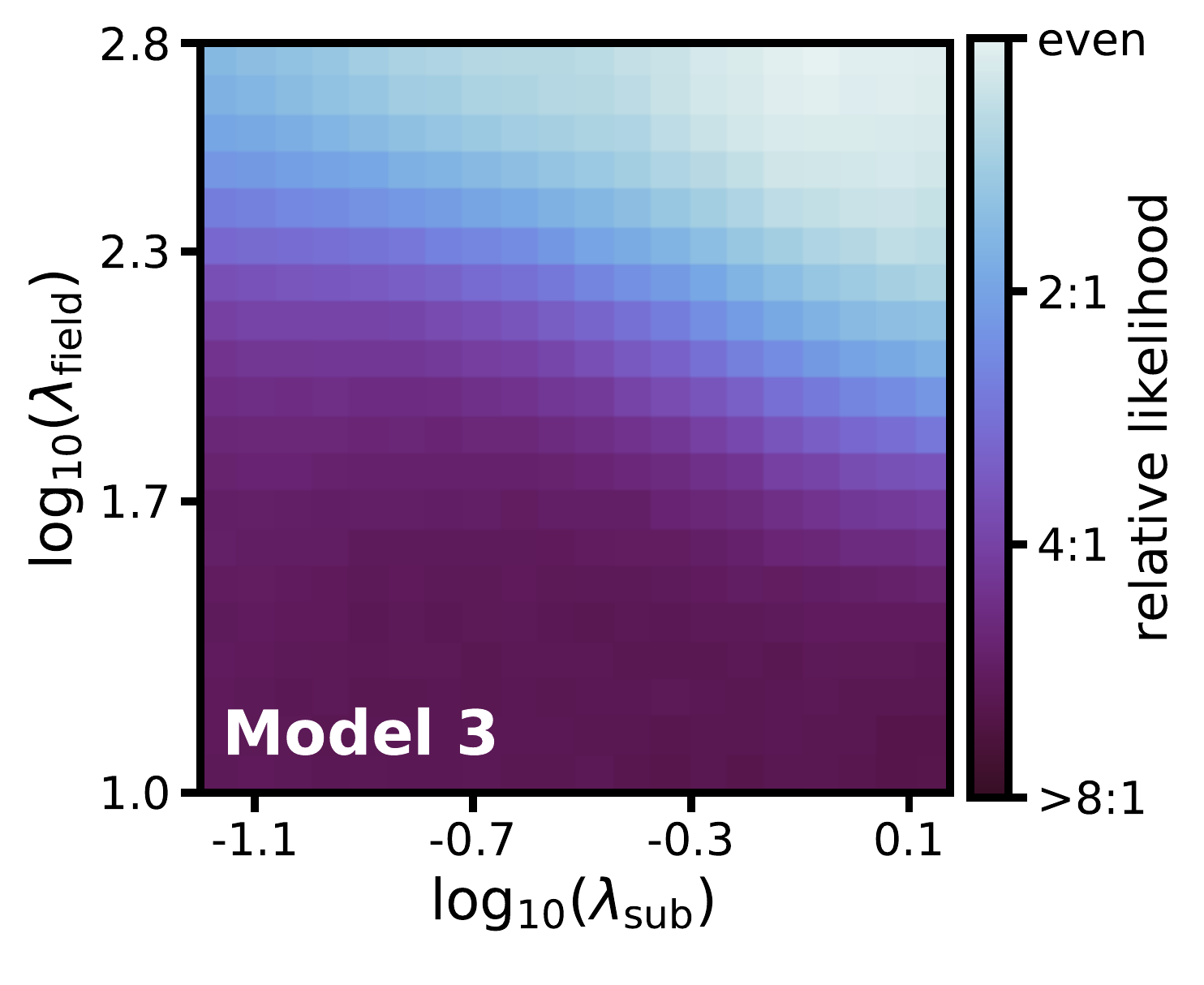}
		\includegraphics[trim=0.cm 0.cm 0.cm
		0.0cm,width=0.45\textwidth]{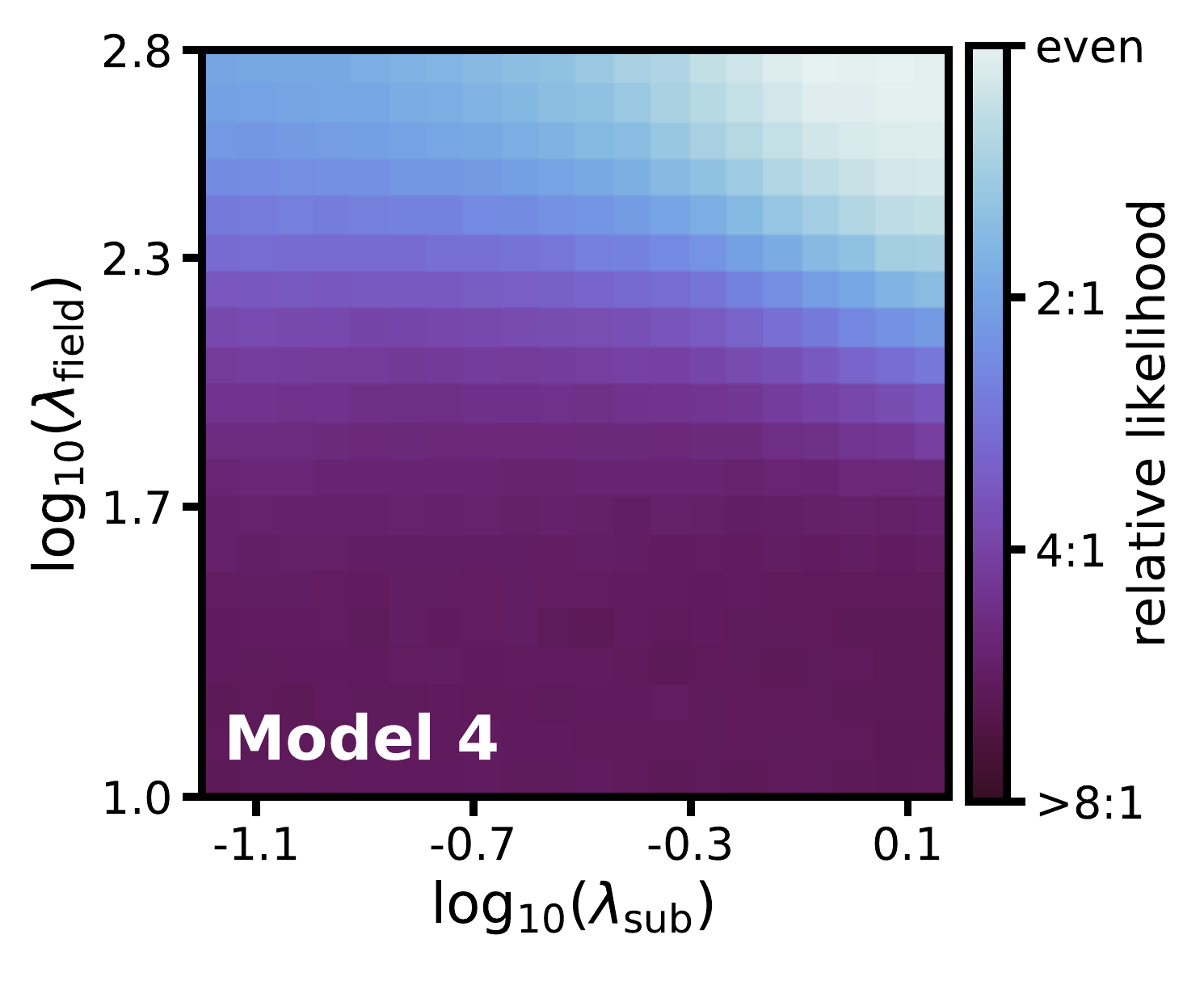}
		\includegraphics[trim=0.cm 0.cm 0.cm
		0.0cm,width=0.45\textwidth]{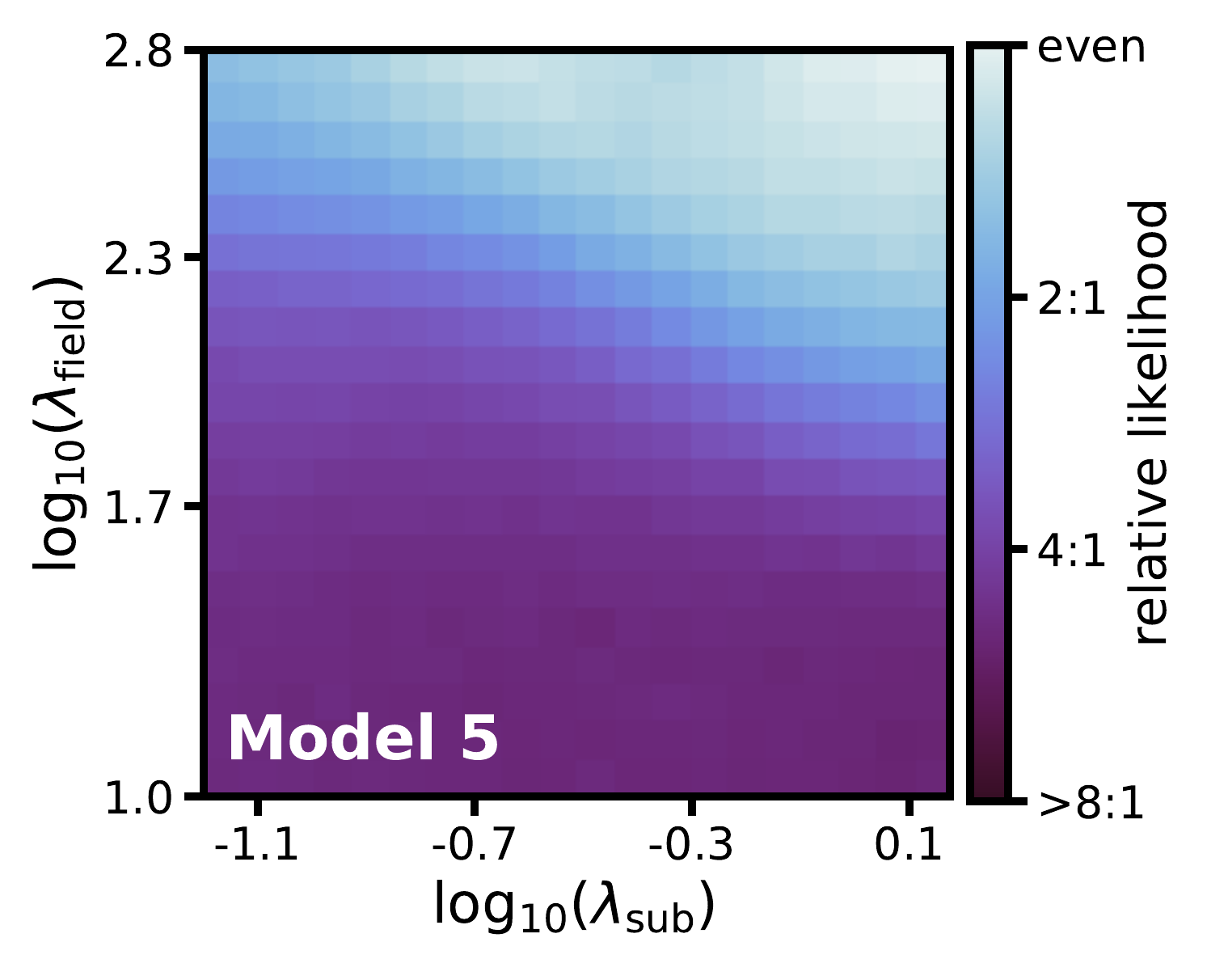}
		\caption{\label{fig:posteriors} The joint likelihood as shown in Figure\ref{fig:model1}, but computed for Models 2-5.  }
	\end{figure*} 
	
	By parameterizing in terms of the fraction of collapsed halos ${\bf{q}}$, we decouple our analysis of the data from assumptions related to how core collapse occurs in low-mass halos. In addition to several advantages from a computational perspective (see Appendix \ref{ssec:inf}), parameterizing in terms of ${\bf{q}}$ allows us to constrain a variety of SIDM models with a single inference, and implement any model for structure formation that predicts how a given cross section causes core collapse in low-mass halos. For this work, we use the structure formation model introduced in the previous section, defined by the vector of hyper-parameters  ${\bf{p}}\equiv \left(\lambda_{\rm{sub}}, \lambda_{\rm{field}}, s_{\rm{sub}}, s_{\rm{field}}\right)$. Writing this model as $p\left({\bf{q}} | \sigma_V, {\bf{p}} \right)$, we recast the likelihood function in terms of ${\bf{p}}$, ${\bf{v}}$ and $\sigma_V$
	\begin{equation}
		\centering
		\mathcal{L}\left({\bf{d}_n} | {\bf{p}}, {\bf{v}}, \sigma_V \right)=\int \mathcal{L}\left({\bf{d}_n} | {\bf{q}},{\bf{v}} \right)  p\left({\bf{q}} | {\bf{p}}, \sigma_V \right) d {\bf{q}},
	\end{equation}
	and use a Monte Carlo method to perform the integral. We generate random samples of ${\bf{p}}$ and ${\bf{v}}$, use the model to $p\left({\bf{q}} | {\bf{p}}, \sigma_V \right)$ to predict the fraction of collapsed halos (${\bf{q}}$) for a given cross section ($\sigma_V$), and then assign each draw of ${\bf{p}}$ and ${\bf{v}}$ an importance weight determined by the likelihood function $\mathcal{L}\left({\bf{d}_n} | {\bf{q}},{\bf{v}}\right)$.  Finally, we obtain the joint likelihood function for the full dataset as a product of the likelihoods computed for each lens
	\begin{equation}
		\mathcal{L}\left({\bf{D}} | {\bf{p}}, {\bf{v}},\sigma_V \right) = \prod_{n=1}^{11} \mathcal{L}\left({\bf{d}_n} | {\bf{p}}, {\bf{v}},\sigma_V \right),
	\end{equation}
	from which we can derive a posterior probability density $p\left({\bf{p}}, {\bf{v}}|\sigma_V, {\bf{D}} \right) \propto \pi\left({\bf{p}}, {\bf{v}}\right) \mathcal{L}\left({\bf{D}} | {\bf{p}}, {\bf{v}},\sigma_V \right)$ by specifying a prior probability density $\pi\left({\bf{p}}, {\bf{v}} \right)$.
	
	Figure \ref{fig:model1} shows the posterior distribution $p\left(\lambda_{\rm{sub}}, \lambda_{\rm{field}}| \sigma_V, {\bf{D}}\right)$ computed for Model 1, and Figure \ref{fig:posteriors} shows the same result for Models 2-5. Figure \ref{fig:model1} includes several annotations related to the physical interpretation of the joint distribution $p\left(\lambda_{\rm{sub}}, \lambda_{\rm{field}}| \sigma_V, {\bf{D}}\right)$, which we discuss in detail in the next sub-section. We display each probability density after marginalizing over ${\bf{v}}$, $s_{\rm{sub}}$, and $s_{\rm{field}}$. When marginalizing over ${\bf{v}}$ (the parameters that describe the logarithmic slope and amplitude of the subhalo and field halo mass functions) we allow for $20\%$ uncertainty in the amplitude of the field halo mass function, and vary the logarithmic slope of the subhalo mass function between $-1.95$ and $-1.85$, spanning the range of theoretical uncertainty predicted by N-body simulations \cite{Springel++08,Fiacconi++16}. We assume uniform priors on $s_{\rm{sub}}$ and $s_{\rm{field}}$ between $0.25 - 1.0 \ \rm{Gyr}$, and a uniform prior on the amplitude of the subhalo mass function that spans the range of theoretical uncertainty for this quantity. We show how our results change with different priors assigned to these parameters in Appendix \ref{ssec:supcollapsescatter} and \ref{ssec:supSHMF}.
	
	\subsection{How should one interpret constraints on $\lambda_{\rm{sub}}$ and $\lambda_{\rm{field}}$?}
	By parameterizing in terms of $\lambda_{\rm{sub}}$ and $\lambda_{\rm{field}}$, we propose that the relevant physics for a strong-lensing analysis of SIDM amounts to mapping the various physical processes relevant for the problem into these two collapse timescales, and their (possible) covariance with other parameters of interest. We emphasize this point through the annotations made on the x and y-axis of Figure \ref{fig:model1}. As shown in the figure, from both idealized and cosmological simulations of SIDM halo evolution, we expect $\lambda_{\rm{field}}$ to be an $\mathcal{O}\left(10^2\right)$ number in elastic SIDM \citep{Nishikawa++20,Outmezguine++22,ShengqiYang++22}, while in models with inelastic scattering $\lambda_{\rm{field}}$ can be an order of magnitude smaller \citep{Essig++19}. While other parameters relevant for our analysis, particularly the amplitude of the subhalo mass function, can also depend on the SIDM cross section, the effect of dark self-interactions on these quantities enters at second order in the likelihood function relative to the range of collapse timescales shown in Figures \ref{fig:model1} and \ref{fig:posteriors} (see Appendix \ref{ssec:supSHMF}). 
	
	For a given value of $\lambda_{\rm{field}}$, moving along the x-axis (i.e. varying $\lambda_{\rm{sub}}$) explores different possible realities for how tidal stripping and evaporation can accelerate or delay the onset of core collapse in subhalos relative to field halos. On the far left, tidal stripping accelerates core collapse in subhalos relative to field halos by over a factor of ten, while higher $\lambda_{\rm{sub}}$ corresponds to a situation in which scattering between subhalo particles and host halo particles counters the effects of tidal stripping and delays collapse. The degree to which tidal stripping or subhalo-host interactions dominate the collapse process depends on the amplitude of $\sigma_{V}$ at a speed comparable to the central velocity dispersion of the host. Thus, it is likely that $\lambda_{\rm{sub}}$ is a function of the cross section amplitude $\sigma_{V}$ evaluated at $v \sim 200 \rm{km} \ \rm{s^{-1}}$, a typical velocity dispersion for a galaxy. From the existing suite of SIDM simulations that examine these processes \cite{Vogelsberger++19,Nishikawa++20,Nadler++21,Correa++22,Zeng++22,Yang++22}, however, it is remains somewhat unclear how exactly to implement the dependence of $\lambda_{\rm{sub}}$ on $\sigma_{V}$, so we have left it as a free parameter.
	
	With the point of view that $\lambda_{\rm{sub}}$ and $\lambda_{\rm{field}}$ encode the relevant physics and meaningful information related to core collapse and SIDM that we can extract from the data, we can use inferences on these parameters to draw conclusions regarding the nature of dark matter. First, a clear detection of a particular combination of $\lambda_{\rm{field}}$ and $\lambda_{\rm{sub}}$ means the data requires the existence of core-collapsed halos. This would constitute strong evidence in support of SIDM, particularly if the inferred collapse timescales match the collapse timescales predicted by N-body simulations or fluid models for the same $\sigma_V$. Among the five benchmark models, we see no evidence for a preferred collapse timescale, and the data disfavors scenarios in which large ($> 80 \%$) fractions of halos collapse for each benchmark model we consider.
	
	Conversely, an SIDM model remains viable only if the data does not rule out the collapse timescale associated with it. To give a concrete example, the region of parameter space with $\lambda_{\rm{field}} < 30$ corresponds core collapse accelerated by a factor of ten, relative to the predictions of SIDM with purely elastic scattering \citep{Nishikawa++20,Outmezguine++22,Yang++22}. Shortened collapse timescales such as these can result from inelastic or dissipative self-interactions \citep{Essig++19,Huo++20}, which can accelerate collapse by a factor up to 1,000 if the specific energy loss per collision matches the square of the velocity dispersion of the (sub)halos. While the particular form of the interaction potentials we consider suppresses dissipative processes \footnote{The ratio between the dissipative cross section and the elastic cross section is about $\sigma(\chi \chi \to \chi \chi \phi)/\sigma(\chi \chi \to \chi \chi) \sim {\alpha_\chi}/{(4\pi)}$ for Yukawa models, or $\sim 10^{-4}$ for the benchmarks we considered. In these cases, elastic scattering dominates the heat conduction \citep{Essig++19}.}, given that our data disfavors scenarios in which a majority of halos collapse, our data would also disfavor a cross section strength exceeding $100\, \rm{cm^2}\,\rm{g^{-1}}$ below $50\, \rm{km} \, \rm{s^{-1}}$ with an efficient energy loss channel that causes halos to collapse in even greater numbers.
	
	As we assign equal prior probability to $\log_{10}\lambda_{\rm{sub}}$ and $\log_{10}\lambda_{\rm{field}}$, the posterior distributions shown in Figure \ref{fig:posteriors} vary in direct proportion with the likelihood $\mathcal{L}\left(\bf{D}| \lambda_{\rm{sub}},\lambda_{\rm{field}}\right)$. Table \ref{tab:models} summarizes the likelihood relative to CDM for various representative combinations of $\lambda_{\rm{sub}}$ and $\lambda_{\rm{field}}$. We define the likelihood of CDM as the mean probability of points in the region $\lambda_{\rm{field}} > 300$ and $\lambda_{\rm{sub}}>1$ (such that no halos collapse). At fixed $\lambda_{\rm{field}}$, different $\lambda_{\rm{sub}}$ correspond to different physical models for core collapse in subhalos. For example, while tidal stripping can accelerate core collapse \citep{Sameie++20,Nishikawa++20}, host halo and subhalo particle interactions, particularly evaporation and tidal heating \citep{Zeng++22,Vogelsberger++19}, can delay collapse. Ram pressure stripping can also lower the amplitude of the subhalo mass function \citep{Nadler++20}, and all of these effects in combination could increase the scatter in collapse times for subhalos, relative to field halos. We investigate these effects in Appendices \ref{ssec:supcollapsescatter} and \ref{ssec:supSHMF} by adding informative priors on the subhalo mass function amplitude and scatter in collapse timescales, and show that they cause $10-30 \%$ differences in the relative likelihoods quoted in Table \ref{tab:models}. In addition to these processes that affect the subhalo mass function amplitude and the scatter in halo collapse times, at fixed $\lambda_{\rm{field}}$, increasing $\lambda_{\rm{sub}}$ significantly impacts the resulting likelihood, underscoring the importance of developing sound theoretical understanding for subhalo evolution in velocity-dependent SIDM. 
	\begin{table*}
		\caption{\label{tab:models} The relative likelihoods of CDM to each SIDM model under different assumptions for the collapse timescales $\lambda_{\rm{sub}}$ and $\lambda_{\rm{field}}$.}
		\begin{ruledtabular}
			\begin{tabular}{ccccccc}
				& &\multicolumn{4}{c}{Relative likelihood (CDM:SIDM)} &  \\ \\
				& $\left(\lambda_{\rm{sub}},\lambda_{\rm{field}}\right)$  &  $\left(\lambda_{\rm{sub}},\lambda_{\rm{field}}\right)$ & $\left(\lambda_{\rm{sub}},\lambda_{\rm{field}}\right)$ & $\left(\lambda_{\rm{sub}},\lambda_{\rm{field}}\right)$ & $\left(\lambda_{\rm{sub}},\lambda_{\rm{field}}\right)$ & $\left(\lambda_{\rm{sub}},\lambda_{\rm{field}}\right)$\\
				& $\left(0.1, 300\right)$& $\left(0.5, 300\right)$ & $\left(0.1, 30\right)$& $\left(0.5, 30\right)$& $\left(1.0, 30\right)$& $\left(0.5, 100\right)$ \\
				\hline \\
				Model 1&  1:1&  1:1& 6:1& 5:1&  4:1& 2:1 \\
				Model 2&  2:1&  1:1&  7:1& 6:1&  6:1& 3:1 \\
				Model 3& 1:1&  1:1&  6:1& 6:1&  5:1&  3:1\\
				Model 4&  2:1&  1:1& 6:1& 6:1&  6:1&  4:1\\
				Model 5& 1:1& 1:1&  4:1& 5:1&  4:1&  3:1\\
			\end{tabular}
		\end{ruledtabular}
	\end{table*} 
	
	\section{Summary and discussion}
	\label{sec:discussion}
	We develop a method to analyze self-interacting dark matter with quadruply-imaged quasars, and apply it to analyze five benchmarks models of a velocity-dependent self-interaction cross section. Using the collective impact of many halos on the magnifications of unresolved lensed images, we infer the fraction of core-collapsed subhalos and field halos in eleven lens systems, and recast this inference as constraints on each cross section using a structure formation model that predicts the fraction of collapsed subhalos and field halos as a function of halo mass for a given cross section. We derive constraints on the core-collapse timescales for subhalos and field halos, and compute the relative likelihood of each cross section to CDM under different assumptions regarding how core collapse proceeds in low-mass halos. The most significant likelihood penalties, between 5:1 to 7:1, apply to scenarios in which a majority of all halos collapse. This can occur, in particular, with dissipative self-interactions \citep{Essig++19,Huo++20}.
	
	A promising avenue for future research involves combining inferences from strong lensing with those from independent probes, breaking covariances inherent to methods when applied individually \cite[e.g.][]{Nadler++21}. An inference of the subhalo mass function amplitude from stellar streams \citep{Banik++21}, for example, is likely not as sensitive to the internal structure of halos as lensing \citep{Bovy++17,Banik++21b}, providing an independent handle on the overall number of subhalos. In the case of dwarf galaxies, resonances in the cross section could cause core collapse within a Hubble time if the velocity dispersion of the galaxy coincides with the velocity scale of the resonance. This process would directly contribute to the diversity of galactic rotation curves. Finally, combining inferences from lensing with analyses of dwarf galaxies \cite[e.g.][]{Kaplinghat++16,Nadler++20b,Correa++20,Kim++21,Dekker++21,Ebisu++22,Silverman++22,Shen++22} would leverage information from an extended range of scales to constrain velocity-dependent cross sections. 
	
	While we have focused on quadruply-imaged quasars, galaxy-scale strong lens systems with extended images and arcs that partially encircle the main deflector can also reveal the presence of low-mass dark matter (sub)halos \cite{Vegetti++12,Vegetti++14,Hezaveh++16,Birrer++17,CaganSengul++21,Minor++21,He++22,Despali++22}. The minimum halo mass accessible with these types of lens systems, and therefore the velocity scale where observations can probe the SIDM cross section, exceeds $10^9 M_{\odot}$ with existing data \cite{Amorisco++22,Despali++22}, roughly two orders of magnitude in halo mass larger than the halo masses impacting image flux ratios. However, these estimations of the minimum halo mass sensitivity for systems with extended images assume NFW profiles. Based on the deformed critical curves in Figure \ref{fig:massdistributions}, a population of core-collapsed subhalos and field halos with masses below $10^9 M_{\odot}$ could possibly imprint measurable signatures on the surface brightness of lensed arcs. Detection or non-detection of this lensing signature would complement constraints from flux ratios at lower halo masses. In addition, searches for lensed radio sources with image separates of $\sim 1 $ m.a.s. could reveal the presence of core-collapsed halos in the field \cite{Casadio++21,Loudas++22}, and probe similar mass and velocity scales as those explored in this work.
	
	\citet{Gilman++21} (hereafter G21) simulate core collapse using an approximate analytic model for the scattering cross section for a weak potential ($\alpha_{\chi} m_{\chi} / m_{\phi} \ll 1$), and make forecasts for strong lensing constraints on SIDM models. G21 conclude that narrow-line flux ratios, like the ones used in this analysis, provide only limiting constraining power over SIDM. However, the cross sections examined by G21 reach a maximum amplitude of $50 \ \rm{cm^2} \ \rm{g^{-1}}$, much lower than the $100-1000 \ \rm{cm^2} \ \rm{g^{-1}}$ cross section strengths below $30 \ \rm{km} \ \rm{s^{-1}}$ examined in this work. Thus, G21 did not consider cross section strengths large enough to allow field halos to core collapse in significant numbers. Our results are broadly consistent with the forecasts by G21, as Table \ref{tab:models} shows that with only eleven lenses, significant relative likelihood penalties apply only to models in which a significant fraction of both field halos and subhalos core collapse. We have argued that both dissipative self-interactions, and resonances in the cross section that significantly increase its amplitude, can trigger collapse in field halos.
	
	With forthcoming data from the James Webb Space Telescope (JWST), the number of lens systems suitable for the analysis we present will triple (JWST GO-2046 \citet{NierenbergJWST}). In addition to expanding the sample size, JWST will measure image fluxes in the mid-infrared, which emanates from a more compact region around the background quasar than the nuclear narrow-line emission we analyze. As shown by G21, the more compact source increases sensitivity to low-mass core-collapsed halos. Improved modeling of SIDM halo profiles \citep{Zeng++22,Jiang++22,Yang++22,ShengqiYang++22,Correa++22,Bhattacharyya++22}, together with these new data, will lead to more stringent constraints on SIDM cross sections.
	
	\begin{acknowledgments}
		We sincerely thank Andrew Benson, Francis-Yan Cyr-Racine, Felix Kahlh\"{o}fer, Annika Peter, Tommaso Treu, Shenqi Yang, Hai-Bo Yu, and Carton Zeng for encouragement and useful discussions. We also than the anonymous referee for feedback that improved the quality of the paper. 
		
		DG was partially supported by a HQP grant from the McDonald Institute (reference number HQP 2019-4-2), and a Schmidt Futures AI in Science Fellowship at the University of Toronto. DG and JB acknowledge financial support from NSERC (funding reference number RGPIN-2020-04712). YZ is supported by the Kavli Institute for Cosmological Physics at the University of Chicago through an endowment from the Kavli Foundation and its founder Fred Kavli. The results presented in this work made use of data collected through HST programs 15177 and 13732.
		
		We performed computations on the Niagara supercomputer at the SciNet HPC Consortium \citep{Loken2010,Ponce2019}. SciNet is funded by: the Canada Foundation for Innovation; the Government of Ontario; Ontario Research Fund -
		Research Excellence; and the University of Toronto. We also used computational and storage services associated with
		the Hoffman2 Shared Cluster provided by the UCLA Institute for Digital Research and Education’s Research Technology Group.
		
		This work made use of the following software packages: {\tt{pyHalo}} (https://github.com/dangilman/pyHalo) for generating realizations of dark matter halos for lensing simulations; {\tt{lenstronomy}} (https://github.com/sibirrer/lenstronomy) \citep{BirrerAmara++18,Birrer++21}, for performing multi-plane ray-tracing computations; {\tt{colossus}} (https://bdiemer.bitbucket.io/colossus/) \citep{Diemer18}, for computations involving the halo mass function and the concentration-mass relation. 
	\end{acknowledgments}
	
	\bibliography{bibliography}

\begin{thebibliography}{133}%
\makeatletter
\providecommand \@ifxundefined [1]{%
 \@ifx{#1\undefined}
}%
\providecommand \@ifnum [1]{%
 \ifnum #1\expandafter \@firstoftwo
 \else \expandafter \@secondoftwo
 \fi
}%
\providecommand \@ifx [1]{%
 \ifx #1\expandafter \@firstoftwo
 \else \expandafter \@secondoftwo
 \fi
}%
\providecommand \natexlab [1]{#1}%
\providecommand \enquote  [1]{``#1''}%
\providecommand \bibnamefont  [1]{#1}%
\providecommand \bibfnamefont [1]{#1}%
\providecommand \citenamefont [1]{#1}%
\providecommand \href@noop [0]{\@secondoftwo}%
\providecommand \href [0]{\begingroup \@sanitize@url \@href}%
\providecommand \@href[1]{\@@startlink{#1}\@@href}%
\providecommand \@@href[1]{\endgroup#1\@@endlink}%
\providecommand \@sanitize@url [0]{\catcode `\\12\catcode `\$12\catcode
  `\&12\catcode `\#12\catcode `\^12\catcode `\_12\catcode `\%12\relax}%
\providecommand \@@startlink[1]{}%
\providecommand \@@endlink[0]{}%
\providecommand \url  [0]{\begingroup\@sanitize@url \@url }%
\providecommand \@url [1]{\endgroup\@href {#1}{\urlprefix }}%
\providecommand \urlprefix  [0]{URL }%
\providecommand \Eprint [0]{\href }%
\providecommand \doibase [0]{https://doi.org/}%
\providecommand \selectlanguage [0]{\@gobble}%
\providecommand \bibinfo  [0]{\@secondoftwo}%
\providecommand \bibfield  [0]{\@secondoftwo}%
\providecommand \translation [1]{[#1]}%
\providecommand \BibitemOpen [0]{}%
\providecommand \bibitemStop [0]{}%
\providecommand \bibitemNoStop [0]{.\EOS\space}%
\providecommand \EOS [0]{\spacefactor3000\relax}%
\providecommand \BibitemShut  [1]{\csname bibitem#1\endcsname}%
\let\auto@bib@innerbib\@empty
\bibitem [{\citenamefont {{Spergel}}\ and\ \citenamefont
  {{Steinhardt}}(2000)}]{Spergel++00}%
  \BibitemOpen
  \bibfield  {author} {\bibinfo {author} {\bibfnamefont {D.~N.}\ \bibnamefont
  {{Spergel}}}\ and\ \bibinfo {author} {\bibfnamefont {P.~J.}\ \bibnamefont
  {{Steinhardt}}},\ }\bibfield  {title} {\bibinfo {title} {{Observational
  Evidence for Self-Interacting Cold Dark Matter}},\ }\href
  {https://doi.org/10.1103/PhysRevLett.84.3760} {\bibfield  {journal} {\bibinfo
   {journal} {\prl}\ }\textbf {\bibinfo {volume} {84}},\ \bibinfo {pages}
  {3760} (\bibinfo {year} {2000})},\ \Eprint
  {https://arxiv.org/abs/astro-ph/9909386} {arXiv:astro-ph/9909386 [astro-ph]}
  \BibitemShut {NoStop}%
\bibitem [{\citenamefont {{Ahn}}\ and\ \citenamefont
  {{Shapiro}}(2005)}]{AhnShapiro05}%
  \BibitemOpen
  \bibfield  {author} {\bibinfo {author} {\bibfnamefont {K.}~\bibnamefont
  {{Ahn}}}\ and\ \bibinfo {author} {\bibfnamefont {P.~R.}\ \bibnamefont
  {{Shapiro}}},\ }\bibfield  {title} {\bibinfo {title} {{Formation and
  evolution of self-interacting dark matter haloes}},\ }\href
  {https://doi.org/10.1111/j.1365-2966.2005.09492.x} {\bibfield  {journal}
  {\bibinfo  {journal} {\mnras}\ }\textbf {\bibinfo {volume} {363}},\ \bibinfo
  {pages} {1092} (\bibinfo {year} {2005})},\ \Eprint
  {https://arxiv.org/abs/astro-ph/0412169} {arXiv:astro-ph/0412169 [astro-ph]}
  \BibitemShut {NoStop}%
\bibitem [{\citenamefont {{Rocha}}\ \emph {et~al.}(2013)\citenamefont
  {{Rocha}}, \citenamefont {{Peter}}, \citenamefont {{Bullock}}, \citenamefont
  {{Kaplinghat}}, \citenamefont {{Garrison-Kimmel}}, \citenamefont
  {{O{\~n}orbe}},\ and\ \citenamefont {{Moustakas}}}]{Rocha++13}%
  \BibitemOpen
  \bibfield  {author} {\bibinfo {author} {\bibfnamefont {M.}~\bibnamefont
  {{Rocha}}}, \bibinfo {author} {\bibfnamefont {A.~H.~G.}\ \bibnamefont
  {{Peter}}}, \bibinfo {author} {\bibfnamefont {J.~S.}\ \bibnamefont
  {{Bullock}}}, \bibinfo {author} {\bibfnamefont {M.}~\bibnamefont
  {{Kaplinghat}}}, \bibinfo {author} {\bibfnamefont {S.}~\bibnamefont
  {{Garrison-Kimmel}}}, \bibinfo {author} {\bibfnamefont {J.}~\bibnamefont
  {{O{\~n}orbe}}},\ and\ \bibinfo {author} {\bibfnamefont {L.~A.}\ \bibnamefont
  {{Moustakas}}},\ }\bibfield  {title} {\bibinfo {title} {{Cosmological
  simulations with self-interacting dark matter - I. Constant-density cores and
  substructure}},\ }\href {https://doi.org/10.1093/mnras/sts514} {\bibfield
  {journal} {\bibinfo  {journal} {\mnras}\ }\textbf {\bibinfo {volume} {430}},\
  \bibinfo {pages} {81} (\bibinfo {year} {2013})},\ \Eprint
  {https://arxiv.org/abs/1208.3025} {arXiv:1208.3025 [astro-ph.CO]}
  \BibitemShut {NoStop}%
\bibitem [{\citenamefont {{Vogelsberger}}\ \emph {et~al.}(2012)\citenamefont
  {{Vogelsberger}}, \citenamefont {{Zavala}},\ and\ \citenamefont
  {{Loeb}}}]{Vogelsberger++12}%
  \BibitemOpen
  \bibfield  {author} {\bibinfo {author} {\bibfnamefont {M.}~\bibnamefont
  {{Vogelsberger}}}, \bibinfo {author} {\bibfnamefont {J.}~\bibnamefont
  {{Zavala}}},\ and\ \bibinfo {author} {\bibfnamefont {A.}~\bibnamefont
  {{Loeb}}},\ }\bibfield  {title} {\bibinfo {title} {{Subhaloes in
  self-interacting galactic dark matter haloes}},\ }\href
  {https://doi.org/10.1111/j.1365-2966.2012.21182.x} {\bibfield  {journal}
  {\bibinfo  {journal} {\mnras}\ }\textbf {\bibinfo {volume} {423}},\ \bibinfo
  {pages} {3740} (\bibinfo {year} {2012})},\ \Eprint
  {https://arxiv.org/abs/1201.5892} {arXiv:1201.5892 [astro-ph.CO]}
  \BibitemShut {NoStop}%
\bibitem [{\citenamefont {{Elbert}}\ \emph {et~al.}(2015)\citenamefont
  {{Elbert}}, \citenamefont {{Bullock}}, \citenamefont {{Garrison-Kimmel}},
  \citenamefont {{Rocha}}, \citenamefont {{O{\~n}orbe}},\ and\ \citenamefont
  {{Peter}}}]{Elbert++15}%
  \BibitemOpen
  \bibfield  {author} {\bibinfo {author} {\bibfnamefont {O.~D.}\ \bibnamefont
  {{Elbert}}}, \bibinfo {author} {\bibfnamefont {J.~S.}\ \bibnamefont
  {{Bullock}}}, \bibinfo {author} {\bibfnamefont {S.}~\bibnamefont
  {{Garrison-Kimmel}}}, \bibinfo {author} {\bibfnamefont {M.}~\bibnamefont
  {{Rocha}}}, \bibinfo {author} {\bibfnamefont {J.}~\bibnamefont
  {{O{\~n}orbe}}},\ and\ \bibinfo {author} {\bibfnamefont {A.~H.~G.}\
  \bibnamefont {{Peter}}},\ }\bibfield  {title} {\bibinfo {title} {{Core
  formation in dwarf haloes with self-interacting dark matter: no fine-tuning
  necessary}},\ }\href {https://doi.org/10.1093/mnras/stv1470} {\bibfield
  {journal} {\bibinfo  {journal} {\mnras}\ }\textbf {\bibinfo {volume} {453}},\
  \bibinfo {pages} {29} (\bibinfo {year} {2015})},\ \Eprint
  {https://arxiv.org/abs/1412.1477} {arXiv:1412.1477 [astro-ph.GA]}
  \BibitemShut {NoStop}%
\bibitem [{\citenamefont {{Lynden-Bell}}\ and\ \citenamefont
  {{Wood}}(1968)}]{LyndenBellWood}%
  \BibitemOpen
  \bibfield  {author} {\bibinfo {author} {\bibfnamefont {D.}~\bibnamefont
  {{Lynden-Bell}}}\ and\ \bibinfo {author} {\bibfnamefont {R.}~\bibnamefont
  {{Wood}}},\ }\bibfield  {title} {\bibinfo {title} {{The gravo-thermal
  catastrophe in isothermal spheres and the onset of red-giant structure for
  stellar systems}},\ }\href {https://doi.org/10.1093/mnras/138.4.495}
  {\bibfield  {journal} {\bibinfo  {journal} {\mnras}\ }\textbf {\bibinfo
  {volume} {138}},\ \bibinfo {pages} {495} (\bibinfo {year}
  {1968})}\BibitemShut {NoStop}%
\bibitem [{\citenamefont {{Lynden-Bell}}\ and\ \citenamefont
  {{Eggleton}}(1980)}]{LyndenBellEggleton80}%
  \BibitemOpen
  \bibfield  {author} {\bibinfo {author} {\bibfnamefont {D.}~\bibnamefont
  {{Lynden-Bell}}}\ and\ \bibinfo {author} {\bibfnamefont {P.~P.}\ \bibnamefont
  {{Eggleton}}},\ }\bibfield  {title} {\bibinfo {title} {{On the consequences
  of the gravothermal catastrophe}},\ }\href
  {https://doi.org/10.1093/mnras/191.3.483} {\bibfield  {journal} {\bibinfo
  {journal} {\mnras}\ }\textbf {\bibinfo {volume} {191}},\ \bibinfo {pages}
  {483} (\bibinfo {year} {1980})}\BibitemShut {NoStop}%
\bibitem [{\citenamefont {{Balberg}}\ \emph {et~al.}(2002)\citenamefont
  {{Balberg}}, \citenamefont {{Shapiro}},\ and\ \citenamefont
  {{Inagaki}}}]{Balberg++02}%
  \BibitemOpen
  \bibfield  {author} {\bibinfo {author} {\bibfnamefont {S.}~\bibnamefont
  {{Balberg}}}, \bibinfo {author} {\bibfnamefont {S.~L.}\ \bibnamefont
  {{Shapiro}}},\ and\ \bibinfo {author} {\bibfnamefont {S.}~\bibnamefont
  {{Inagaki}}},\ }\bibfield  {title} {\bibinfo {title} {{Self-Interacting Dark
  Matter Halos and the Gravothermal Catastrophe}},\ }\href
  {https://doi.org/10.1086/339038} {\bibfield  {journal} {\bibinfo  {journal}
  {\apj}\ }\textbf {\bibinfo {volume} {568}},\ \bibinfo {pages} {475} (\bibinfo
  {year} {2002})},\ \Eprint {https://arxiv.org/abs/astro-ph/0110561}
  {arXiv:astro-ph/0110561 [astro-ph]} \BibitemShut {NoStop}%
\bibitem [{\citenamefont {{Nishikawa}}\ \emph {et~al.}(2020)\citenamefont
  {{Nishikawa}}, \citenamefont {{Boddy}},\ and\ \citenamefont
  {{Kaplinghat}}}]{Nishikawa++20}%
  \BibitemOpen
  \bibfield  {author} {\bibinfo {author} {\bibfnamefont {H.}~\bibnamefont
  {{Nishikawa}}}, \bibinfo {author} {\bibfnamefont {K.~K.}\ \bibnamefont
  {{Boddy}}},\ and\ \bibinfo {author} {\bibfnamefont {M.}~\bibnamefont
  {{Kaplinghat}}},\ }\bibfield  {title} {\bibinfo {title} {{Accelerated core
  collapse in tidally stripped self-interacting dark matter halos}},\ }\href
  {https://doi.org/10.1103/PhysRevD.101.063009} {\bibfield  {journal} {\bibinfo
   {journal} {\prd}\ }\textbf {\bibinfo {volume} {101}},\ \bibinfo {eid}
  {063009} (\bibinfo {year} {2020})},\ \Eprint
  {https://arxiv.org/abs/1901.00499} {arXiv:1901.00499 [astro-ph.GA]}
  \BibitemShut {NoStop}%
\bibitem [{\citenamefont {{Sameie}}\ \emph {et~al.}(2020)\citenamefont
  {{Sameie}}, \citenamefont {{Yu}}, \citenamefont {{Sales}}, \citenamefont
  {{Vogelsberger}},\ and\ \citenamefont {{Zavala}}}]{Sameie++20}%
  \BibitemOpen
  \bibfield  {author} {\bibinfo {author} {\bibfnamefont {O.}~\bibnamefont
  {{Sameie}}}, \bibinfo {author} {\bibfnamefont {H.-B.}\ \bibnamefont {{Yu}}},
  \bibinfo {author} {\bibfnamefont {L.~V.}\ \bibnamefont {{Sales}}}, \bibinfo
  {author} {\bibfnamefont {M.}~\bibnamefont {{Vogelsberger}}},\ and\ \bibinfo
  {author} {\bibfnamefont {J.}~\bibnamefont {{Zavala}}},\ }\bibfield  {title}
  {\bibinfo {title} {{Self-Interacting Dark Matter Subhalos in the Milky Way's
  Tides}},\ }\href {https://doi.org/10.1103/PhysRevLett.124.141102} {\bibfield
  {journal} {\bibinfo  {journal} {\prl}\ }\textbf {\bibinfo {volume} {124}},\
  \bibinfo {eid} {141102} (\bibinfo {year} {2020})},\ \Eprint
  {https://arxiv.org/abs/1904.07872} {arXiv:1904.07872 [astro-ph.GA]}
  \BibitemShut {NoStop}%
\bibitem [{\citenamefont {{Turner}}\ \emph {et~al.}(2021)\citenamefont
  {{Turner}}, \citenamefont {{Lovell}}, \citenamefont {{Zavala}},\ and\
  \citenamefont {{Vogelsberger}}}]{Turner++21}%
  \BibitemOpen
  \bibfield  {author} {\bibinfo {author} {\bibfnamefont {H.~C.}\ \bibnamefont
  {{Turner}}}, \bibinfo {author} {\bibfnamefont {M.~R.}\ \bibnamefont
  {{Lovell}}}, \bibinfo {author} {\bibfnamefont {J.}~\bibnamefont {{Zavala}}},\
  and\ \bibinfo {author} {\bibfnamefont {M.}~\bibnamefont {{Vogelsberger}}},\
  }\bibfield  {title} {\bibinfo {title} {{The onset of gravothermal core
  collapse in velocity-dependent self-interacting dark matter subhaloes}},\
  }\href {https://doi.org/10.1093/mnras/stab1725} {\bibfield  {journal}
  {\bibinfo  {journal} {\mnras}\ }\textbf {\bibinfo {volume} {505}},\ \bibinfo
  {pages} {5327} (\bibinfo {year} {2021})},\ \Eprint
  {https://arxiv.org/abs/2010.02924} {arXiv:2010.02924 [astro-ph.GA]}
  \BibitemShut {NoStop}%
\bibitem [{\citenamefont {{Correa}}(2021)}]{Correa++20}%
  \BibitemOpen
  \bibfield  {author} {\bibinfo {author} {\bibfnamefont {C.~A.}\ \bibnamefont
  {{Correa}}},\ }\bibfield  {title} {\bibinfo {title} {{Constraining
  velocity-dependent self-interacting dark matter with the Milky Way's dwarf
  spheroidal galaxies}},\ }\href {https://doi.org/10.1093/mnras/stab506}
  {\bibfield  {journal} {\bibinfo  {journal} {\mnras}\ }\textbf {\bibinfo
  {volume} {503}},\ \bibinfo {pages} {920} (\bibinfo {year} {2021})},\ \Eprint
  {https://arxiv.org/abs/2007.02958} {arXiv:2007.02958 [astro-ph.GA]}
  \BibitemShut {NoStop}%
\bibitem [{\citenamefont {{Correa}}\ \emph {et~al.}(2022)\citenamefont
  {{Correa}}, \citenamefont {{Schaller}}, \citenamefont {{Ploeckinger}},
  \citenamefont {{Anau Montel}}, \citenamefont {{Weniger}},\ and\ \citenamefont
  {{Ando}}}]{Correa++22}%
  \BibitemOpen
  \bibfield  {author} {\bibinfo {author} {\bibfnamefont {C.~A.}\ \bibnamefont
  {{Correa}}}, \bibinfo {author} {\bibfnamefont {M.}~\bibnamefont
  {{Schaller}}}, \bibinfo {author} {\bibfnamefont {S.}~\bibnamefont
  {{Ploeckinger}}}, \bibinfo {author} {\bibfnamefont {N.}~\bibnamefont {{Anau
  Montel}}}, \bibinfo {author} {\bibfnamefont {C.}~\bibnamefont {{Weniger}}},\
  and\ \bibinfo {author} {\bibfnamefont {S.}~\bibnamefont {{Ando}}},\
  }\bibfield  {title} {\bibinfo {title} {{TangoSIDM: Tantalizing models of
  Self-Interacting Dark Matter}},\ }\href@noop {} {\bibfield  {journal}
  {\bibinfo  {journal} {arXiv e-prints}\ ,\ \bibinfo {eid} {arXiv:2206.11298}}
  (\bibinfo {year} {2022})},\ \Eprint {https://arxiv.org/abs/2206.11298}
  {arXiv:2206.11298 [astro-ph.GA]} \BibitemShut {NoStop}%
\bibitem [{\citenamefont {{Oman}}\ \emph {et~al.}(2015)\citenamefont {{Oman}},
  \citenamefont {{Navarro}}, \citenamefont {{Fattahi}}, \citenamefont
  {{Frenk}}, \citenamefont {{Sawala}}, \citenamefont {{White}}, \citenamefont
  {{Bower}}, \citenamefont {{Crain}}, \citenamefont {{Furlong}}, \citenamefont
  {{Schaller}}, \citenamefont {{Schaye}},\ and\ \citenamefont
  {{Theuns}}}]{Oman++15}%
  \BibitemOpen
  \bibfield  {author} {\bibinfo {author} {\bibfnamefont {K.~A.}\ \bibnamefont
  {{Oman}}}, \bibinfo {author} {\bibfnamefont {J.~F.}\ \bibnamefont
  {{Navarro}}}, \bibinfo {author} {\bibfnamefont {A.}~\bibnamefont
  {{Fattahi}}}, \bibinfo {author} {\bibfnamefont {C.~S.}\ \bibnamefont
  {{Frenk}}}, \bibinfo {author} {\bibfnamefont {T.}~\bibnamefont {{Sawala}}},
  \bibinfo {author} {\bibfnamefont {S.~D.~M.}\ \bibnamefont {{White}}},
  \bibinfo {author} {\bibfnamefont {R.}~\bibnamefont {{Bower}}}, \bibinfo
  {author} {\bibfnamefont {R.~A.}\ \bibnamefont {{Crain}}}, \bibinfo {author}
  {\bibfnamefont {M.}~\bibnamefont {{Furlong}}}, \bibinfo {author}
  {\bibfnamefont {M.}~\bibnamefont {{Schaller}}}, \bibinfo {author}
  {\bibfnamefont {J.}~\bibnamefont {{Schaye}}},\ and\ \bibinfo {author}
  {\bibfnamefont {T.}~\bibnamefont {{Theuns}}},\ }\bibfield  {title} {\bibinfo
  {title} {{The unexpected diversity of dwarf galaxy rotation curves}},\ }\href
  {https://doi.org/10.1093/mnras/stv1504} {\bibfield  {journal} {\bibinfo
  {journal} {\mnras}\ }\textbf {\bibinfo {volume} {452}},\ \bibinfo {pages}
  {3650} (\bibinfo {year} {2015})},\ \Eprint {https://arxiv.org/abs/1504.01437}
  {arXiv:1504.01437 [astro-ph.GA]} \BibitemShut {NoStop}%
\bibitem [{\citenamefont {{Kaplinghat}}\ \emph {et~al.}(2016)\citenamefont
  {{Kaplinghat}}, \citenamefont {{Tulin}},\ and\ \citenamefont
  {{Yu}}}]{Kaplinghat++16}%
  \BibitemOpen
  \bibfield  {author} {\bibinfo {author} {\bibfnamefont {M.}~\bibnamefont
  {{Kaplinghat}}}, \bibinfo {author} {\bibfnamefont {S.}~\bibnamefont
  {{Tulin}}},\ and\ \bibinfo {author} {\bibfnamefont {H.-B.}\ \bibnamefont
  {{Yu}}},\ }\bibfield  {title} {\bibinfo {title} {{Dark Matter Halos as
  Particle Colliders: Unified Solution to Small-Scale Structure Puzzles from
  Dwarfs to Clusters}},\ }\href
  {https://doi.org/10.1103/PhysRevLett.116.041302} {\bibfield  {journal}
  {\bibinfo  {journal} {\prl}\ }\textbf {\bibinfo {volume} {116}},\ \bibinfo
  {eid} {041302} (\bibinfo {year} {2016})},\ \Eprint
  {https://arxiv.org/abs/1508.03339} {arXiv:1508.03339 [astro-ph.CO]}
  \BibitemShut {NoStop}%
\bibitem [{\citenamefont {{Kamada}}\ \emph {et~al.}(2017)\citenamefont
  {{Kamada}}, \citenamefont {{Kaplinghat}}, \citenamefont {{Pace}},\ and\
  \citenamefont {{Yu}}}]{Kamada++17}%
  \BibitemOpen
  \bibfield  {author} {\bibinfo {author} {\bibfnamefont {A.}~\bibnamefont
  {{Kamada}}}, \bibinfo {author} {\bibfnamefont {M.}~\bibnamefont
  {{Kaplinghat}}}, \bibinfo {author} {\bibfnamefont {A.~B.}\ \bibnamefont
  {{Pace}}},\ and\ \bibinfo {author} {\bibfnamefont {H.-B.}\ \bibnamefont
  {{Yu}}},\ }\bibfield  {title} {\bibinfo {title} {{Self-Interacting Dark
  Matter Can Explain Diverse Galactic Rotation Curves}},\ }\href
  {https://doi.org/10.1103/PhysRevLett.119.111102} {\bibfield  {journal}
  {\bibinfo  {journal} {\prl}\ }\textbf {\bibinfo {volume} {119}},\ \bibinfo
  {eid} {111102} (\bibinfo {year} {2017})},\ \Eprint
  {https://arxiv.org/abs/1611.02716} {arXiv:1611.02716 [astro-ph.GA]}
  \BibitemShut {NoStop}%
\bibitem [{\citenamefont {{Tulin}}\ and\ \citenamefont
  {{Yu}}(2018)}]{Tulin++18}%
  \BibitemOpen
  \bibfield  {author} {\bibinfo {author} {\bibfnamefont {S.}~\bibnamefont
  {{Tulin}}}\ and\ \bibinfo {author} {\bibfnamefont {H.-B.}\ \bibnamefont
  {{Yu}}},\ }\bibfield  {title} {\bibinfo {title} {{Dark matter
  self-interactions and small scale structure}},\ }\href
  {https://doi.org/10.1016/j.physrep.2017.11.004} {\bibfield  {journal}
  {\bibinfo  {journal} {\physrep}\ }\textbf {\bibinfo {volume} {730}},\
  \bibinfo {pages} {1} (\bibinfo {year} {2018})},\ \Eprint
  {https://arxiv.org/abs/1705.02358} {arXiv:1705.02358 [hep-ph]} \BibitemShut
  {NoStop}%
\bibitem [{\citenamefont {{Kahlhoefer}}\ \emph {et~al.}(2019)\citenamefont
  {{Kahlhoefer}}, \citenamefont {{Kaplinghat}}, \citenamefont {{Slatyer}},\
  and\ \citenamefont {{Wu}}}]{Kahlhoefer++19}%
  \BibitemOpen
  \bibfield  {author} {\bibinfo {author} {\bibfnamefont {F.}~\bibnamefont
  {{Kahlhoefer}}}, \bibinfo {author} {\bibfnamefont {M.}~\bibnamefont
  {{Kaplinghat}}}, \bibinfo {author} {\bibfnamefont {T.~R.}\ \bibnamefont
  {{Slatyer}}},\ and\ \bibinfo {author} {\bibfnamefont {C.-L.}\ \bibnamefont
  {{Wu}}},\ }\bibfield  {title} {\bibinfo {title} {{Diversity in density
  profiles of self-interacting dark matter satellite halos}},\ }\href
  {https://doi.org/10.1088/1475-7516/2019/12/010} {\bibfield  {journal}
  {\bibinfo  {journal} {\jcap}\ }\textbf {\bibinfo {volume} {2019}},\ \bibinfo
  {eid} {010} (\bibinfo {year} {2019})},\ \Eprint
  {https://arxiv.org/abs/1904.10539} {arXiv:1904.10539 [astro-ph.GA]}
  \BibitemShut {NoStop}%
\bibitem [{\citenamefont {{Zavala}}\ \emph {et~al.}(2019)\citenamefont
  {{Zavala}}, \citenamefont {{Lovell}}, \citenamefont {{Vogelsberger}},\ and\
  \citenamefont {{Burger}}}]{Zavala++19}%
  \BibitemOpen
  \bibfield  {author} {\bibinfo {author} {\bibfnamefont {J.}~\bibnamefont
  {{Zavala}}}, \bibinfo {author} {\bibfnamefont {M.~R.}\ \bibnamefont
  {{Lovell}}}, \bibinfo {author} {\bibfnamefont {M.}~\bibnamefont
  {{Vogelsberger}}},\ and\ \bibinfo {author} {\bibfnamefont {J.~D.}\
  \bibnamefont {{Burger}}},\ }\bibfield  {title} {\bibinfo {title} {{Diverse
  dark matter density at sub-kiloparsec scales in Milky Way satellites:
  Implications for the nature of dark matter}},\ }\href
  {https://doi.org/10.1103/PhysRevD.100.063007} {\bibfield  {journal} {\bibinfo
   {journal} {\prd}\ }\textbf {\bibinfo {volume} {100}},\ \bibinfo {eid}
  {063007} (\bibinfo {year} {2019})},\ \Eprint
  {https://arxiv.org/abs/1904.09998} {arXiv:1904.09998 [astro-ph.GA]}
  \BibitemShut {NoStop}%
\bibitem [{\citenamefont {{Silverman}}\ \emph {et~al.}(2022)\citenamefont
  {{Silverman}}, \citenamefont {{Bullock}}, \citenamefont {{Kaplinghat}},
  \citenamefont {{Robles}},\ and\ \citenamefont {{Valli}}}]{Silverman++22}%
  \BibitemOpen
  \bibfield  {author} {\bibinfo {author} {\bibfnamefont {M.}~\bibnamefont
  {{Silverman}}}, \bibinfo {author} {\bibfnamefont {J.~S.}\ \bibnamefont
  {{Bullock}}}, \bibinfo {author} {\bibfnamefont {M.}~\bibnamefont
  {{Kaplinghat}}}, \bibinfo {author} {\bibfnamefont {V.~H.}\ \bibnamefont
  {{Robles}}},\ and\ \bibinfo {author} {\bibfnamefont {M.}~\bibnamefont
  {{Valli}}},\ }\bibfield  {title} {\bibinfo {title} {{Motivations for a Large
  Self-Interacting Dark Matter Cross Section from Milky Way Satellites}},\
  }\href@noop {} {\bibfield  {journal} {\bibinfo  {journal} {arXiv e-prints}\
  ,\ \bibinfo {eid} {arXiv:2203.10104}} (\bibinfo {year} {2022})},\ \Eprint
  {https://arxiv.org/abs/2203.10104} {arXiv:2203.10104 [astro-ph.GA]}
  \BibitemShut {NoStop}%
\bibitem [{\citenamefont {{Adhikari}}\ \emph {et~al.}(2022)\citenamefont
  {{Adhikari}}, \citenamefont {{Banerjee}}, \citenamefont {{Boddy}},
  \citenamefont {{Cyr-Racine}}, \citenamefont {{Desmond}}, \citenamefont
  {{Dvorkin}}, \citenamefont {{Jain}}, \citenamefont {{Kahlhoefer}},
  \citenamefont {{Kaplinghat}}, \citenamefont {{Nierenberg}}, \citenamefont
  {{Peter}}, \citenamefont {{Robertson}}, \citenamefont {{Sakstein}},\ and\
  \citenamefont {{Zavala}}}]{Adhikari++22}%
  \BibitemOpen
  \bibfield  {author} {\bibinfo {author} {\bibfnamefont {S.}~\bibnamefont
  {{Adhikari}}}, \bibinfo {author} {\bibfnamefont {A.}~\bibnamefont
  {{Banerjee}}}, \bibinfo {author} {\bibfnamefont {K.~K.}\ \bibnamefont
  {{Boddy}}}, \bibinfo {author} {\bibfnamefont {F.-Y.}\ \bibnamefont
  {{Cyr-Racine}}}, \bibinfo {author} {\bibfnamefont {H.}~\bibnamefont
  {{Desmond}}}, \bibinfo {author} {\bibfnamefont {C.}~\bibnamefont
  {{Dvorkin}}}, \bibinfo {author} {\bibfnamefont {B.}~\bibnamefont {{Jain}}},
  \bibinfo {author} {\bibfnamefont {F.}~\bibnamefont {{Kahlhoefer}}}, \bibinfo
  {author} {\bibfnamefont {M.}~\bibnamefont {{Kaplinghat}}}, \bibinfo {author}
  {\bibfnamefont {A.}~\bibnamefont {{Nierenberg}}}, \bibinfo {author}
  {\bibfnamefont {A.~H.~G.}\ \bibnamefont {{Peter}}}, \bibinfo {author}
  {\bibfnamefont {A.}~\bibnamefont {{Robertson}}}, \bibinfo {author}
  {\bibfnamefont {J.}~\bibnamefont {{Sakstein}}},\ and\ \bibinfo {author}
  {\bibfnamefont {J.}~\bibnamefont {{Zavala}}},\ }\bibfield  {title} {\bibinfo
  {title} {{Astrophysical Tests of Dark Matter Self-Interactions}},\
  }\href@noop {} {\bibfield  {journal} {\bibinfo  {journal} {arXiv e-prints}\
  ,\ \bibinfo {eid} {arXiv:2207.10638}} (\bibinfo {year} {2022})},\ \Eprint
  {https://arxiv.org/abs/2207.10638} {arXiv:2207.10638 [astro-ph.CO]}
  \BibitemShut {NoStop}%
\bibitem [{\citenamefont {{Slone}}\ \emph {et~al.}(2023)\citenamefont
  {{Slone}}, \citenamefont {{Jiang}}, \citenamefont {{Lisanti}},\ and\
  \citenamefont {{Kaplinghat}}}]{Slone++23}%
  \BibitemOpen
  \bibfield  {author} {\bibinfo {author} {\bibfnamefont {O.}~\bibnamefont
  {{Slone}}}, \bibinfo {author} {\bibfnamefont {F.}~\bibnamefont {{Jiang}}},
  \bibinfo {author} {\bibfnamefont {M.}~\bibnamefont {{Lisanti}}},\ and\
  \bibinfo {author} {\bibfnamefont {M.}~\bibnamefont {{Kaplinghat}}},\
  }\bibfield  {title} {\bibinfo {title} {{Orbital evolution of satellite
  galaxies in self-interacting dark matter models}},\ }\href
  {https://doi.org/10.1103/PhysRevD.107.043014} {\bibfield  {journal} {\bibinfo
   {journal} {\prd}\ }\textbf {\bibinfo {volume} {107}},\ \bibinfo {eid}
  {043014} (\bibinfo {year} {2023})}\BibitemShut {NoStop}%
\bibitem [{\citenamefont {{Randall}}\ \emph {et~al.}(2008)\citenamefont
  {{Randall}}, \citenamefont {{Markevitch}}, \citenamefont {{Clowe}},
  \citenamefont {{Gonzalez}},\ and\ \citenamefont
  {{Brada{\v{c}}}}}]{Randall++08}%
  \BibitemOpen
  \bibfield  {author} {\bibinfo {author} {\bibfnamefont {S.~W.}\ \bibnamefont
  {{Randall}}}, \bibinfo {author} {\bibfnamefont {M.}~\bibnamefont
  {{Markevitch}}}, \bibinfo {author} {\bibfnamefont {D.}~\bibnamefont
  {{Clowe}}}, \bibinfo {author} {\bibfnamefont {A.~H.}\ \bibnamefont
  {{Gonzalez}}},\ and\ \bibinfo {author} {\bibfnamefont {M.}~\bibnamefont
  {{Brada{\v{c}}}}},\ }\bibfield  {title} {\bibinfo {title} {{Constraints on
  the Self-Interaction Cross Section of Dark Matter from Numerical Simulations
  of the Merging Galaxy Cluster 1E 0657-56}},\ }\href
  {https://doi.org/10.1086/587859} {\bibfield  {journal} {\bibinfo  {journal}
  {\apj}\ }\textbf {\bibinfo {volume} {679}},\ \bibinfo {pages} {1173}
  (\bibinfo {year} {2008})},\ \Eprint {https://arxiv.org/abs/0704.0261}
  {arXiv:0704.0261 [astro-ph]} \BibitemShut {NoStop}%
\bibitem [{\citenamefont {{Peter}}\ \emph {et~al.}(2013)\citenamefont
  {{Peter}}, \citenamefont {{Rocha}}, \citenamefont {{Bullock}},\ and\
  \citenamefont {{Kaplinghat}}}]{Peter++13}%
  \BibitemOpen
  \bibfield  {author} {\bibinfo {author} {\bibfnamefont {A.~H.~G.}\
  \bibnamefont {{Peter}}}, \bibinfo {author} {\bibfnamefont {M.}~\bibnamefont
  {{Rocha}}}, \bibinfo {author} {\bibfnamefont {J.~S.}\ \bibnamefont
  {{Bullock}}},\ and\ \bibinfo {author} {\bibfnamefont {M.}~\bibnamefont
  {{Kaplinghat}}},\ }\bibfield  {title} {\bibinfo {title} {{Cosmological
  simulations with self-interacting dark matter - II. Halo shapes versus
  observations}},\ }\href {https://doi.org/10.1093/mnras/sts535} {\bibfield
  {journal} {\bibinfo  {journal} {\mnras}\ }\textbf {\bibinfo {volume} {430}},\
  \bibinfo {pages} {105} (\bibinfo {year} {2013})},\ \Eprint
  {https://arxiv.org/abs/1208.3026} {arXiv:1208.3026 [astro-ph.CO]}
  \BibitemShut {NoStop}%
\bibitem [{\citenamefont {{Harvey}}\ \emph {et~al.}(2019)\citenamefont
  {{Harvey}}, \citenamefont {{Robertson}}, \citenamefont {{Massey}},\ and\
  \citenamefont {{McCarthy}}}]{Harvey++19}%
  \BibitemOpen
  \bibfield  {author} {\bibinfo {author} {\bibfnamefont {D.}~\bibnamefont
  {{Harvey}}}, \bibinfo {author} {\bibfnamefont {A.}~\bibnamefont
  {{Robertson}}}, \bibinfo {author} {\bibfnamefont {R.}~\bibnamefont
  {{Massey}}},\ and\ \bibinfo {author} {\bibfnamefont {I.~G.}\ \bibnamefont
  {{McCarthy}}},\ }\bibfield  {title} {\bibinfo {title} {{Observable tests of
  self-interacting dark matter in galaxy clusters: BCG wobbles in a constant
  density core}},\ }\href {https://doi.org/10.1093/mnras/stz1816} {\bibfield
  {journal} {\bibinfo  {journal} {\mnras}\ }\textbf {\bibinfo {volume} {488}},\
  \bibinfo {pages} {1572} (\bibinfo {year} {2019})},\ \Eprint
  {https://arxiv.org/abs/1812.06981} {arXiv:1812.06981 [astro-ph.CO]}
  \BibitemShut {NoStop}%
\bibitem [{\citenamefont {{Sagunski}}\ \emph {et~al.}(2021)\citenamefont
  {{Sagunski}}, \citenamefont {{Gad-Nasr}}, \citenamefont {{Colquhoun}},
  \citenamefont {{Robertson}},\ and\ \citenamefont {{Tulin}}}]{Sagunski++21}%
  \BibitemOpen
  \bibfield  {author} {\bibinfo {author} {\bibfnamefont {L.}~\bibnamefont
  {{Sagunski}}}, \bibinfo {author} {\bibfnamefont {S.}~\bibnamefont
  {{Gad-Nasr}}}, \bibinfo {author} {\bibfnamefont {B.}~\bibnamefont
  {{Colquhoun}}}, \bibinfo {author} {\bibfnamefont {A.}~\bibnamefont
  {{Robertson}}},\ and\ \bibinfo {author} {\bibfnamefont {S.}~\bibnamefont
  {{Tulin}}},\ }\bibfield  {title} {\bibinfo {title} {{Velocity-dependent
  self-interacting dark matter from groups and clusters of galaxies}},\ }\href
  {https://doi.org/10.1088/1475-7516/2021/01/024} {\bibfield  {journal}
  {\bibinfo  {journal} {\jcap}\ }\textbf {\bibinfo {volume} {2021}},\ \bibinfo
  {eid} {024} (\bibinfo {year} {2021})},\ \Eprint
  {https://arxiv.org/abs/2006.12515} {arXiv:2006.12515 [astro-ph.CO]}
  \BibitemShut {NoStop}%
\bibitem [{\citenamefont {{Andrade}}\ \emph {et~al.}(2022)\citenamefont
  {{Andrade}}, \citenamefont {{Fuson}}, \citenamefont {{Gad-Nasr}},
  \citenamefont {{Kong}}, \citenamefont {{Minor}}, \citenamefont {{Roberts}},\
  and\ \citenamefont {{Kaplinghat}}}]{Andrade++22}%
  \BibitemOpen
  \bibfield  {author} {\bibinfo {author} {\bibfnamefont {K.~E.}\ \bibnamefont
  {{Andrade}}}, \bibinfo {author} {\bibfnamefont {J.}~\bibnamefont {{Fuson}}},
  \bibinfo {author} {\bibfnamefont {S.}~\bibnamefont {{Gad-Nasr}}}, \bibinfo
  {author} {\bibfnamefont {D.}~\bibnamefont {{Kong}}}, \bibinfo {author}
  {\bibfnamefont {Q.}~\bibnamefont {{Minor}}}, \bibinfo {author} {\bibfnamefont
  {M.~G.}\ \bibnamefont {{Roberts}}},\ and\ \bibinfo {author} {\bibfnamefont
  {M.}~\bibnamefont {{Kaplinghat}}},\ }\bibfield  {title} {\bibinfo {title} {{A
  stringent upper limit on dark matter self-interaction cross-section from
  cluster strong lensing}},\ }\href {https://doi.org/10.1093/mnras/stab3241}
  {\bibfield  {journal} {\bibinfo  {journal} {\mnras}\ }\textbf {\bibinfo
  {volume} {510}},\ \bibinfo {pages} {54} (\bibinfo {year} {2022})},\ \Eprint
  {https://arxiv.org/abs/2012.06611} {arXiv:2012.06611 [astro-ph.CO]}
  \BibitemShut {NoStop}%
\bibitem [{\citenamefont {{Yoshida}}\ \emph {et~al.}(2000)\citenamefont
  {{Yoshida}}, \citenamefont {{Springel}}, \citenamefont {{White}},\ and\
  \citenamefont {{Tormen}}}]{Yoshida2000}%
  \BibitemOpen
  \bibfield  {author} {\bibinfo {author} {\bibfnamefont {N.}~\bibnamefont
  {{Yoshida}}}, \bibinfo {author} {\bibfnamefont {V.}~\bibnamefont
  {{Springel}}}, \bibinfo {author} {\bibfnamefont {S.~D.~M.}\ \bibnamefont
  {{White}}},\ and\ \bibinfo {author} {\bibfnamefont {G.}~\bibnamefont
  {{Tormen}}},\ }\bibfield  {title} {\bibinfo {title} {{Weakly Self-interacting
  Dark Matter and the Structure of Dark Halos}},\ }\href
  {https://doi.org/10.1086/317306} {\bibfield  {journal} {\bibinfo  {journal}
  {\apjl}\ }\textbf {\bibinfo {volume} {544}},\ \bibinfo {pages} {L87}
  (\bibinfo {year} {2000})},\ \Eprint {https://arxiv.org/abs/astro-ph/0006134}
  {arXiv:astro-ph/0006134 [astro-ph]} \BibitemShut {NoStop}%
\bibitem [{\citenamefont {{Arkani-Hamed}}\ \emph {et~al.}(2009)\citenamefont
  {{Arkani-Hamed}}, \citenamefont {{Finkbeiner}}, \citenamefont {{Slatyer}},\
  and\ \citenamefont {{Weiner}}}]{ArkaniHamed++09}%
  \BibitemOpen
  \bibfield  {author} {\bibinfo {author} {\bibfnamefont {N.}~\bibnamefont
  {{Arkani-Hamed}}}, \bibinfo {author} {\bibfnamefont {D.~P.}\ \bibnamefont
  {{Finkbeiner}}}, \bibinfo {author} {\bibfnamefont {T.~R.}\ \bibnamefont
  {{Slatyer}}},\ and\ \bibinfo {author} {\bibfnamefont {N.}~\bibnamefont
  {{Weiner}}},\ }\bibfield  {title} {\bibinfo {title} {{A theory of dark
  matter}},\ }\href {https://doi.org/10.1103/PhysRevD.79.015014} {\bibfield
  {journal} {\bibinfo  {journal} {\prd}\ }\textbf {\bibinfo {volume} {79}},\
  \bibinfo {eid} {015014} (\bibinfo {year} {2009})},\ \Eprint
  {https://arxiv.org/abs/0810.0713} {arXiv:0810.0713 [hep-ph]} \BibitemShut
  {NoStop}%
\bibitem [{\citenamefont {{Buckley}}\ and\ \citenamefont
  {{Fox}}(2010)}]{BuckleyFox10}%
  \BibitemOpen
  \bibfield  {author} {\bibinfo {author} {\bibfnamefont {M.~R.}\ \bibnamefont
  {{Buckley}}}\ and\ \bibinfo {author} {\bibfnamefont {P.~J.}\ \bibnamefont
  {{Fox}}},\ }\bibfield  {title} {\bibinfo {title} {{Dark matter
  self-interactions and light force carriers}},\ }\href
  {https://doi.org/10.1103/PhysRevD.81.083522} {\bibfield  {journal} {\bibinfo
  {journal} {\prd}\ }\textbf {\bibinfo {volume} {81}},\ \bibinfo {eid} {083522}
  (\bibinfo {year} {2010})},\ \Eprint {https://arxiv.org/abs/0911.3898}
  {arXiv:0911.3898 [hep-ph]} \BibitemShut {NoStop}%
\bibitem [{\citenamefont {{Foot}}\ and\ \citenamefont
  {{Vagnozzi}}(2015)}]{Foot++15}%
  \BibitemOpen
  \bibfield  {author} {\bibinfo {author} {\bibfnamefont {R.}~\bibnamefont
  {{Foot}}}\ and\ \bibinfo {author} {\bibfnamefont {S.}~\bibnamefont
  {{Vagnozzi}}},\ }\bibfield  {title} {\bibinfo {title} {{Dissipative hidden
  sector dark matter}},\ }\href {https://doi.org/10.1103/PhysRevD.91.023512}
  {\bibfield  {journal} {\bibinfo  {journal} {\prd}\ }\textbf {\bibinfo
  {volume} {91}},\ \bibinfo {eid} {023512} (\bibinfo {year} {2015})},\ \Eprint
  {https://arxiv.org/abs/1409.7174} {arXiv:1409.7174 [hep-ph]} \BibitemShut
  {NoStop}%
\bibitem [{\citenamefont {{Boddy}}\ \emph {et~al.}(2016)\citenamefont
  {{Boddy}}, \citenamefont {{Kaplinghat}}, \citenamefont {{Kwa}},\ and\
  \citenamefont {{Peter}}}]{Boddy++16}%
  \BibitemOpen
  \bibfield  {author} {\bibinfo {author} {\bibfnamefont {K.~K.}\ \bibnamefont
  {{Boddy}}}, \bibinfo {author} {\bibfnamefont {M.}~\bibnamefont
  {{Kaplinghat}}}, \bibinfo {author} {\bibfnamefont {A.}~\bibnamefont
  {{Kwa}}},\ and\ \bibinfo {author} {\bibfnamefont {A.~H.~G.}\ \bibnamefont
  {{Peter}}},\ }\bibfield  {title} {\bibinfo {title} {{Hidden sector hydrogen
  as dark matter: Small-scale structure formation predictions and the
  importance of hyperfine interactions}},\ }\href
  {https://doi.org/10.1103/PhysRevD.94.123017} {\bibfield  {journal} {\bibinfo
  {journal} {\prd}\ }\textbf {\bibinfo {volume} {94}},\ \bibinfo {eid} {123017}
  (\bibinfo {year} {2016})},\ \Eprint {https://arxiv.org/abs/1609.03592}
  {arXiv:1609.03592 [hep-ph]} \BibitemShut {NoStop}%
\bibitem [{\citenamefont {{Kahlhoefer}}\ \emph {et~al.}(2017)\citenamefont
  {{Kahlhoefer}}, \citenamefont {{Schmidt-Hoberg}},\ and\ \citenamefont
  {{Wild}}}]{Kahlhoefer++17}%
  \BibitemOpen
  \bibfield  {author} {\bibinfo {author} {\bibfnamefont {F.}~\bibnamefont
  {{Kahlhoefer}}}, \bibinfo {author} {\bibfnamefont {K.}~\bibnamefont
  {{Schmidt-Hoberg}}},\ and\ \bibinfo {author} {\bibfnamefont {S.}~\bibnamefont
  {{Wild}}},\ }\bibfield  {title} {\bibinfo {title} {{Dark matter
  self-interactions from a general spin-0 mediator}},\ }\href
  {https://doi.org/10.1088/1475-7516/2017/08/003} {\bibfield  {journal}
  {\bibinfo  {journal} {\jcap}\ }\textbf {\bibinfo {volume} {2017}},\ \bibinfo
  {eid} {003} (\bibinfo {year} {2017})},\ \Eprint
  {https://arxiv.org/abs/1704.02149} {arXiv:1704.02149 [hep-ph]} \BibitemShut
  {NoStop}%
\bibitem [{\citenamefont {{Cyr-Racine}}\ \emph
  {et~al.}(2016{\natexlab{a}})\citenamefont {{Cyr-Racine}}, \citenamefont
  {{Sigurdson}}, \citenamefont {{Zavala}}, \citenamefont {{Bringmann}},
  \citenamefont {{Vogelsberger}},\ and\ \citenamefont
  {{Pfrommer}}}]{Cyr-Racine++16}%
  \BibitemOpen
  \bibfield  {author} {\bibinfo {author} {\bibfnamefont {F.-Y.}\ \bibnamefont
  {{Cyr-Racine}}}, \bibinfo {author} {\bibfnamefont {K.}~\bibnamefont
  {{Sigurdson}}}, \bibinfo {author} {\bibfnamefont {J.}~\bibnamefont
  {{Zavala}}}, \bibinfo {author} {\bibfnamefont {T.}~\bibnamefont
  {{Bringmann}}}, \bibinfo {author} {\bibfnamefont {M.}~\bibnamefont
  {{Vogelsberger}}},\ and\ \bibinfo {author} {\bibfnamefont {C.}~\bibnamefont
  {{Pfrommer}}},\ }\bibfield  {title} {\bibinfo {title} {{ETHOS{\textemdash}an
  effective theory of structure formation: From dark particle physics to the
  matter distribution of the Universe}},\ }\href
  {https://doi.org/10.1103/PhysRevD.93.123527} {\bibfield  {journal} {\bibinfo
  {journal} {\prd}\ }\textbf {\bibinfo {volume} {93}},\ \bibinfo {eid} {123527}
  (\bibinfo {year} {2016}{\natexlab{a}})},\ \Eprint
  {https://arxiv.org/abs/1512.05344} {arXiv:1512.05344 [astro-ph.CO]}
  \BibitemShut {NoStop}%
\bibitem [{\citenamefont {{Ryan}}\ \emph {et~al.}(2021)\citenamefont {{Ryan}},
  \citenamefont {{Shandera}}, \citenamefont {{Gurian}},\ and\ \citenamefont
  {{Jeong}}}]{Ryan++21}%
  \BibitemOpen
  \bibfield  {author} {\bibinfo {author} {\bibfnamefont {M.}~\bibnamefont
  {{Ryan}}}, \bibinfo {author} {\bibfnamefont {S.}~\bibnamefont {{Shandera}}},
  \bibinfo {author} {\bibfnamefont {J.}~\bibnamefont {{Gurian}}},\ and\
  \bibinfo {author} {\bibfnamefont {D.}~\bibnamefont {{Jeong}}},\ }\bibfield
  {title} {\bibinfo {title} {{Molecular Chemistry for Dark Matter III:
  DarkKROME}},\ }\href@noop {} {\bibfield  {journal} {\bibinfo  {journal}
  {arXiv e-prints}\ ,\ \bibinfo {eid} {arXiv:2110.11971}} (\bibinfo {year}
  {2021})},\ \Eprint {https://arxiv.org/abs/2110.11971} {arXiv:2110.11971
  [astro-ph.CO]} \BibitemShut {NoStop}%
\bibitem [{\citenamefont {{Gilman}}\ \emph {et~al.}(2021)\citenamefont
  {{Gilman}}, \citenamefont {{Bovy}}, \citenamefont {{Treu}}, \citenamefont
  {{Nierenberg}}, \citenamefont {{Birrer}}, \citenamefont {{Benson}},\ and\
  \citenamefont {{Sameie}}}]{Gilman++21}%
  \BibitemOpen
  \bibfield  {author} {\bibinfo {author} {\bibfnamefont {D.}~\bibnamefont
  {{Gilman}}}, \bibinfo {author} {\bibfnamefont {J.}~\bibnamefont {{Bovy}}},
  \bibinfo {author} {\bibfnamefont {T.}~\bibnamefont {{Treu}}}, \bibinfo
  {author} {\bibfnamefont {A.}~\bibnamefont {{Nierenberg}}}, \bibinfo {author}
  {\bibfnamefont {S.}~\bibnamefont {{Birrer}}}, \bibinfo {author}
  {\bibfnamefont {A.}~\bibnamefont {{Benson}}},\ and\ \bibinfo {author}
  {\bibfnamefont {O.}~\bibnamefont {{Sameie}}},\ }\bibfield  {title} {\bibinfo
  {title} {{Strong lensing signatures of self-interacting dark matter in
  low-mass haloes}},\ }\href {https://doi.org/10.1093/mnras/stab2335}
  {\bibfield  {journal} {\bibinfo  {journal} {\mnras}\ }\textbf {\bibinfo
  {volume} {507}},\ \bibinfo {pages} {2432} (\bibinfo {year} {2021})},\ \Eprint
  {https://arxiv.org/abs/2105.05259} {arXiv:2105.05259 [astro-ph.CO]}
  \BibitemShut {NoStop}%
\bibitem [{\citenamefont {{Yang}}\ and\ \citenamefont {{Yu}}(2021)}]{Yang++21}%
  \BibitemOpen
  \bibfield  {author} {\bibinfo {author} {\bibfnamefont {D.}~\bibnamefont
  {{Yang}}}\ and\ \bibinfo {author} {\bibfnamefont {H.-B.}\ \bibnamefont
  {{Yu}}},\ }\bibfield  {title} {\bibinfo {title} {{Self-interacting dark
  matter and small-scale gravitational lenses in galaxy clusters}},\ }\href
  {https://doi.org/10.1103/PhysRevD.104.103031} {\bibfield  {journal} {\bibinfo
   {journal} {\prd}\ }\textbf {\bibinfo {volume} {104}},\ \bibinfo {eid}
  {103031} (\bibinfo {year} {2021})},\ \Eprint
  {https://arxiv.org/abs/2102.02375} {arXiv:2102.02375 [astro-ph.GA]}
  \BibitemShut {NoStop}%
\bibitem [{\citenamefont {{Dalal}}\ and\ \citenamefont
  {{Kochanek}}(2002)}]{Dalal++02}%
  \BibitemOpen
  \bibfield  {author} {\bibinfo {author} {\bibfnamefont {N.}~\bibnamefont
  {{Dalal}}}\ and\ \bibinfo {author} {\bibfnamefont {C.~S.}\ \bibnamefont
  {{Kochanek}}},\ }\bibfield  {title} {\bibinfo {title} {{Direct Detection of
  Cold Dark Matter Substructure}},\ }\href {https://doi.org/10.1086/340303}
  {\bibfield  {journal} {\bibinfo  {journal} {\apj}\ }\textbf {\bibinfo
  {volume} {572}},\ \bibinfo {pages} {25} (\bibinfo {year} {2002})},\ \Eprint
  {https://arxiv.org/abs/astro-ph/0111456} {arXiv:astro-ph/0111456 [astro-ph]}
  \BibitemShut {NoStop}%
\bibitem [{\citenamefont {{Nierenberg}}\ \emph {et~al.}(2014)\citenamefont
  {{Nierenberg}}, \citenamefont {{Treu}}, \citenamefont {{Wright}},
  \citenamefont {{Fassnacht}},\ and\ \citenamefont {{Auger}}}]{Nierenberg++14}%
  \BibitemOpen
  \bibfield  {author} {\bibinfo {author} {\bibfnamefont {A.~M.}\ \bibnamefont
  {{Nierenberg}}}, \bibinfo {author} {\bibfnamefont {T.}~\bibnamefont
  {{Treu}}}, \bibinfo {author} {\bibfnamefont {S.~A.}\ \bibnamefont
  {{Wright}}}, \bibinfo {author} {\bibfnamefont {C.~D.}\ \bibnamefont
  {{Fassnacht}}},\ and\ \bibinfo {author} {\bibfnamefont {M.~W.}\ \bibnamefont
  {{Auger}}},\ }\bibfield  {title} {\bibinfo {title} {{Detection of
  substructure with adaptive optics integral field spectroscopy of the
  gravitational lens B1422+231}},\ }\href
  {https://doi.org/10.1093/mnras/stu862} {\bibfield  {journal} {\bibinfo
  {journal} {\mnras}\ }\textbf {\bibinfo {volume} {442}},\ \bibinfo {pages}
  {2434} (\bibinfo {year} {2014})},\ \Eprint {https://arxiv.org/abs/1402.1496}
  {arXiv:1402.1496 [astro-ph.GA]} \BibitemShut {NoStop}%
\bibitem [{\citenamefont {{Nierenberg}}\ \emph {et~al.}(2017)\citenamefont
  {{Nierenberg}}, \citenamefont {{Treu}}, \citenamefont {{Brammer}},
  \citenamefont {{Peter}}, \citenamefont {{Fassnacht}}, \citenamefont
  {{Keeton}}, \citenamefont {{Kochanek}}, \citenamefont {{Schmidt}},
  \citenamefont {{Sluse}},\ and\ \citenamefont {{Wright}}}]{Nierenberg++17}%
  \BibitemOpen
  \bibfield  {author} {\bibinfo {author} {\bibfnamefont {A.~M.}\ \bibnamefont
  {{Nierenberg}}}, \bibinfo {author} {\bibfnamefont {T.}~\bibnamefont
  {{Treu}}}, \bibinfo {author} {\bibfnamefont {G.}~\bibnamefont {{Brammer}}},
  \bibinfo {author} {\bibfnamefont {A.~H.~G.}\ \bibnamefont {{Peter}}},
  \bibinfo {author} {\bibfnamefont {C.~D.}\ \bibnamefont {{Fassnacht}}},
  \bibinfo {author} {\bibfnamefont {C.~R.}\ \bibnamefont {{Keeton}}}, \bibinfo
  {author} {\bibfnamefont {C.~S.}\ \bibnamefont {{Kochanek}}}, \bibinfo
  {author} {\bibfnamefont {K.~B.}\ \bibnamefont {{Schmidt}}}, \bibinfo {author}
  {\bibfnamefont {D.}~\bibnamefont {{Sluse}}},\ and\ \bibinfo {author}
  {\bibfnamefont {S.~A.}\ \bibnamefont {{Wright}}},\ }\bibfield  {title}
  {\bibinfo {title} {{Probing dark matter substructure in the gravitational
  lens HE 0435-1223 with the WFC3 grism}},\ }\href
  {https://doi.org/10.1093/mnras/stx1400} {\bibfield  {journal} {\bibinfo
  {journal} {\mnras}\ }\textbf {\bibinfo {volume} {471}},\ \bibinfo {pages}
  {2224} (\bibinfo {year} {2017})},\ \Eprint {https://arxiv.org/abs/1701.05188}
  {arXiv:1701.05188 [astro-ph.CO]} \BibitemShut {NoStop}%
\bibitem [{\citenamefont {{Gilman}}\ \emph
  {et~al.}(2020{\natexlab{a}})\citenamefont {{Gilman}}, \citenamefont
  {{Birrer}}, \citenamefont {{Nierenberg}}, \citenamefont {{Treu}},
  \citenamefont {{Du}},\ and\ \citenamefont {{Benson}}}]{Gilman++20}%
  \BibitemOpen
  \bibfield  {author} {\bibinfo {author} {\bibfnamefont {D.}~\bibnamefont
  {{Gilman}}}, \bibinfo {author} {\bibfnamefont {S.}~\bibnamefont {{Birrer}}},
  \bibinfo {author} {\bibfnamefont {A.}~\bibnamefont {{Nierenberg}}}, \bibinfo
  {author} {\bibfnamefont {T.}~\bibnamefont {{Treu}}}, \bibinfo {author}
  {\bibfnamefont {X.}~\bibnamefont {{Du}}},\ and\ \bibinfo {author}
  {\bibfnamefont {A.}~\bibnamefont {{Benson}}},\ }\bibfield  {title} {\bibinfo
  {title} {{Warm dark matter chills out: constraints on the halo mass function
  and the free-streaming length of dark matter with eight quadruple-image
  strong gravitational lenses}},\ }\href
  {https://doi.org/10.1093/mnras/stz3480} {\bibfield  {journal} {\bibinfo
  {journal} {\mnras}\ }\textbf {\bibinfo {volume} {491}},\ \bibinfo {pages}
  {6077} (\bibinfo {year} {2020}{\natexlab{a}})},\ \Eprint
  {https://arxiv.org/abs/1908.06983} {arXiv:1908.06983 [astro-ph.CO]}
  \BibitemShut {NoStop}%
\bibitem [{\citenamefont {{Gilman}}\ \emph
  {et~al.}(2020{\natexlab{b}})\citenamefont {{Gilman}}, \citenamefont {{Du}},
  \citenamefont {{Benson}}, \citenamefont {{Birrer}}, \citenamefont
  {{Nierenberg}},\ and\ \citenamefont {{Treu}}}]{Gilman++20b}%
  \BibitemOpen
  \bibfield  {author} {\bibinfo {author} {\bibfnamefont {D.}~\bibnamefont
  {{Gilman}}}, \bibinfo {author} {\bibfnamefont {X.}~\bibnamefont {{Du}}},
  \bibinfo {author} {\bibfnamefont {A.}~\bibnamefont {{Benson}}}, \bibinfo
  {author} {\bibfnamefont {S.}~\bibnamefont {{Birrer}}}, \bibinfo {author}
  {\bibfnamefont {A.}~\bibnamefont {{Nierenberg}}},\ and\ \bibinfo {author}
  {\bibfnamefont {T.}~\bibnamefont {{Treu}}},\ }\bibfield  {title} {\bibinfo
  {title} {{Constraints on the mass-concentration relation of cold dark matter
  halos with 11 strong gravitational lenses}},\ }\href
  {https://doi.org/10.1093/mnrasl/slz173} {\bibfield  {journal} {\bibinfo
  {journal} {\mnras}\ }\textbf {\bibinfo {volume} {492}},\ \bibinfo {pages}
  {L12} (\bibinfo {year} {2020}{\natexlab{b}})},\ \Eprint
  {https://arxiv.org/abs/1909.02573} {arXiv:1909.02573 [astro-ph.CO]}
  \BibitemShut {NoStop}%
\bibitem [{\citenamefont {{Hsueh}}\ \emph {et~al.}(2020)\citenamefont
  {{Hsueh}}, \citenamefont {{Enzi}}, \citenamefont {{Vegetti}}, \citenamefont
  {{Auger}}, \citenamefont {{Fassnacht}}, \citenamefont {{Despali}},
  \citenamefont {{Koopmans}},\ and\ \citenamefont {{McKean}}}]{Hsueh++20}%
  \BibitemOpen
  \bibfield  {author} {\bibinfo {author} {\bibfnamefont {J.~W.}\ \bibnamefont
  {{Hsueh}}}, \bibinfo {author} {\bibfnamefont {W.}~\bibnamefont {{Enzi}}},
  \bibinfo {author} {\bibfnamefont {S.}~\bibnamefont {{Vegetti}}}, \bibinfo
  {author} {\bibfnamefont {M.~W.}\ \bibnamefont {{Auger}}}, \bibinfo {author}
  {\bibfnamefont {C.~D.}\ \bibnamefont {{Fassnacht}}}, \bibinfo {author}
  {\bibfnamefont {G.}~\bibnamefont {{Despali}}}, \bibinfo {author}
  {\bibfnamefont {L.~V.~E.}\ \bibnamefont {{Koopmans}}},\ and\ \bibinfo
  {author} {\bibfnamefont {J.~P.}\ \bibnamefont {{McKean}}},\ }\bibfield
  {title} {\bibinfo {title} {{SHARP - VII. New constraints on the dark matter
  free-streaming properties and substructure abundance from gravitationally
  lensed quasars}},\ }\href {https://doi.org/10.1093/mnras/stz3177} {\bibfield
  {journal} {\bibinfo  {journal} {\mnras}\ }\textbf {\bibinfo {volume} {492}},\
  \bibinfo {pages} {3047} (\bibinfo {year} {2020})},\ \Eprint
  {https://arxiv.org/abs/1905.04182} {arXiv:1905.04182 [astro-ph.CO]}
  \BibitemShut {NoStop}%
\bibitem [{\citenamefont {{Gilman}}\ \emph {et~al.}(2022)\citenamefont
  {{Gilman}}, \citenamefont {{Benson}}, \citenamefont {{Bovy}}, \citenamefont
  {{Birrer}}, \citenamefont {{Treu}},\ and\ \citenamefont
  {{Nierenberg}}}]{Gilman++22}%
  \BibitemOpen
  \bibfield  {author} {\bibinfo {author} {\bibfnamefont {D.}~\bibnamefont
  {{Gilman}}}, \bibinfo {author} {\bibfnamefont {A.}~\bibnamefont {{Benson}}},
  \bibinfo {author} {\bibfnamefont {J.}~\bibnamefont {{Bovy}}}, \bibinfo
  {author} {\bibfnamefont {S.}~\bibnamefont {{Birrer}}}, \bibinfo {author}
  {\bibfnamefont {T.}~\bibnamefont {{Treu}}},\ and\ \bibinfo {author}
  {\bibfnamefont {A.}~\bibnamefont {{Nierenberg}}},\ }\bibfield  {title}
  {\bibinfo {title} {{The primordial matter power spectrum on sub-galactic
  scales}},\ }\href {https://doi.org/10.1093/mnras/stac670} {\bibfield
  {journal} {\bibinfo  {journal} {\mnras}\ }\textbf {\bibinfo {volume} {512}},\
  \bibinfo {pages} {3163} (\bibinfo {year} {2022})},\ \Eprint
  {https://arxiv.org/abs/2112.03293} {arXiv:2112.03293 [astro-ph.CO]}
  \BibitemShut {NoStop}%
\bibitem [{\citenamefont {{Laroche}}\ \emph {et~al.}(2022)\citenamefont
  {{Laroche}}, \citenamefont {{Gilman}}, \citenamefont {{Li}}, \citenamefont
  {{Bovy}},\ and\ \citenamefont {{Du}}}]{Laroche++22}%
  \BibitemOpen
  \bibfield  {author} {\bibinfo {author} {\bibfnamefont {A.}~\bibnamefont
  {{Laroche}}}, \bibinfo {author} {\bibfnamefont {D.}~\bibnamefont {{Gilman}}},
  \bibinfo {author} {\bibfnamefont {X.}~\bibnamefont {{Li}}}, \bibinfo {author}
  {\bibfnamefont {J.}~\bibnamefont {{Bovy}}},\ and\ \bibinfo {author}
  {\bibfnamefont {X.}~\bibnamefont {{Du}}},\ }\bibfield  {title} {\bibinfo
  {title} {{Quantum fluctuations masquerade as halos: Bounds on ultra-light
  dark matter from quadruply-imaged quasars}},\ }\href@noop {} {\bibfield
  {journal} {\bibinfo  {journal} {arXiv e-prints}\ ,\ \bibinfo {eid}
  {arXiv:2206.11269}} (\bibinfo {year} {2022})},\ \Eprint
  {https://arxiv.org/abs/2206.11269} {arXiv:2206.11269 [astro-ph.CO]}
  \BibitemShut {NoStop}%
\bibitem [{\citenamefont {{Pontzen}}\ and\ \citenamefont
  {{Governato}}(2012)}]{Pontzen++12}%
  \BibitemOpen
  \bibfield  {author} {\bibinfo {author} {\bibfnamefont {A.}~\bibnamefont
  {{Pontzen}}}\ and\ \bibinfo {author} {\bibfnamefont {F.}~\bibnamefont
  {{Governato}}},\ }\bibfield  {title} {\bibinfo {title} {{How supernova
  feedback turns dark matter cusps into cores}},\ }\href
  {https://doi.org/10.1111/j.1365-2966.2012.20571.x} {\bibfield  {journal}
  {\bibinfo  {journal} {\mnras}\ }\textbf {\bibinfo {volume} {421}},\ \bibinfo
  {pages} {3464} (\bibinfo {year} {2012})},\ \Eprint
  {https://arxiv.org/abs/1106.0499} {arXiv:1106.0499 [astro-ph.CO]}
  \BibitemShut {NoStop}%
\bibitem [{\citenamefont {{Chan}}\ \emph {et~al.}(2015)\citenamefont {{Chan}},
  \citenamefont {{Kere{\v{s}}}}, \citenamefont {{O{\~n}orbe}}, \citenamefont
  {{Hopkins}}, \citenamefont {{Muratov}}, \citenamefont
  {{Faucher-Gigu{\`e}re}},\ and\ \citenamefont {{Quataert}}}]{Chan++15}%
  \BibitemOpen
  \bibfield  {author} {\bibinfo {author} {\bibfnamefont {T.~K.}\ \bibnamefont
  {{Chan}}}, \bibinfo {author} {\bibfnamefont {D.}~\bibnamefont
  {{Kere{\v{s}}}}}, \bibinfo {author} {\bibfnamefont {J.}~\bibnamefont
  {{O{\~n}orbe}}}, \bibinfo {author} {\bibfnamefont {P.~F.}\ \bibnamefont
  {{Hopkins}}}, \bibinfo {author} {\bibfnamefont {A.~L.}\ \bibnamefont
  {{Muratov}}}, \bibinfo {author} {\bibfnamefont {C.~A.}\ \bibnamefont
  {{Faucher-Gigu{\`e}re}}},\ and\ \bibinfo {author} {\bibfnamefont
  {E.}~\bibnamefont {{Quataert}}},\ }\bibfield  {title} {\bibinfo {title} {{The
  impact of baryonic physics on the structure of dark matter haloes: the view
  from the FIRE cosmological simulations}},\ }\href
  {https://doi.org/10.1093/mnras/stv2165} {\bibfield  {journal} {\bibinfo
  {journal} {\mnras}\ }\textbf {\bibinfo {volume} {454}},\ \bibinfo {pages}
  {2981} (\bibinfo {year} {2015})},\ \Eprint {https://arxiv.org/abs/1507.02282}
  {arXiv:1507.02282 [astro-ph.GA]} \BibitemShut {NoStop}%
\bibitem [{\citenamefont {{Sawala}}\ \emph {et~al.}(2016)\citenamefont
  {{Sawala}}, \citenamefont {{Frenk}}, \citenamefont {{Fattahi}}, \citenamefont
  {{Navarro}}, \citenamefont {{Bower}}, \citenamefont {{Crain}}, \citenamefont
  {{Dalla Vecchia}}, \citenamefont {{Furlong}}, \citenamefont {{Helly}},
  \citenamefont {{Jenkins}}, \citenamefont {{Oman}}, \citenamefont
  {{Schaller}}, \citenamefont {{Schaye}}, \citenamefont {{Theuns}},
  \citenamefont {{Trayford}},\ and\ \citenamefont {{White}}}]{Sawala++16}%
  \BibitemOpen
  \bibfield  {author} {\bibinfo {author} {\bibfnamefont {T.}~\bibnamefont
  {{Sawala}}}, \bibinfo {author} {\bibfnamefont {C.~S.}\ \bibnamefont
  {{Frenk}}}, \bibinfo {author} {\bibfnamefont {A.}~\bibnamefont {{Fattahi}}},
  \bibinfo {author} {\bibfnamefont {J.~F.}\ \bibnamefont {{Navarro}}}, \bibinfo
  {author} {\bibfnamefont {R.~G.}\ \bibnamefont {{Bower}}}, \bibinfo {author}
  {\bibfnamefont {R.~A.}\ \bibnamefont {{Crain}}}, \bibinfo {author}
  {\bibfnamefont {C.}~\bibnamefont {{Dalla Vecchia}}}, \bibinfo {author}
  {\bibfnamefont {M.}~\bibnamefont {{Furlong}}}, \bibinfo {author}
  {\bibfnamefont {J.~C.}\ \bibnamefont {{Helly}}}, \bibinfo {author}
  {\bibfnamefont {A.}~\bibnamefont {{Jenkins}}}, \bibinfo {author}
  {\bibfnamefont {K.~A.}\ \bibnamefont {{Oman}}}, \bibinfo {author}
  {\bibfnamefont {M.}~\bibnamefont {{Schaller}}}, \bibinfo {author}
  {\bibfnamefont {J.}~\bibnamefont {{Schaye}}}, \bibinfo {author}
  {\bibfnamefont {T.}~\bibnamefont {{Theuns}}}, \bibinfo {author}
  {\bibfnamefont {J.}~\bibnamefont {{Trayford}}},\ and\ \bibinfo {author}
  {\bibfnamefont {S.~D.~M.}\ \bibnamefont {{White}}},\ }\bibfield  {title}
  {\bibinfo {title} {{The APOSTLE simulations: solutions to the Local Group's
  cosmic puzzles}},\ }\href {https://doi.org/10.1093/mnras/stw145} {\bibfield
  {journal} {\bibinfo  {journal} {\mnras}\ }\textbf {\bibinfo {volume} {457}},\
  \bibinfo {pages} {1931} (\bibinfo {year} {2016})},\ \Eprint
  {https://arxiv.org/abs/1511.01098} {arXiv:1511.01098 [astro-ph.GA]}
  \BibitemShut {NoStop}%
\bibitem [{\citenamefont {{Robles}}\ \emph {et~al.}(2019)\citenamefont
  {{Robles}}, \citenamefont {{Kelley}}, \citenamefont {{Bullock}},\ and\
  \citenamefont {{Kaplinghat}}}]{Robles++19}%
  \BibitemOpen
  \bibfield  {author} {\bibinfo {author} {\bibfnamefont {V.~H.}\ \bibnamefont
  {{Robles}}}, \bibinfo {author} {\bibfnamefont {T.}~\bibnamefont {{Kelley}}},
  \bibinfo {author} {\bibfnamefont {J.~S.}\ \bibnamefont {{Bullock}}},\ and\
  \bibinfo {author} {\bibfnamefont {M.}~\bibnamefont {{Kaplinghat}}},\
  }\bibfield  {title} {\bibinfo {title} {{The Milky Way's halo and subhaloes in
  self-interacting dark matter}},\ }\href
  {https://doi.org/10.1093/mnras/stz2345} {\bibfield  {journal} {\bibinfo
  {journal} {\mnras}\ }\textbf {\bibinfo {volume} {490}},\ \bibinfo {pages}
  {2117} (\bibinfo {year} {2019})},\ \Eprint {https://arxiv.org/abs/1903.01469}
  {arXiv:1903.01469 [astro-ph.GA]} \BibitemShut {NoStop}%
\bibitem [{\citenamefont {{Fitts}}\ \emph {et~al.}(2019)\citenamefont
  {{Fitts}}, \citenamefont {{Boylan-Kolchin}}, \citenamefont {{Bozek}},
  \citenamefont {{Bullock}}, \citenamefont {{Graus}}, \citenamefont {{Robles}},
  \citenamefont {{Hopkins}}, \citenamefont {{El-Badry}}, \citenamefont
  {{Garrison-Kimmel}}, \citenamefont {{Faucher-Gigu{\`e}re}}, \citenamefont
  {{Wetzel}},\ and\ \citenamefont {{Kere{\v{s}}}}}]{Fitts++19}%
  \BibitemOpen
  \bibfield  {author} {\bibinfo {author} {\bibfnamefont {A.}~\bibnamefont
  {{Fitts}}}, \bibinfo {author} {\bibfnamefont {M.}~\bibnamefont
  {{Boylan-Kolchin}}}, \bibinfo {author} {\bibfnamefont {B.}~\bibnamefont
  {{Bozek}}}, \bibinfo {author} {\bibfnamefont {J.~S.}\ \bibnamefont
  {{Bullock}}}, \bibinfo {author} {\bibfnamefont {A.}~\bibnamefont {{Graus}}},
  \bibinfo {author} {\bibfnamefont {V.}~\bibnamefont {{Robles}}}, \bibinfo
  {author} {\bibfnamefont {P.~F.}\ \bibnamefont {{Hopkins}}}, \bibinfo {author}
  {\bibfnamefont {K.}~\bibnamefont {{El-Badry}}}, \bibinfo {author}
  {\bibfnamefont {S.}~\bibnamefont {{Garrison-Kimmel}}}, \bibinfo {author}
  {\bibfnamefont {C.-A.}\ \bibnamefont {{Faucher-Gigu{\`e}re}}}, \bibinfo
  {author} {\bibfnamefont {A.}~\bibnamefont {{Wetzel}}},\ and\ \bibinfo
  {author} {\bibfnamefont {D.}~\bibnamefont {{Kere{\v{s}}}}},\ }\bibfield
  {title} {\bibinfo {title} {{Dwarf galaxies in CDM, WDM, and SIDM:
  disentangling baryons and dark matter physics}},\ }\href
  {https://doi.org/10.1093/mnras/stz2613} {\bibfield  {journal} {\bibinfo
  {journal} {\mnras}\ }\textbf {\bibinfo {volume} {490}},\ \bibinfo {pages}
  {962} (\bibinfo {year} {2019})},\ \Eprint {https://arxiv.org/abs/1811.11791}
  {arXiv:1811.11791 [astro-ph.GA]} \BibitemShut {NoStop}%
\bibitem [{\citenamefont {{Kaplinghat}}\ \emph {et~al.}(2020)\citenamefont
  {{Kaplinghat}}, \citenamefont {{Ren}},\ and\ \citenamefont
  {{Yu}}}]{Kaplinghat++20}%
  \BibitemOpen
  \bibfield  {author} {\bibinfo {author} {\bibfnamefont {M.}~\bibnamefont
  {{Kaplinghat}}}, \bibinfo {author} {\bibfnamefont {T.}~\bibnamefont
  {{Ren}}},\ and\ \bibinfo {author} {\bibfnamefont {H.-B.}\ \bibnamefont
  {{Yu}}},\ }\bibfield  {title} {\bibinfo {title} {{Dark matter cores and cusps
  in spiral galaxies and their explanations}},\ }\href
  {https://doi.org/10.1088/1475-7516/2020/06/027} {\bibfield  {journal}
  {\bibinfo  {journal} {\jcap}\ }\textbf {\bibinfo {volume} {2020}},\ \bibinfo
  {eid} {027} (\bibinfo {year} {2020})},\ \Eprint
  {https://arxiv.org/abs/1911.00544} {arXiv:1911.00544 [astro-ph.GA]}
  \BibitemShut {NoStop}%
\bibitem [{\citenamefont {{Sameie}}\ \emph {et~al.}(2021)\citenamefont
  {{Sameie}}, \citenamefont {{Boylan-Kolchin}}, \citenamefont {{Sanderson}},
  \citenamefont {{Vargya}}, \citenamefont {{Hopkins}}, \citenamefont
  {{Wetzel}}, \citenamefont {{Bullock}}, \citenamefont {{Graus}},\ and\
  \citenamefont {{Robles}}}]{Sameie++21}%
  \BibitemOpen
  \bibfield  {author} {\bibinfo {author} {\bibfnamefont {O.}~\bibnamefont
  {{Sameie}}}, \bibinfo {author} {\bibfnamefont {M.}~\bibnamefont
  {{Boylan-Kolchin}}}, \bibinfo {author} {\bibfnamefont {R.}~\bibnamefont
  {{Sanderson}}}, \bibinfo {author} {\bibfnamefont {D.}~\bibnamefont
  {{Vargya}}}, \bibinfo {author} {\bibfnamefont {P.~F.}\ \bibnamefont
  {{Hopkins}}}, \bibinfo {author} {\bibfnamefont {A.}~\bibnamefont {{Wetzel}}},
  \bibinfo {author} {\bibfnamefont {J.}~\bibnamefont {{Bullock}}}, \bibinfo
  {author} {\bibfnamefont {A.}~\bibnamefont {{Graus}}},\ and\ \bibinfo {author}
  {\bibfnamefont {V.~H.}\ \bibnamefont {{Robles}}},\ }\bibfield  {title}
  {\bibinfo {title} {{The central densities of Milky Way-mass galaxies in cold
  and self-interacting dark matter models}},\ }\href
  {https://doi.org/10.1093/mnras/stab2173} {\bibfield  {journal} {\bibinfo
  {journal} {\mnras}\ }\textbf {\bibinfo {volume} {507}},\ \bibinfo {pages}
  {720} (\bibinfo {year} {2021})},\ \Eprint {https://arxiv.org/abs/2102.12480}
  {arXiv:2102.12480 [astro-ph.GA]} \BibitemShut {NoStop}%
\bibitem [{\citenamefont {{Zentner}}\ \emph {et~al.}(2022)\citenamefont
  {{Zentner}}, \citenamefont {{Dandavate}}, \citenamefont {{Slone}},\ and\
  \citenamefont {{Lisanti}}}]{Zentner++22}%
  \BibitemOpen
  \bibfield  {author} {\bibinfo {author} {\bibfnamefont {A.}~\bibnamefont
  {{Zentner}}}, \bibinfo {author} {\bibfnamefont {S.}~\bibnamefont
  {{Dandavate}}}, \bibinfo {author} {\bibfnamefont {O.}~\bibnamefont
  {{Slone}}},\ and\ \bibinfo {author} {\bibfnamefont {M.}~\bibnamefont
  {{Lisanti}}},\ }\bibfield  {title} {\bibinfo {title} {{A Critical Assessment
  of Solutions to the Galaxy Diversity Problem}},\ }\href@noop {} {\bibfield
  {journal} {\bibinfo  {journal} {arXiv e-prints}\ ,\ \bibinfo {eid}
  {arXiv:2202.00012}} (\bibinfo {year} {2022})},\ \Eprint
  {https://arxiv.org/abs/2202.00012} {arXiv:2202.00012 [astro-ph.GA]}
  \BibitemShut {NoStop}%
\bibitem [{\citenamefont {{Roper}}\ \emph {et~al.}(2022)\citenamefont
  {{Roper}}, \citenamefont {{Oman}}, \citenamefont {{Frenk}}, \citenamefont
  {{Ben{\'\i}tez-Llambay}}, \citenamefont {{Navarro}},\ and\ \citenamefont
  {{Santos-Santos}}}]{Roper++22}%
  \BibitemOpen
  \bibfield  {author} {\bibinfo {author} {\bibfnamefont {F.~A.}\ \bibnamefont
  {{Roper}}}, \bibinfo {author} {\bibfnamefont {K.~A.}\ \bibnamefont {{Oman}}},
  \bibinfo {author} {\bibfnamefont {C.~S.}\ \bibnamefont {{Frenk}}}, \bibinfo
  {author} {\bibfnamefont {A.}~\bibnamefont {{Ben{\'\i}tez-Llambay}}}, \bibinfo
  {author} {\bibfnamefont {J.~F.}\ \bibnamefont {{Navarro}}},\ and\ \bibinfo
  {author} {\bibfnamefont {I.~M.~E.}\ \bibnamefont {{Santos-Santos}}},\
  }\bibfield  {title} {\bibinfo {title} {{The diversity of rotation curves of
  simulated galaxies with cusps and cores}},\ }\href@noop {} {\bibfield
  {journal} {\bibinfo  {journal} {arXiv e-prints}\ ,\ \bibinfo {eid}
  {arXiv:2203.16652}} (\bibinfo {year} {2022})},\ \Eprint
  {https://arxiv.org/abs/2203.16652} {arXiv:2203.16652 [astro-ph.GA]}
  \BibitemShut {NoStop}%
\bibitem [{\citenamefont {{Robles}}\ \emph {et~al.}(2017)\citenamefont
  {{Robles}}, \citenamefont {{Bullock}}, \citenamefont {{Elbert}},
  \citenamefont {{Fitts}}, \citenamefont {{Gonz{\'a}lez-Samaniego}},
  \citenamefont {{Boylan-Kolchin}}, \citenamefont {{Hopkins}}, \citenamefont
  {{Faucher-Gigu{\`e}re}}, \citenamefont {{Kere{\v{s}}}},\ and\ \citenamefont
  {{Hayward}}}]{Robles++17}%
  \BibitemOpen
  \bibfield  {author} {\bibinfo {author} {\bibfnamefont {V.~H.}\ \bibnamefont
  {{Robles}}}, \bibinfo {author} {\bibfnamefont {J.~S.}\ \bibnamefont
  {{Bullock}}}, \bibinfo {author} {\bibfnamefont {O.~D.}\ \bibnamefont
  {{Elbert}}}, \bibinfo {author} {\bibfnamefont {A.}~\bibnamefont {{Fitts}}},
  \bibinfo {author} {\bibfnamefont {A.}~\bibnamefont
  {{Gonz{\'a}lez-Samaniego}}}, \bibinfo {author} {\bibfnamefont
  {M.}~\bibnamefont {{Boylan-Kolchin}}}, \bibinfo {author} {\bibfnamefont
  {P.~F.}\ \bibnamefont {{Hopkins}}}, \bibinfo {author} {\bibfnamefont {C.-A.}\
  \bibnamefont {{Faucher-Gigu{\`e}re}}}, \bibinfo {author} {\bibfnamefont
  {D.}~\bibnamefont {{Kere{\v{s}}}}},\ and\ \bibinfo {author} {\bibfnamefont
  {C.~C.}\ \bibnamefont {{Hayward}}},\ }\bibfield  {title} {\bibinfo {title}
  {{SIDM on FIRE: hydrodynamical self-interacting dark matter simulations of
  low-mass dwarf galaxies}},\ }\href {https://doi.org/10.1093/mnras/stx2253}
  {\bibfield  {journal} {\bibinfo  {journal} {\mnras}\ }\textbf {\bibinfo
  {volume} {472}},\ \bibinfo {pages} {2945} (\bibinfo {year} {2017})},\ \Eprint
  {https://arxiv.org/abs/1706.07514} {arXiv:1706.07514 [astro-ph.GA]}
  \BibitemShut {NoStop}%
\bibitem [{\citenamefont {{Lazar}}\ \emph {et~al.}(2020)\citenamefont
  {{Lazar}}, \citenamefont {{Bullock}}, \citenamefont {{Boylan-Kolchin}},
  \citenamefont {{Chan}}, \citenamefont {{Hopkins}}, \citenamefont {{Graus}},
  \citenamefont {{Wetzel}}, \citenamefont {{El-Badry}}, \citenamefont
  {{Wheeler}}, \citenamefont {{Straight}}, \citenamefont {{Kere{\v{s}}}},
  \citenamefont {{Faucher-Gigu{\`e}re}}, \citenamefont {{Fitts}},\ and\
  \citenamefont {{Garrison-Kimmel}}}]{Lazar++20}%
  \BibitemOpen
  \bibfield  {author} {\bibinfo {author} {\bibfnamefont {A.}~\bibnamefont
  {{Lazar}}}, \bibinfo {author} {\bibfnamefont {J.~S.}\ \bibnamefont
  {{Bullock}}}, \bibinfo {author} {\bibfnamefont {M.}~\bibnamefont
  {{Boylan-Kolchin}}}, \bibinfo {author} {\bibfnamefont {T.~K.}\ \bibnamefont
  {{Chan}}}, \bibinfo {author} {\bibfnamefont {P.~F.}\ \bibnamefont
  {{Hopkins}}}, \bibinfo {author} {\bibfnamefont {A.~S.}\ \bibnamefont
  {{Graus}}}, \bibinfo {author} {\bibfnamefont {A.}~\bibnamefont {{Wetzel}}},
  \bibinfo {author} {\bibfnamefont {K.}~\bibnamefont {{El-Badry}}}, \bibinfo
  {author} {\bibfnamefont {C.}~\bibnamefont {{Wheeler}}}, \bibinfo {author}
  {\bibfnamefont {M.~C.}\ \bibnamefont {{Straight}}}, \bibinfo {author}
  {\bibfnamefont {D.}~\bibnamefont {{Kere{\v{s}}}}}, \bibinfo {author}
  {\bibfnamefont {C.-A.}\ \bibnamefont {{Faucher-Gigu{\`e}re}}}, \bibinfo
  {author} {\bibfnamefont {A.}~\bibnamefont {{Fitts}}},\ and\ \bibinfo {author}
  {\bibfnamefont {S.}~\bibnamefont {{Garrison-Kimmel}}},\ }\bibfield  {title}
  {\bibinfo {title} {{A dark matter profile to model diverse feedback-induced
  core sizes of {\ensuremath{\Lambda}}CDM haloes}},\ }\href
  {https://doi.org/10.1093/mnras/staa2101} {\bibfield  {journal} {\bibinfo
  {journal} {\mnras}\ }\textbf {\bibinfo {volume} {497}},\ \bibinfo {pages}
  {2393} (\bibinfo {year} {2020})},\ \Eprint {https://arxiv.org/abs/2004.10817}
  {arXiv:2004.10817 [astro-ph.GA]} \BibitemShut {NoStop}%
\bibitem [{\citenamefont {{Rose}}\ \emph {et~al.}(2022)\citenamefont {{Rose}},
  \citenamefont {{Torrey}}, \citenamefont {{Vogelsberger}},\ and\ \citenamefont
  {{O'Neil}}}]{Rose++22}%
  \BibitemOpen
  \bibfield  {author} {\bibinfo {author} {\bibfnamefont {J.~C.}\ \bibnamefont
  {{Rose}}}, \bibinfo {author} {\bibfnamefont {P.}~\bibnamefont {{Torrey}}},
  \bibinfo {author} {\bibfnamefont {M.}~\bibnamefont {{Vogelsberger}}},\ and\
  \bibinfo {author} {\bibfnamefont {S.}~\bibnamefont {{O'Neil}}},\ }\bibfield
  {title} {\bibinfo {title} {{Unraveling the interplay between SIDM and baryons
  in MW halos: defining where baryons dictate heat transfer}},\ }\href@noop {}
  {\bibfield  {journal} {\bibinfo  {journal} {arXiv e-prints}\ ,\ \bibinfo
  {eid} {arXiv:2206.14830}} (\bibinfo {year} {2022})},\ \Eprint
  {https://arxiv.org/abs/2206.14830} {arXiv:2206.14830 [astro-ph.GA]}
  \BibitemShut {NoStop}%
\bibitem [{\citenamefont {{Gilman}}\ \emph {et~al.}(2018)\citenamefont
  {{Gilman}}, \citenamefont {{Birrer}}, \citenamefont {{Treu}}, \citenamefont
  {{Keeton}},\ and\ \citenamefont {{Nierenberg}}}]{Gilman++18}%
  \BibitemOpen
  \bibfield  {author} {\bibinfo {author} {\bibfnamefont {D.}~\bibnamefont
  {{Gilman}}}, \bibinfo {author} {\bibfnamefont {S.}~\bibnamefont {{Birrer}}},
  \bibinfo {author} {\bibfnamefont {T.}~\bibnamefont {{Treu}}}, \bibinfo
  {author} {\bibfnamefont {C.~R.}\ \bibnamefont {{Keeton}}},\ and\ \bibinfo
  {author} {\bibfnamefont {A.}~\bibnamefont {{Nierenberg}}},\ }\bibfield
  {title} {\bibinfo {title} {{Probing the nature of dark matter by forward
  modelling flux ratios in strong gravitational lenses}},\ }\href
  {https://doi.org/10.1093/mnras/sty2261} {\bibfield  {journal} {\bibinfo
  {journal} {\mnras}\ }\textbf {\bibinfo {volume} {481}},\ \bibinfo {pages}
  {819} (\bibinfo {year} {2018})},\ \Eprint {https://arxiv.org/abs/1712.04945}
  {arXiv:1712.04945 [astro-ph.CO]} \BibitemShut {NoStop}%
\bibitem [{\citenamefont {{Gilman}}\ \emph {et~al.}(2019)\citenamefont
  {{Gilman}}, \citenamefont {{Birrer}}, \citenamefont {{Treu}}, \citenamefont
  {{Nierenberg}},\ and\ \citenamefont {{Benson}}}]{Gilman++19}%
  \BibitemOpen
  \bibfield  {author} {\bibinfo {author} {\bibfnamefont {D.}~\bibnamefont
  {{Gilman}}}, \bibinfo {author} {\bibfnamefont {S.}~\bibnamefont {{Birrer}}},
  \bibinfo {author} {\bibfnamefont {T.}~\bibnamefont {{Treu}}}, \bibinfo
  {author} {\bibfnamefont {A.}~\bibnamefont {{Nierenberg}}},\ and\ \bibinfo
  {author} {\bibfnamefont {A.}~\bibnamefont {{Benson}}},\ }\bibfield  {title}
  {\bibinfo {title} {{Probing dark matter structure down to {}10$^{7}$ solar
  masses: flux ratio statistics in gravitational lenses with line-of-sight
  haloes}},\ }\href {https://doi.org/10.1093/mnras/stz1593} {\bibfield
  {journal} {\bibinfo  {journal} {\mnras}\ }\textbf {\bibinfo {volume} {487}},\
  \bibinfo {pages} {5721} (\bibinfo {year} {2019})},\ \Eprint
  {https://arxiv.org/abs/1901.11031} {arXiv:1901.11031 [astro-ph.CO]}
  \BibitemShut {NoStop}%
\bibitem [{\citenamefont {{Chiba}}\ \emph {et~al.}(2005)\citenamefont
  {{Chiba}}, \citenamefont {{Minezaki}}, \citenamefont {{Kashikawa}},
  \citenamefont {{Kataza}},\ and\ \citenamefont {{Inoue}}}]{Chiba++05}%
  \BibitemOpen
  \bibfield  {author} {\bibinfo {author} {\bibfnamefont {M.}~\bibnamefont
  {{Chiba}}}, \bibinfo {author} {\bibfnamefont {T.}~\bibnamefont {{Minezaki}}},
  \bibinfo {author} {\bibfnamefont {N.}~\bibnamefont {{Kashikawa}}}, \bibinfo
  {author} {\bibfnamefont {H.}~\bibnamefont {{Kataza}}},\ and\ \bibinfo
  {author} {\bibfnamefont {K.~T.}\ \bibnamefont {{Inoue}}},\ }\bibfield
  {title} {\bibinfo {title} {{Subaru Mid-Infrared Imaging of the Quadruple
  Lenses PG 1115+080 and B1422+231: Limits on Substructure Lensing}},\ }\href
  {https://doi.org/10.1086/430403} {\bibfield  {journal} {\bibinfo  {journal}
  {\apj}\ }\textbf {\bibinfo {volume} {627}},\ \bibinfo {pages} {53} (\bibinfo
  {year} {2005})},\ \Eprint {https://arxiv.org/abs/astro-ph/0503487}
  {arXiv:astro-ph/0503487 [astro-ph]} \BibitemShut {NoStop}%
\bibitem [{\citenamefont {{Sugai}}\ \emph {et~al.}(2007)\citenamefont
  {{Sugai}}, \citenamefont {{Kawai}}, \citenamefont {{Shimono}}, \citenamefont
  {{Hattori}}, \citenamefont {{Kosugi}}, \citenamefont {{Kashikawa}},
  \citenamefont {{Inoue}},\ and\ \citenamefont {{Chiba}}}]{Sugai++07}%
  \BibitemOpen
  \bibfield  {author} {\bibinfo {author} {\bibfnamefont {H.}~\bibnamefont
  {{Sugai}}}, \bibinfo {author} {\bibfnamefont {A.}~\bibnamefont {{Kawai}}},
  \bibinfo {author} {\bibfnamefont {A.}~\bibnamefont {{Shimono}}}, \bibinfo
  {author} {\bibfnamefont {T.}~\bibnamefont {{Hattori}}}, \bibinfo {author}
  {\bibfnamefont {G.}~\bibnamefont {{Kosugi}}}, \bibinfo {author}
  {\bibfnamefont {N.}~\bibnamefont {{Kashikawa}}}, \bibinfo {author}
  {\bibfnamefont {K.~T.}\ \bibnamefont {{Inoue}}},\ and\ \bibinfo {author}
  {\bibfnamefont {M.}~\bibnamefont {{Chiba}}},\ }\bibfield  {title} {\bibinfo
  {title} {{Integral Field Spectroscopy of the Quadruply Lensed Quasar 1RXS
  J1131-1231: New Light on Lens Substructures}},\ }\href
  {https://doi.org/10.1086/513731} {\bibfield  {journal} {\bibinfo  {journal}
  {\apj}\ }\textbf {\bibinfo {volume} {660}},\ \bibinfo {pages} {1016}
  (\bibinfo {year} {2007})},\ \Eprint {https://arxiv.org/abs/astro-ph/0702392}
  {arXiv:astro-ph/0702392 [astro-ph]} \BibitemShut {NoStop}%
\bibitem [{\citenamefont {{Stacey}}\ and\ \citenamefont
  {{McKean}}(2018)}]{Stacey++18}%
  \BibitemOpen
  \bibfield  {author} {\bibinfo {author} {\bibfnamefont {H.~R.}\ \bibnamefont
  {{Stacey}}}\ and\ \bibinfo {author} {\bibfnamefont {J.~P.}\ \bibnamefont
  {{McKean}}},\ }\bibfield  {title} {\bibinfo {title} {{A flux-ratio anomaly in
  the CO spectral line emission from gravitationally lensed quasar MG
  J0414+0534}},\ }\href {https://doi.org/10.1093/mnrasl/sly153} {\bibfield
  {journal} {\bibinfo  {journal} {\mnras}\ }\textbf {\bibinfo {volume} {481}},\
  \bibinfo {pages} {L40} (\bibinfo {year} {2018})},\ \Eprint
  {https://arxiv.org/abs/1808.05571} {arXiv:1808.05571 [astro-ph.GA]}
  \BibitemShut {NoStop}%
\bibitem [{\citenamefont {{Nierenberg}}\ \emph {et~al.}(2020)\citenamefont
  {{Nierenberg}}, \citenamefont {{Gilman}}, \citenamefont {{Treu}},
  \citenamefont {{Brammer}}, \citenamefont {{Birrer}}, \citenamefont
  {{Moustakas}}, \citenamefont {{Agnello}}, \citenamefont {{Anguita}},
  \citenamefont {{Fassnacht}}, \citenamefont {{Motta}}, \citenamefont
  {{Peter}},\ and\ \citenamefont {{Sluse}}}]{Nierenberg++20}%
  \BibitemOpen
  \bibfield  {author} {\bibinfo {author} {\bibfnamefont {A.~M.}\ \bibnamefont
  {{Nierenberg}}}, \bibinfo {author} {\bibfnamefont {D.}~\bibnamefont
  {{Gilman}}}, \bibinfo {author} {\bibfnamefont {T.}~\bibnamefont {{Treu}}},
  \bibinfo {author} {\bibfnamefont {G.}~\bibnamefont {{Brammer}}}, \bibinfo
  {author} {\bibfnamefont {S.}~\bibnamefont {{Birrer}}}, \bibinfo {author}
  {\bibfnamefont {L.}~\bibnamefont {{Moustakas}}}, \bibinfo {author}
  {\bibfnamefont {A.}~\bibnamefont {{Agnello}}}, \bibinfo {author}
  {\bibfnamefont {T.}~\bibnamefont {{Anguita}}}, \bibinfo {author}
  {\bibfnamefont {C.~D.}\ \bibnamefont {{Fassnacht}}}, \bibinfo {author}
  {\bibfnamefont {V.}~\bibnamefont {{Motta}}}, \bibinfo {author} {\bibfnamefont
  {A.~H.~G.}\ \bibnamefont {{Peter}}},\ and\ \bibinfo {author} {\bibfnamefont
  {D.}~\bibnamefont {{Sluse}}},\ }\bibfield  {title} {\bibinfo {title} {{Double
  dark matter vision: twice the number of compact-source lenses with
  narrow-line lensing and the WFC3 grism}},\ }\href
  {https://doi.org/10.1093/mnras/stz3588} {\bibfield  {journal} {\bibinfo
  {journal} {\mnras}\ }\textbf {\bibinfo {volume} {492}},\ \bibinfo {pages}
  {5314} (\bibinfo {year} {2020})},\ \Eprint {https://arxiv.org/abs/1908.06344}
  {arXiv:1908.06344 [astro-ph.GA]} \BibitemShut {NoStop}%
\bibitem [{\citenamefont {{Tulin}}\ \emph
  {et~al.}(2013{\natexlab{a}})\citenamefont {{Tulin}}, \citenamefont {{Yu}},\
  and\ \citenamefont {{Zurek}}}]{Tulin++13}%
  \BibitemOpen
  \bibfield  {author} {\bibinfo {author} {\bibfnamefont {S.}~\bibnamefont
  {{Tulin}}}, \bibinfo {author} {\bibfnamefont {H.-B.}\ \bibnamefont {{Yu}}},\
  and\ \bibinfo {author} {\bibfnamefont {K.~M.}\ \bibnamefont {{Zurek}}},\
  }\bibfield  {title} {\bibinfo {title} {{Beyond collisionless dark matter:
  Particle physics dynamics for dark matter halo structure}},\ }\href
  {https://doi.org/10.1103/PhysRevD.87.115007} {\bibfield  {journal} {\bibinfo
  {journal} {\prd}\ }\textbf {\bibinfo {volume} {87}},\ \bibinfo {eid} {115007}
  (\bibinfo {year} {2013}{\natexlab{a}})},\ \Eprint
  {https://arxiv.org/abs/1302.3898} {arXiv:1302.3898 [hep-ph]} \BibitemShut
  {NoStop}%
\bibitem [{\citenamefont {{Colquhoun}}\ \emph {et~al.}(2021)\citenamefont
  {{Colquhoun}}, \citenamefont {{Heeba}}, \citenamefont {{Kahlhoefer}},
  \citenamefont {{Sagunski}},\ and\ \citenamefont {{Tulin}}}]{Colquhoun++21}%
  \BibitemOpen
  \bibfield  {author} {\bibinfo {author} {\bibfnamefont {B.}~\bibnamefont
  {{Colquhoun}}}, \bibinfo {author} {\bibfnamefont {S.}~\bibnamefont
  {{Heeba}}}, \bibinfo {author} {\bibfnamefont {F.}~\bibnamefont
  {{Kahlhoefer}}}, \bibinfo {author} {\bibfnamefont {L.}~\bibnamefont
  {{Sagunski}}},\ and\ \bibinfo {author} {\bibfnamefont {S.}~\bibnamefont
  {{Tulin}}},\ }\bibfield  {title} {\bibinfo {title} {{Semiclassical regime for
  dark matter self-interactions}},\ }\href
  {https://doi.org/10.1103/PhysRevD.103.035006} {\bibfield  {journal} {\bibinfo
   {journal} {\prd}\ }\textbf {\bibinfo {volume} {103}},\ \bibinfo {eid}
  {035006} (\bibinfo {year} {2021})},\ \Eprint
  {https://arxiv.org/abs/2011.04679} {arXiv:2011.04679 [hep-ph]} \BibitemShut
  {NoStop}%
\bibitem [{\citenamefont {{Yang}}\ \emph
  {et~al.}(2022{\natexlab{a}})\citenamefont {{Yang}}, \citenamefont
  {{Nadler}},\ and\ \citenamefont {{Yu}}}]{Yang++22}%
  \BibitemOpen
  \bibfield  {author} {\bibinfo {author} {\bibfnamefont {D.}~\bibnamefont
  {{Yang}}}, \bibinfo {author} {\bibfnamefont {E.~O.}\ \bibnamefont
  {{Nadler}}},\ and\ \bibinfo {author} {\bibfnamefont {H.-b.}\ \bibnamefont
  {{Yu}}},\ }\bibfield  {title} {\bibinfo {title} {{Strong Dark Matter
  Self-interactions Diversify Halo Populations Within and Surrounding the Milky
  Way}},\ }\href {https://doi.org/10.48550/arXiv.2211.13768} {\bibfield
  {journal} {\bibinfo  {journal} {arXiv e-prints}\ ,\ \bibinfo {eid}
  {arXiv:2211.13768}} (\bibinfo {year} {2022}{\natexlab{a}})},\ \Eprint
  {https://arxiv.org/abs/2211.13768} {arXiv:2211.13768 [astro-ph.GA]}
  \BibitemShut {NoStop}%
\bibitem [{\citenamefont {{Chu}}\ \emph {et~al.}(2020)\citenamefont {{Chu}},
  \citenamefont {{Garcia-Cely}},\ and\ \citenamefont {{Murayama}}}]{Chu++20}%
  \BibitemOpen
  \bibfield  {author} {\bibinfo {author} {\bibfnamefont {X.}~\bibnamefont
  {{Chu}}}, \bibinfo {author} {\bibfnamefont {C.}~\bibnamefont
  {{Garcia-Cely}}},\ and\ \bibinfo {author} {\bibfnamefont {H.}~\bibnamefont
  {{Murayama}}},\ }\bibfield  {title} {\bibinfo {title} {{A practical and
  consistent parametrization of dark matter self-interactions}},\ }\href
  {https://doi.org/10.1088/1475-7516/2020/06/043} {\bibfield  {journal}
  {\bibinfo  {journal} {\jcap}\ }\textbf {\bibinfo {volume} {2020}},\ \bibinfo
  {eid} {043} (\bibinfo {year} {2020})},\ \Eprint
  {https://arxiv.org/abs/1908.06067} {arXiv:1908.06067 [hep-ph]} \BibitemShut
  {NoStop}%
\bibitem [{\citenamefont {{Tulin}}\ \emph
  {et~al.}(2013{\natexlab{b}})\citenamefont {{Tulin}}, \citenamefont {{Yu}},\
  and\ \citenamefont {{Zurek}}}]{Tulin++13a}%
  \BibitemOpen
  \bibfield  {author} {\bibinfo {author} {\bibfnamefont {S.}~\bibnamefont
  {{Tulin}}}, \bibinfo {author} {\bibfnamefont {H.-B.}\ \bibnamefont {{Yu}}},\
  and\ \bibinfo {author} {\bibfnamefont {K.~M.}\ \bibnamefont {{Zurek}}},\
  }\bibfield  {title} {\bibinfo {title} {{Resonant Dark Forces and Small-Scale
  Structure}},\ }\href {https://doi.org/10.1103/PhysRevLett.110.111301}
  {\bibfield  {journal} {\bibinfo  {journal} {\prl}\ }\textbf {\bibinfo
  {volume} {110}},\ \bibinfo {eid} {111301} (\bibinfo {year}
  {2013}{\natexlab{b}})},\ \Eprint {https://arxiv.org/abs/1210.0900}
  {arXiv:1210.0900 [hep-ph]} \BibitemShut {NoStop}%
\bibitem [{\citenamefont {{Chu}}\ \emph {et~al.}(2019)\citenamefont {{Chu}},
  \citenamefont {{Garcia-Cely}},\ and\ \citenamefont {{Murayama}}}]{Chu++19}%
  \BibitemOpen
  \bibfield  {author} {\bibinfo {author} {\bibfnamefont {X.}~\bibnamefont
  {{Chu}}}, \bibinfo {author} {\bibfnamefont {C.}~\bibnamefont
  {{Garcia-Cely}}},\ and\ \bibinfo {author} {\bibfnamefont {H.}~\bibnamefont
  {{Murayama}}},\ }\bibfield  {title} {\bibinfo {title} {{Velocity Dependence
  from Resonant Self-Interacting Dark Matter}},\ }\href
  {https://doi.org/10.1103/PhysRevLett.122.071103} {\bibfield  {journal}
  {\bibinfo  {journal} {\prl}\ }\textbf {\bibinfo {volume} {122}},\ \bibinfo
  {eid} {071103} (\bibinfo {year} {2019})},\ \Eprint
  {https://arxiv.org/abs/1810.04709} {arXiv:1810.04709 [hep-ph]} \BibitemShut
  {NoStop}%
\bibitem [{\citenamefont {{Navarro}}\ \emph {et~al.}(1997)\citenamefont
  {{Navarro}}, \citenamefont {{Frenk}},\ and\ \citenamefont
  {{White}}}]{Navarro++97}%
  \BibitemOpen
  \bibfield  {author} {\bibinfo {author} {\bibfnamefont {J.~F.}\ \bibnamefont
  {{Navarro}}}, \bibinfo {author} {\bibfnamefont {C.~S.}\ \bibnamefont
  {{Frenk}}},\ and\ \bibinfo {author} {\bibfnamefont {S.~D.~M.}\ \bibnamefont
  {{White}}},\ }\bibfield  {title} {\bibinfo {title} {{A Universal Density
  Profile from Hierarchical Clustering}},\ }\href
  {https://doi.org/10.1086/304888} {\bibfield  {journal} {\bibinfo  {journal}
  {\apj}\ }\textbf {\bibinfo {volume} {490}},\ \bibinfo {pages} {493} (\bibinfo
  {year} {1997})},\ \Eprint {https://arxiv.org/abs/astro-ph/9611107}
  {arXiv:astro-ph/9611107 [astro-ph]} \BibitemShut {NoStop}%
\bibitem [{\citenamefont {{Yang}}\ \emph
  {et~al.}(2022{\natexlab{b}})\citenamefont {{Yang}}, \citenamefont {{Du}},
  \citenamefont {{Carton Zeng}}, \citenamefont {{Benson}}, \citenamefont
  {{Jiang}}, \citenamefont {{Nadler}},\ and\ \citenamefont
  {{Peter}}}]{ShengqiYang++22}%
  \BibitemOpen
  \bibfield  {author} {\bibinfo {author} {\bibfnamefont {S.}~\bibnamefont
  {{Yang}}}, \bibinfo {author} {\bibfnamefont {X.}~\bibnamefont {{Du}}},
  \bibinfo {author} {\bibfnamefont {Z.}~\bibnamefont {{Carton Zeng}}}, \bibinfo
  {author} {\bibfnamefont {A.}~\bibnamefont {{Benson}}}, \bibinfo {author}
  {\bibfnamefont {F.}~\bibnamefont {{Jiang}}}, \bibinfo {author} {\bibfnamefont
  {E.~O.}\ \bibnamefont {{Nadler}}},\ and\ \bibinfo {author} {\bibfnamefont
  {A.~H.~G.}\ \bibnamefont {{Peter}}},\ }\bibfield  {title} {\bibinfo {title}
  {{Gravothermal solutions of SIDM halos: mapping from constant to
  velocity-dependent cross section}},\ }\href@noop {} {\bibfield  {journal}
  {\bibinfo  {journal} {arXiv e-prints}\ ,\ \bibinfo {eid} {arXiv:2205.02957}}
  (\bibinfo {year} {2022}{\natexlab{b}})},\ \Eprint
  {https://arxiv.org/abs/2205.02957} {arXiv:2205.02957 [astro-ph.CO]}
  \BibitemShut {NoStop}%
\bibitem [{\citenamefont {{Lifshitz}}\ and\ \citenamefont
  {{Pitaevskii}}(1981)}]{LifshitzPitaevskii81}%
  \BibitemOpen
  \bibfield  {author} {\bibinfo {author} {\bibfnamefont {E.~M.}\ \bibnamefont
  {{Lifshitz}}}\ and\ \bibinfo {author} {\bibfnamefont {L.~P.}\ \bibnamefont
  {{Pitaevskii}}},\ }\href@noop {} {\emph {\bibinfo {title} {{Physical
  kinetics}}}}\ (\bibinfo {year} {1981})\BibitemShut {NoStop}%
\bibitem [{\citenamefont {{Outmezguine}}\ \emph {et~al.}(2022)\citenamefont
  {{Outmezguine}}, \citenamefont {{Boddy}}, \citenamefont {{Gad-Nasr}},
  \citenamefont {{Kaplinghat}},\ and\ \citenamefont
  {{Sagunski}}}]{Outmezguine++22}%
  \BibitemOpen
  \bibfield  {author} {\bibinfo {author} {\bibfnamefont {N.~J.}\ \bibnamefont
  {{Outmezguine}}}, \bibinfo {author} {\bibfnamefont {K.~K.}\ \bibnamefont
  {{Boddy}}}, \bibinfo {author} {\bibfnamefont {S.}~\bibnamefont {{Gad-Nasr}}},
  \bibinfo {author} {\bibfnamefont {M.}~\bibnamefont {{Kaplinghat}}},\ and\
  \bibinfo {author} {\bibfnamefont {L.}~\bibnamefont {{Sagunski}}},\ }\bibfield
   {title} {\bibinfo {title} {{Universal gravothermal evolution of isolated
  self-interacting dark matter halos for velocity-dependent cross sections}},\
  }\href@noop {} {\bibfield  {journal} {\bibinfo  {journal} {arXiv e-prints}\
  ,\ \bibinfo {eid} {arXiv:2204.06568}} (\bibinfo {year} {2022})},\ \Eprint
  {https://arxiv.org/abs/2204.06568} {arXiv:2204.06568 [astro-ph.GA]}
  \BibitemShut {NoStop}%
\bibitem [{\citenamefont {{Diemer}}\ and\ \citenamefont
  {{Joyce}}(2019)}]{Diemer++19}%
  \BibitemOpen
  \bibfield  {author} {\bibinfo {author} {\bibfnamefont {B.}~\bibnamefont
  {{Diemer}}}\ and\ \bibinfo {author} {\bibfnamefont {M.}~\bibnamefont
  {{Joyce}}},\ }\bibfield  {title} {\bibinfo {title} {{An Accurate Physical
  Model for Halo Concentrations}},\ }\href
  {https://doi.org/10.3847/1538-4357/aafad6} {\bibfield  {journal} {\bibinfo
  {journal} {\apj}\ }\textbf {\bibinfo {volume} {871}},\ \bibinfo {eid} {168}
  (\bibinfo {year} {2019})},\ \Eprint {https://arxiv.org/abs/1809.07326}
  {arXiv:1809.07326 [astro-ph.CO]} \BibitemShut {NoStop}%
\bibitem [{\citenamefont {{Nadler}}\ \emph {et~al.}(2021)\citenamefont
  {{Nadler}}, \citenamefont {{Birrer}}, \citenamefont {{Gilman}}, \citenamefont
  {{Wechsler}}, \citenamefont {{Du}}, \citenamefont {{Benson}}, \citenamefont
  {{Nierenberg}},\ and\ \citenamefont {{Treu}}}]{Nadler++21}%
  \BibitemOpen
  \bibfield  {author} {\bibinfo {author} {\bibfnamefont {E.~O.}\ \bibnamefont
  {{Nadler}}}, \bibinfo {author} {\bibfnamefont {S.}~\bibnamefont {{Birrer}}},
  \bibinfo {author} {\bibfnamefont {D.}~\bibnamefont {{Gilman}}}, \bibinfo
  {author} {\bibfnamefont {R.~H.}\ \bibnamefont {{Wechsler}}}, \bibinfo
  {author} {\bibfnamefont {X.}~\bibnamefont {{Du}}}, \bibinfo {author}
  {\bibfnamefont {A.}~\bibnamefont {{Benson}}}, \bibinfo {author}
  {\bibfnamefont {A.~M.}\ \bibnamefont {{Nierenberg}}},\ and\ \bibinfo {author}
  {\bibfnamefont {T.}~\bibnamefont {{Treu}}},\ }\bibfield  {title} {\bibinfo
  {title} {{Dark Matter Constraints from a Unified Analysis of Strong
  Gravitational Lenses and Milky Way Satellite Galaxies}},\ }\href
  {https://doi.org/10.3847/1538-4357/abf9a3} {\bibfield  {journal} {\bibinfo
  {journal} {\apj}\ }\textbf {\bibinfo {volume} {917}},\ \bibinfo {eid} {7}
  (\bibinfo {year} {2021})},\ \Eprint {https://arxiv.org/abs/2101.07810}
  {arXiv:2101.07810 [astro-ph.CO]} \BibitemShut {NoStop}%
\bibitem [{\citenamefont {{Vogelsberger}}\ \emph {et~al.}(2019)\citenamefont
  {{Vogelsberger}}, \citenamefont {{Zavala}}, \citenamefont {{Schutz}},\ and\
  \citenamefont {{Slatyer}}}]{Vogelsberger++19}%
  \BibitemOpen
  \bibfield  {author} {\bibinfo {author} {\bibfnamefont {M.}~\bibnamefont
  {{Vogelsberger}}}, \bibinfo {author} {\bibfnamefont {J.}~\bibnamefont
  {{Zavala}}}, \bibinfo {author} {\bibfnamefont {K.}~\bibnamefont {{Schutz}}},\
  and\ \bibinfo {author} {\bibfnamefont {T.~R.}\ \bibnamefont {{Slatyer}}},\
  }\bibfield  {title} {\bibinfo {title} {{Evaporating the Milky Way halo and
  its satellites with inelastic self-interacting dark matter}},\ }\href
  {https://doi.org/10.1093/mnras/stz340} {\bibfield  {journal} {\bibinfo
  {journal} {\mnras}\ }\textbf {\bibinfo {volume} {484}},\ \bibinfo {pages}
  {5437} (\bibinfo {year} {2019})},\ \Eprint {https://arxiv.org/abs/1805.03203}
  {arXiv:1805.03203 [astro-ph.GA]} \BibitemShut {NoStop}%
\bibitem [{\citenamefont {{Zeng}}\ \emph {et~al.}(2022)\citenamefont {{Zeng}},
  \citenamefont {{Peter}}, \citenamefont {{Du}}, \citenamefont {{Benson}},
  \citenamefont {{Kim}}, \citenamefont {{Jiang}}, \citenamefont
  {{Cyr-Racine}},\ and\ \citenamefont {{Vogelsberger}}}]{Zeng++22}%
  \BibitemOpen
  \bibfield  {author} {\bibinfo {author} {\bibfnamefont {Z.~C.}\ \bibnamefont
  {{Zeng}}}, \bibinfo {author} {\bibfnamefont {A.~H.~G.}\ \bibnamefont
  {{Peter}}}, \bibinfo {author} {\bibfnamefont {X.}~\bibnamefont {{Du}}},
  \bibinfo {author} {\bibfnamefont {A.}~\bibnamefont {{Benson}}}, \bibinfo
  {author} {\bibfnamefont {S.}~\bibnamefont {{Kim}}}, \bibinfo {author}
  {\bibfnamefont {F.}~\bibnamefont {{Jiang}}}, \bibinfo {author} {\bibfnamefont
  {F.-Y.}\ \bibnamefont {{Cyr-Racine}}},\ and\ \bibinfo {author} {\bibfnamefont
  {M.}~\bibnamefont {{Vogelsberger}}},\ }\bibfield  {title} {\bibinfo {title}
  {{Core-collapse, evaporation, and tidal effects: the life story of a
  self-interacting dark matter subhalo}},\ }\href
  {https://doi.org/10.1093/mnras/stac1094} {\bibfield  {journal} {\bibinfo
  {journal} {\mnras}\ }\textbf {\bibinfo {volume} {513}},\ \bibinfo {pages}
  {4845} (\bibinfo {year} {2022})}\BibitemShut {NoStop}%
\bibitem [{\citenamefont {{Nadler}}\ \emph
  {et~al.}(2020{\natexlab{a}})\citenamefont {{Nadler}}, \citenamefont
  {{Banerjee}}, \citenamefont {{Adhikari}}, \citenamefont {{Mao}},\ and\
  \citenamefont {{Wechsler}}}]{Nadler++20}%
  \BibitemOpen
  \bibfield  {author} {\bibinfo {author} {\bibfnamefont {E.~O.}\ \bibnamefont
  {{Nadler}}}, \bibinfo {author} {\bibfnamefont {A.}~\bibnamefont
  {{Banerjee}}}, \bibinfo {author} {\bibfnamefont {S.}~\bibnamefont
  {{Adhikari}}}, \bibinfo {author} {\bibfnamefont {Y.-Y.}\ \bibnamefont
  {{Mao}}},\ and\ \bibinfo {author} {\bibfnamefont {R.~H.}\ \bibnamefont
  {{Wechsler}}},\ }\bibfield  {title} {\bibinfo {title} {{Signatures of
  Velocity-dependent Dark Matter Self-interactions in Milky Way-mass Halos}},\
  }\href {https://doi.org/10.3847/1538-4357/ab94b0} {\bibfield  {journal}
  {\bibinfo  {journal} {\apj}\ }\textbf {\bibinfo {volume} {896}},\ \bibinfo
  {eid} {112} (\bibinfo {year} {2020}{\natexlab{a}})},\ \Eprint
  {https://arxiv.org/abs/2001.08754} {arXiv:2001.08754 [astro-ph.CO]}
  \BibitemShut {NoStop}%
\bibitem [{\citenamefont {{Vogelsberger}}\ \emph {et~al.}(2016)\citenamefont
  {{Vogelsberger}}, \citenamefont {{Zavala}}, \citenamefont {{Cyr-Racine}},
  \citenamefont {{Pfrommer}}, \citenamefont {{Bringmann}},\ and\ \citenamefont
  {{Sigurdson}}}]{Vogelsberger++16}%
  \BibitemOpen
  \bibfield  {author} {\bibinfo {author} {\bibfnamefont {M.}~\bibnamefont
  {{Vogelsberger}}}, \bibinfo {author} {\bibfnamefont {J.}~\bibnamefont
  {{Zavala}}}, \bibinfo {author} {\bibfnamefont {F.-Y.}\ \bibnamefont
  {{Cyr-Racine}}}, \bibinfo {author} {\bibfnamefont {C.}~\bibnamefont
  {{Pfrommer}}}, \bibinfo {author} {\bibfnamefont {T.}~\bibnamefont
  {{Bringmann}}},\ and\ \bibinfo {author} {\bibfnamefont {K.}~\bibnamefont
  {{Sigurdson}}},\ }\bibfield  {title} {\bibinfo {title} {{ETHOS - an effective
  theory of structure formation: dark matter physics as a possible explanation
  of the small-scale CDM problems}},\ }\href
  {https://doi.org/10.1093/mnras/stw1076} {\bibfield  {journal} {\bibinfo
  {journal} {\mnras}\ }\textbf {\bibinfo {volume} {460}},\ \bibinfo {pages}
  {1399} (\bibinfo {year} {2016})},\ \Eprint {https://arxiv.org/abs/1512.05349}
  {arXiv:1512.05349 [astro-ph.CO]} \BibitemShut {NoStop}%
\bibitem [{Note1()}]{Note1}%
  \BibitemOpen
  \bibinfo {note} {We can easily extend our analysis to models with suppressed
  small-scale power, given a model for the halo mass function and
  concentration-mass relation in these scenarios.}\BibitemShut {Stop}%
\bibitem [{\citenamefont {{Jiang}}\ \emph {et~al.}(2022)\citenamefont
  {{Jiang}}, \citenamefont {{Benson}}, \citenamefont {{Hopkins}}, \citenamefont
  {{Slone}}, \citenamefont {{Lisanti}}, \citenamefont {{Kaplinghat}},
  \citenamefont {{Peter}}, \citenamefont {{Carton Zeng}}, \citenamefont {{Du}},
  \citenamefont {{Yang}},\ and\ \citenamefont {{Shen}}}]{Jiang++22}%
  \BibitemOpen
  \bibfield  {author} {\bibinfo {author} {\bibfnamefont {F.}~\bibnamefont
  {{Jiang}}}, \bibinfo {author} {\bibfnamefont {A.}~\bibnamefont {{Benson}}},
  \bibinfo {author} {\bibfnamefont {P.~F.}\ \bibnamefont {{Hopkins}}}, \bibinfo
  {author} {\bibfnamefont {O.}~\bibnamefont {{Slone}}}, \bibinfo {author}
  {\bibfnamefont {M.}~\bibnamefont {{Lisanti}}}, \bibinfo {author}
  {\bibfnamefont {M.}~\bibnamefont {{Kaplinghat}}}, \bibinfo {author}
  {\bibfnamefont {A.~H.~G.}\ \bibnamefont {{Peter}}}, \bibinfo {author}
  {\bibfnamefont {Z.}~\bibnamefont {{Carton Zeng}}}, \bibinfo {author}
  {\bibfnamefont {X.}~\bibnamefont {{Du}}}, \bibinfo {author} {\bibfnamefont
  {S.}~\bibnamefont {{Yang}}},\ and\ \bibinfo {author} {\bibfnamefont
  {X.}~\bibnamefont {{Shen}}},\ }\bibfield  {title} {\bibinfo {title} {{A
  semi-analytic study of self-interacting dark-matter haloes with baryons}},\
  }\href@noop {} {\bibfield  {journal} {\bibinfo  {journal} {arXiv e-prints}\
  ,\ \bibinfo {eid} {arXiv:2206.12425}} (\bibinfo {year} {2022})},\ \Eprint
  {https://arxiv.org/abs/2206.12425} {arXiv:2206.12425 [astro-ph.CO]}
  \BibitemShut {NoStop}%
\bibitem [{\citenamefont {Koda}\ and\ \citenamefont
  {Shapiro}(2011)}]{Koda2011}%
  \BibitemOpen
  \bibfield  {author} {\bibinfo {author} {\bibfnamefont {J.}~\bibnamefont
  {Koda}}\ and\ \bibinfo {author} {\bibfnamefont {P.~R.}\ \bibnamefont
  {Shapiro}},\ }\bibfield  {title} {\bibinfo {title} {{Gravothermal collapse of
  isolated self-interacting dark matter haloes: N-body simulation versus the
  fluid model}},\ }\href {https://doi.org/10.1111/j.1365-2966.2011.18684.x}
  {\bibfield  {journal} {\bibinfo  {journal} {Mon. Not. Roy. Astron. Soc.}\
  }\textbf {\bibinfo {volume} {415}},\ \bibinfo {pages} {1125} (\bibinfo {year}
  {2011})},\ \Eprint {https://arxiv.org/abs/1101.3097} {arXiv:1101.3097
  [astro-ph.CO]} \BibitemShut {NoStop}%
\bibitem [{\citenamefont {{Essig}}\ \emph {et~al.}(2019)\citenamefont
  {{Essig}}, \citenamefont {{McDermott}}, \citenamefont {{Yu}},\ and\
  \citenamefont {{Zhong}}}]{Essig++19}%
  \BibitemOpen
  \bibfield  {author} {\bibinfo {author} {\bibfnamefont {R.}~\bibnamefont
  {{Essig}}}, \bibinfo {author} {\bibfnamefont {S.~D.}\ \bibnamefont
  {{McDermott}}}, \bibinfo {author} {\bibfnamefont {H.-B.}\ \bibnamefont
  {{Yu}}},\ and\ \bibinfo {author} {\bibfnamefont {Y.-M.}\ \bibnamefont
  {{Zhong}}},\ }\bibfield  {title} {\bibinfo {title} {{Constraining Dissipative
  Dark Matter Self-Interactions}},\ }\href
  {https://doi.org/10.1103/PhysRevLett.123.121102} {\bibfield  {journal}
  {\bibinfo  {journal} {\prl}\ }\textbf {\bibinfo {volume} {123}},\ \bibinfo
  {eid} {121102} (\bibinfo {year} {2019})},\ \Eprint
  {https://arxiv.org/abs/1809.01144} {arXiv:1809.01144 [hep-ph]} \BibitemShut
  {NoStop}%
\bibitem [{\citenamefont {{Gilman}}\ \emph {et~al.}(2017)\citenamefont
  {{Gilman}}, \citenamefont {{Agnello}}, \citenamefont {{Treu}}, \citenamefont
  {{Keeton}},\ and\ \citenamefont {{Nierenberg}}}]{Gilman++17}%
  \BibitemOpen
  \bibfield  {author} {\bibinfo {author} {\bibfnamefont {D.}~\bibnamefont
  {{Gilman}}}, \bibinfo {author} {\bibfnamefont {A.}~\bibnamefont {{Agnello}}},
  \bibinfo {author} {\bibfnamefont {T.}~\bibnamefont {{Treu}}}, \bibinfo
  {author} {\bibfnamefont {C.~R.}\ \bibnamefont {{Keeton}}},\ and\ \bibinfo
  {author} {\bibfnamefont {A.~M.}\ \bibnamefont {{Nierenberg}}},\ }\bibfield
  {title} {\bibinfo {title} {{Strong lensing signatures of luminous structure
  and substructure in early-type galaxies}},\ }\href
  {https://doi.org/10.1093/mnras/stx158} {\bibfield  {journal} {\bibinfo
  {journal} {\mnras}\ }\textbf {\bibinfo {volume} {467}},\ \bibinfo {pages}
  {3970} (\bibinfo {year} {2017})},\ \Eprint {https://arxiv.org/abs/1610.08525}
  {arXiv:1610.08525 [astro-ph.CO]} \BibitemShut {NoStop}%
\bibitem [{\citenamefont {{Hsueh}}\ \emph {et~al.}(2017)\citenamefont
  {{Hsueh}}, \citenamefont {{Oldham}}, \citenamefont {{Spingola}},
  \citenamefont {{Vegetti}}, \citenamefont {{Fassnacht}}, \citenamefont
  {{Auger}}, \citenamefont {{Koopmans}}, \citenamefont {{McKean}},\ and\
  \citenamefont {{Lagattuta}}}]{Hsueh++17}%
  \BibitemOpen
  \bibfield  {author} {\bibinfo {author} {\bibfnamefont {J.~W.}\ \bibnamefont
  {{Hsueh}}}, \bibinfo {author} {\bibfnamefont {L.}~\bibnamefont {{Oldham}}},
  \bibinfo {author} {\bibfnamefont {C.}~\bibnamefont {{Spingola}}}, \bibinfo
  {author} {\bibfnamefont {S.}~\bibnamefont {{Vegetti}}}, \bibinfo {author}
  {\bibfnamefont {C.~D.}\ \bibnamefont {{Fassnacht}}}, \bibinfo {author}
  {\bibfnamefont {M.~W.}\ \bibnamefont {{Auger}}}, \bibinfo {author}
  {\bibfnamefont {L.~V.~E.}\ \bibnamefont {{Koopmans}}}, \bibinfo {author}
  {\bibfnamefont {J.~P.}\ \bibnamefont {{McKean}}},\ and\ \bibinfo {author}
  {\bibfnamefont {D.~J.}\ \bibnamefont {{Lagattuta}}},\ }\bibfield  {title}
  {\bibinfo {title} {{SHARP - IV. An apparent flux-ratio anomaly resolved by
  the edge-on disc in B0712+472}},\ }\href
  {https://doi.org/10.1093/mnras/stx1082} {\bibfield  {journal} {\bibinfo
  {journal} {\mnras}\ }\textbf {\bibinfo {volume} {469}},\ \bibinfo {pages}
  {3713} (\bibinfo {year} {2017})},\ \Eprint {https://arxiv.org/abs/1701.06575}
  {arXiv:1701.06575 [astro-ph.GA]} \BibitemShut {NoStop}%
\bibitem [{Note2()}]{Note2}%
  \BibitemOpen
  \bibinfo {note} {This is the relevant mass range for substructure lensing
  because the size of the background source removes signal from less massive
  halos, while halos more massive than $10^{10} M_{\odot }$ are rare, and
  likely host a luminous galaxy, in which case we would explicitly model
  them.}\BibitemShut {Stop}%
\bibitem [{\citenamefont {{Springel}}\ \emph {et~al.}(2008)\citenamefont
  {{Springel}}, \citenamefont {{Wang}}, \citenamefont {{Vogelsberger}},
  \citenamefont {{Ludlow}}, \citenamefont {{Jenkins}}, \citenamefont {{Helmi}},
  \citenamefont {{Navarro}}, \citenamefont {{Frenk}},\ and\ \citenamefont
  {{White}}}]{Springel++08}%
  \BibitemOpen
  \bibfield  {author} {\bibinfo {author} {\bibfnamefont {V.}~\bibnamefont
  {{Springel}}}, \bibinfo {author} {\bibfnamefont {J.}~\bibnamefont {{Wang}}},
  \bibinfo {author} {\bibfnamefont {M.}~\bibnamefont {{Vogelsberger}}},
  \bibinfo {author} {\bibfnamefont {A.}~\bibnamefont {{Ludlow}}}, \bibinfo
  {author} {\bibfnamefont {A.}~\bibnamefont {{Jenkins}}}, \bibinfo {author}
  {\bibfnamefont {A.}~\bibnamefont {{Helmi}}}, \bibinfo {author} {\bibfnamefont
  {J.~F.}\ \bibnamefont {{Navarro}}}, \bibinfo {author} {\bibfnamefont {C.~S.}\
  \bibnamefont {{Frenk}}},\ and\ \bibinfo {author} {\bibfnamefont {S.~D.~M.}\
  \bibnamefont {{White}}},\ }\bibfield  {title} {\bibinfo {title} {{The
  Aquarius Project: the subhaloes of galactic haloes}},\ }\href
  {https://doi.org/10.1111/j.1365-2966.2008.14066.x} {\bibfield  {journal}
  {\bibinfo  {journal} {\mnras}\ }\textbf {\bibinfo {volume} {391}},\ \bibinfo
  {pages} {1685} (\bibinfo {year} {2008})},\ \Eprint
  {https://arxiv.org/abs/0809.0898} {arXiv:0809.0898 [astro-ph]} \BibitemShut
  {NoStop}%
\bibitem [{\citenamefont {{Fiacconi}}\ \emph {et~al.}(2016)\citenamefont
  {{Fiacconi}}, \citenamefont {{Madau}}, \citenamefont {{Potter}},\ and\
  \citenamefont {{Stadel}}}]{Fiacconi++16}%
  \BibitemOpen
  \bibfield  {author} {\bibinfo {author} {\bibfnamefont {D.}~\bibnamefont
  {{Fiacconi}}}, \bibinfo {author} {\bibfnamefont {P.}~\bibnamefont {{Madau}}},
  \bibinfo {author} {\bibfnamefont {D.}~\bibnamefont {{Potter}}},\ and\
  \bibinfo {author} {\bibfnamefont {J.}~\bibnamefont {{Stadel}}},\ }\bibfield
  {title} {\bibinfo {title} {{Cold Dark Matter Substructures in Early-type
  Galaxy Halos}},\ }\href {https://doi.org/10.3847/0004-637X/824/2/144}
  {\bibfield  {journal} {\bibinfo  {journal} {\apj}\ }\textbf {\bibinfo
  {volume} {824}},\ \bibinfo {eid} {144} (\bibinfo {year} {2016})},\ \Eprint
  {https://arxiv.org/abs/1602.03526} {arXiv:1602.03526 [astro-ph.GA]}
  \BibitemShut {NoStop}%
\bibitem [{\citenamefont {{Huo}}\ \emph {et~al.}(2020)\citenamefont {{Huo}},
  \citenamefont {{Yu}},\ and\ \citenamefont {{Zhong}}}]{Huo++20}%
  \BibitemOpen
  \bibfield  {author} {\bibinfo {author} {\bibfnamefont {R.}~\bibnamefont
  {{Huo}}}, \bibinfo {author} {\bibfnamefont {H.-B.}\ \bibnamefont {{Yu}}},\
  and\ \bibinfo {author} {\bibfnamefont {Y.-M.}\ \bibnamefont {{Zhong}}},\
  }\bibfield  {title} {\bibinfo {title} {{The structure of dissipative dark
  matter halos}},\ }\href {https://doi.org/10.1088/1475-7516/2020/06/051}
  {\bibfield  {journal} {\bibinfo  {journal} {\jcap}\ }\textbf {\bibinfo
  {volume} {2020}},\ \bibinfo {eid} {051} (\bibinfo {year} {2020})},\ \Eprint
  {https://arxiv.org/abs/1912.06757} {arXiv:1912.06757 [astro-ph.CO]}
  \BibitemShut {NoStop}%
\bibitem [{Note3()}]{Note3}%
  \BibitemOpen
  \bibinfo {note} {The ratio between the dissipative cross section and the
  elastic cross section is about $\sigma (\chi \chi \to \chi \chi \phi )/\sigma
  (\chi \chi \to \chi \chi ) \sim {\alpha _\chi }/{(4\pi )}$ for Yukawa models,
  or $\sim 10^{-4}$ for the benchmarks we considered. In these cases, elastic
  scattering dominates the heat conduction \protect \citep
  {Essig++19}.}\BibitemShut {Stop}%
\bibitem [{\citenamefont {{Banik}}\ \emph {et~al.}(2021)\citenamefont
  {{Banik}}, \citenamefont {{Bovy}}, \citenamefont {{Bertone}}, \citenamefont
  {{Erkal}},\ and\ \citenamefont {{de Boer}}}]{Banik++21}%
  \BibitemOpen
  \bibfield  {author} {\bibinfo {author} {\bibfnamefont {N.}~\bibnamefont
  {{Banik}}}, \bibinfo {author} {\bibfnamefont {J.}~\bibnamefont {{Bovy}}},
  \bibinfo {author} {\bibfnamefont {G.}~\bibnamefont {{Bertone}}}, \bibinfo
  {author} {\bibfnamefont {D.}~\bibnamefont {{Erkal}}},\ and\ \bibinfo {author}
  {\bibfnamefont {T.~J.~L.}\ \bibnamefont {{de Boer}}},\ }\bibfield  {title}
  {\bibinfo {title} {{Evidence of a population of dark subhaloes from Gaia and
  Pan-STARRS observations of the GD-1 stream}},\ }\href
  {https://doi.org/10.1093/mnras/stab210} {\bibfield  {journal} {\bibinfo
  {journal} {\mnras}\ }\textbf {\bibinfo {volume} {502}},\ \bibinfo {pages}
  {2364} (\bibinfo {year} {2021})},\ \Eprint {https://arxiv.org/abs/1911.02662}
  {arXiv:1911.02662 [astro-ph.GA]} \BibitemShut {NoStop}%
\bibitem [{\citenamefont {{Bovy}}\ \emph {et~al.}(2017)\citenamefont {{Bovy}},
  \citenamefont {{Erkal}},\ and\ \citenamefont {{Sanders}}}]{Bovy++17}%
  \BibitemOpen
  \bibfield  {author} {\bibinfo {author} {\bibfnamefont {J.}~\bibnamefont
  {{Bovy}}}, \bibinfo {author} {\bibfnamefont {D.}~\bibnamefont {{Erkal}}},\
  and\ \bibinfo {author} {\bibfnamefont {J.~L.}\ \bibnamefont {{Sanders}}},\
  }\bibfield  {title} {\bibinfo {title} {{Linear perturbation theory for tidal
  streams and the small-scale CDM power spectrum}},\ }\href
  {https://doi.org/10.1093/mnras/stw3067} {\bibfield  {journal} {\bibinfo
  {journal} {\mnras}\ }\textbf {\bibinfo {volume} {466}},\ \bibinfo {pages}
  {628} (\bibinfo {year} {2017})},\ \Eprint {https://arxiv.org/abs/1606.03470}
  {arXiv:1606.03470 [astro-ph.GA]} \BibitemShut {NoStop}%
\bibitem [{\citenamefont {{Banik}}\ and\ \citenamefont
  {{Bovy}}(2021)}]{Banik++21b}%
  \BibitemOpen
  \bibfield  {author} {\bibinfo {author} {\bibfnamefont {N.}~\bibnamefont
  {{Banik}}}\ and\ \bibinfo {author} {\bibfnamefont {J.}~\bibnamefont
  {{Bovy}}},\ }\bibfield  {title} {\bibinfo {title} {{On N-body simulations of
  globular cluster streams}},\ }\href {https://doi.org/10.1093/mnras/stab886}
  {\bibfield  {journal} {\bibinfo  {journal} {\mnras}\ }\textbf {\bibinfo
  {volume} {504}},\ \bibinfo {pages} {648} (\bibinfo {year} {2021})},\ \Eprint
  {https://arxiv.org/abs/2101.12201} {arXiv:2101.12201 [astro-ph.GA]}
  \BibitemShut {NoStop}%
\bibitem [{\citenamefont {{Nadler}}\ \emph
  {et~al.}(2020{\natexlab{b}})\citenamefont {{Nadler}}, \citenamefont
  {{Wechsler}}, \citenamefont {{Bechtol}}, \citenamefont {{Mao}}, \citenamefont
  {{Green}}, \citenamefont {{Drlica-Wagner}}, \citenamefont {{McNanna}},
  \citenamefont {{Mau}}, \citenamefont {{Pace}}, \citenamefont {{Simon}},
  \citenamefont {{Kravtsov}}, \citenamefont {{Dodelson}}, \citenamefont {{Li}},
  \citenamefont {{Riley}}, \citenamefont {{Wang}}, \citenamefont {{Abbott}},
  \citenamefont {{Aguena}}, \citenamefont {{Allam}}, \citenamefont {{Annis}},
  \citenamefont {{Avila}}, \citenamefont {{Bernstein}}, \citenamefont
  {{Bertin}}, \citenamefont {{Brooks}}, \citenamefont {{Burke}}, \citenamefont
  {{Rosell}}, \citenamefont {{Kind}}, \citenamefont {{Carretero}},
  \citenamefont {{Costanzi}}, \citenamefont {{da Costa}}, \citenamefont {{De
  Vicente}}, \citenamefont {{Desai}}, \citenamefont {{Evrard}}, \citenamefont
  {{Flaugher}}, \citenamefont {{Fosalba}}, \citenamefont {{Frieman}},
  \citenamefont {{Garc{\'\i}a-Bellido}}, \citenamefont {{Gaztanaga}},
  \citenamefont {{Gerdes}}, \citenamefont {{Gruen}}, \citenamefont
  {{Gschwend}}, \citenamefont {{Gutierrez}}, \citenamefont {{Hartley}},
  \citenamefont {{Hinton}}, \citenamefont {{Honscheid}}, \citenamefont
  {{Krause}}, \citenamefont {{Kuehn}}, \citenamefont {{Kuropatkin}},
  \citenamefont {{Lahav}}, \citenamefont {{Maia}}, \citenamefont {{Marshall}},
  \citenamefont {{Menanteau}}, \citenamefont {{Miquel}}, \citenamefont
  {{Palmese}}, \citenamefont {{Paz-Chinch{\'o}n}}, \citenamefont {{Plazas}},
  \citenamefont {{Romer}}, \citenamefont {{Sanchez}}, \citenamefont
  {{Santiago}}, \citenamefont {{Scarpine}}, \citenamefont {{Serrano}},
  \citenamefont {{Smith}}, \citenamefont {{Soares-Santos}}, \citenamefont
  {{Suchyta}}, \citenamefont {{Tarle}}, \citenamefont {{Thomas}}, \citenamefont
  {{Varga}}, \citenamefont {{Walker}},\ and\ \citenamefont {{DES
  Collaboration}}}]{Nadler++20b}%
  \BibitemOpen
  \bibfield  {author} {\bibinfo {author} {\bibfnamefont {E.~O.}\ \bibnamefont
  {{Nadler}}}, \bibinfo {author} {\bibfnamefont {R.~H.}\ \bibnamefont
  {{Wechsler}}}, \bibinfo {author} {\bibfnamefont {K.}~\bibnamefont
  {{Bechtol}}}, \bibinfo {author} {\bibfnamefont {Y.~Y.}\ \bibnamefont
  {{Mao}}}, \bibinfo {author} {\bibfnamefont {G.}~\bibnamefont {{Green}}},
  \bibinfo {author} {\bibfnamefont {A.}~\bibnamefont {{Drlica-Wagner}}},
  \bibinfo {author} {\bibfnamefont {M.}~\bibnamefont {{McNanna}}}, \bibinfo
  {author} {\bibfnamefont {S.}~\bibnamefont {{Mau}}}, \bibinfo {author}
  {\bibfnamefont {A.~B.}\ \bibnamefont {{Pace}}}, \bibinfo {author}
  {\bibfnamefont {J.~D.}\ \bibnamefont {{Simon}}}, \bibinfo {author}
  {\bibfnamefont {A.}~\bibnamefont {{Kravtsov}}}, \bibinfo {author}
  {\bibfnamefont {S.}~\bibnamefont {{Dodelson}}}, \bibinfo {author}
  {\bibfnamefont {T.~S.}\ \bibnamefont {{Li}}}, \bibinfo {author}
  {\bibfnamefont {A.~H.}\ \bibnamefont {{Riley}}}, \bibinfo {author}
  {\bibfnamefont {M.~Y.}\ \bibnamefont {{Wang}}}, \bibinfo {author}
  {\bibfnamefont {T.~M.~C.}\ \bibnamefont {{Abbott}}}, \bibinfo {author}
  {\bibfnamefont {M.}~\bibnamefont {{Aguena}}}, \bibinfo {author}
  {\bibfnamefont {S.}~\bibnamefont {{Allam}}}, \bibinfo {author} {\bibfnamefont
  {J.}~\bibnamefont {{Annis}}}, \bibinfo {author} {\bibfnamefont
  {S.}~\bibnamefont {{Avila}}}, \bibinfo {author} {\bibfnamefont {G.~M.}\
  \bibnamefont {{Bernstein}}}, \bibinfo {author} {\bibfnamefont
  {E.}~\bibnamefont {{Bertin}}}, \bibinfo {author} {\bibfnamefont
  {D.}~\bibnamefont {{Brooks}}}, \bibinfo {author} {\bibfnamefont {D.~L.}\
  \bibnamefont {{Burke}}}, \bibinfo {author} {\bibfnamefont {A.~C.}\
  \bibnamefont {{Rosell}}}, \bibinfo {author} {\bibfnamefont {M.~C.}\
  \bibnamefont {{Kind}}}, \bibinfo {author} {\bibfnamefont {J.}~\bibnamefont
  {{Carretero}}}, \bibinfo {author} {\bibfnamefont {M.}~\bibnamefont
  {{Costanzi}}}, \bibinfo {author} {\bibfnamefont {L.~N.}\ \bibnamefont {{da
  Costa}}}, \bibinfo {author} {\bibfnamefont {J.}~\bibnamefont {{De Vicente}}},
  \bibinfo {author} {\bibfnamefont {S.}~\bibnamefont {{Desai}}}, \bibinfo
  {author} {\bibfnamefont {A.~E.}\ \bibnamefont {{Evrard}}}, \bibinfo {author}
  {\bibfnamefont {B.}~\bibnamefont {{Flaugher}}}, \bibinfo {author}
  {\bibfnamefont {P.}~\bibnamefont {{Fosalba}}}, \bibinfo {author}
  {\bibfnamefont {J.}~\bibnamefont {{Frieman}}}, \bibinfo {author}
  {\bibfnamefont {J.}~\bibnamefont {{Garc{\'\i}a-Bellido}}}, \bibinfo {author}
  {\bibfnamefont {E.}~\bibnamefont {{Gaztanaga}}}, \bibinfo {author}
  {\bibfnamefont {D.~W.}\ \bibnamefont {{Gerdes}}}, \bibinfo {author}
  {\bibfnamefont {D.}~\bibnamefont {{Gruen}}}, \bibinfo {author} {\bibfnamefont
  {J.}~\bibnamefont {{Gschwend}}}, \bibinfo {author} {\bibfnamefont
  {G.}~\bibnamefont {{Gutierrez}}}, \bibinfo {author} {\bibfnamefont {W.~G.}\
  \bibnamefont {{Hartley}}}, \bibinfo {author} {\bibfnamefont {S.~R.}\
  \bibnamefont {{Hinton}}}, \bibinfo {author} {\bibfnamefont {K.}~\bibnamefont
  {{Honscheid}}}, \bibinfo {author} {\bibfnamefont {E.}~\bibnamefont
  {{Krause}}}, \bibinfo {author} {\bibfnamefont {K.}~\bibnamefont {{Kuehn}}},
  \bibinfo {author} {\bibfnamefont {N.}~\bibnamefont {{Kuropatkin}}}, \bibinfo
  {author} {\bibfnamefont {O.}~\bibnamefont {{Lahav}}}, \bibinfo {author}
  {\bibfnamefont {M.~A.~G.}\ \bibnamefont {{Maia}}}, \bibinfo {author}
  {\bibfnamefont {J.~L.}\ \bibnamefont {{Marshall}}}, \bibinfo {author}
  {\bibfnamefont {F.}~\bibnamefont {{Menanteau}}}, \bibinfo {author}
  {\bibfnamefont {R.}~\bibnamefont {{Miquel}}}, \bibinfo {author}
  {\bibfnamefont {A.}~\bibnamefont {{Palmese}}}, \bibinfo {author}
  {\bibfnamefont {F.}~\bibnamefont {{Paz-Chinch{\'o}n}}}, \bibinfo {author}
  {\bibfnamefont {A.~A.}\ \bibnamefont {{Plazas}}}, \bibinfo {author}
  {\bibfnamefont {A.~K.}\ \bibnamefont {{Romer}}}, \bibinfo {author}
  {\bibfnamefont {E.}~\bibnamefont {{Sanchez}}}, \bibinfo {author}
  {\bibfnamefont {B.}~\bibnamefont {{Santiago}}}, \bibinfo {author}
  {\bibfnamefont {V.}~\bibnamefont {{Scarpine}}}, \bibinfo {author}
  {\bibfnamefont {S.}~\bibnamefont {{Serrano}}}, \bibinfo {author}
  {\bibfnamefont {M.}~\bibnamefont {{Smith}}}, \bibinfo {author} {\bibfnamefont
  {M.}~\bibnamefont {{Soares-Santos}}}, \bibinfo {author} {\bibfnamefont
  {E.}~\bibnamefont {{Suchyta}}}, \bibinfo {author} {\bibfnamefont
  {G.}~\bibnamefont {{Tarle}}}, \bibinfo {author} {\bibfnamefont
  {D.}~\bibnamefont {{Thomas}}}, \bibinfo {author} {\bibfnamefont {T.~N.}\
  \bibnamefont {{Varga}}}, \bibinfo {author} {\bibfnamefont {A.~R.}\
  \bibnamefont {{Walker}}},\ and\ \bibinfo {author} {\bibnamefont {{DES
  Collaboration}}},\ }\bibfield  {title} {\bibinfo {title} {{Milky Way
  Satellite Census. II. Galaxy-Halo Connection Constraints Including the Impact
  of the Large Magellanic Cloud}},\ }\href
  {https://doi.org/10.3847/1538-4357/ab846a} {\bibfield  {journal} {\bibinfo
  {journal} {\apj}\ }\textbf {\bibinfo {volume} {893}},\ \bibinfo {eid} {48}
  (\bibinfo {year} {2020}{\natexlab{b}})},\ \Eprint
  {https://arxiv.org/abs/1912.03303} {arXiv:1912.03303 [astro-ph.GA]}
  \BibitemShut {NoStop}%
\bibitem [{\citenamefont {{Kim}}\ and\ \citenamefont
  {{Peter}}(2021)}]{Kim++21}%
  \BibitemOpen
  \bibfield  {author} {\bibinfo {author} {\bibfnamefont {S.~Y.}\ \bibnamefont
  {{Kim}}}\ and\ \bibinfo {author} {\bibfnamefont {A.~H.~G.}\ \bibnamefont
  {{Peter}}},\ }\bibfield  {title} {\bibinfo {title} {{The Milky Way satellite
  velocity function is a sharp probe of small-scale structure problems}},\
  }\href@noop {} {\bibfield  {journal} {\bibinfo  {journal} {arXiv e-prints}\
  ,\ \bibinfo {eid} {arXiv:2106.09050}} (\bibinfo {year} {2021})},\ \Eprint
  {https://arxiv.org/abs/2106.09050} {arXiv:2106.09050 [astro-ph.GA]}
  \BibitemShut {NoStop}%
\bibitem [{\citenamefont {{Dekker}}\ \emph {et~al.}(2021)\citenamefont
  {{Dekker}}, \citenamefont {{Ando}}, \citenamefont {{Correa}},\ and\
  \citenamefont {{Ng}}}]{Dekker++21}%
  \BibitemOpen
  \bibfield  {author} {\bibinfo {author} {\bibfnamefont {A.}~\bibnamefont
  {{Dekker}}}, \bibinfo {author} {\bibfnamefont {S.}~\bibnamefont {{Ando}}},
  \bibinfo {author} {\bibfnamefont {C.~A.}\ \bibnamefont {{Correa}}},\ and\
  \bibinfo {author} {\bibfnamefont {K.~C.~Y.}\ \bibnamefont {{Ng}}},\
  }\bibfield  {title} {\bibinfo {title} {{Warm Dark Matter Constraints Using
  Milky-Way Satellite Observations and Subhalo Evolution Modeling}},\
  }\href@noop {} {\bibfield  {journal} {\bibinfo  {journal} {arXiv e-prints}\
  ,\ \bibinfo {eid} {arXiv:2111.13137}} (\bibinfo {year} {2021})},\ \Eprint
  {https://arxiv.org/abs/2111.13137} {arXiv:2111.13137 [astro-ph.CO]}
  \BibitemShut {NoStop}%
\bibitem [{\citenamefont {{Ebisu}}\ \emph {et~al.}(2022)\citenamefont
  {{Ebisu}}, \citenamefont {{Ishiyama}},\ and\ \citenamefont
  {{Hayashi}}}]{Ebisu++22}%
  \BibitemOpen
  \bibfield  {author} {\bibinfo {author} {\bibfnamefont {T.}~\bibnamefont
  {{Ebisu}}}, \bibinfo {author} {\bibfnamefont {T.}~\bibnamefont
  {{Ishiyama}}},\ and\ \bibinfo {author} {\bibfnamefont {K.}~\bibnamefont
  {{Hayashi}}},\ }\bibfield  {title} {\bibinfo {title} {{Constraining
  self-interacting dark matter with dwarf spheroidal galaxies and
  high-resolution cosmological N -body simulations}},\ }\href
  {https://doi.org/10.1103/PhysRevD.105.023016} {\bibfield  {journal} {\bibinfo
   {journal} {\prd}\ }\textbf {\bibinfo {volume} {105}},\ \bibinfo {eid}
  {023016} (\bibinfo {year} {2022})},\ \Eprint
  {https://arxiv.org/abs/2107.05967} {arXiv:2107.05967 [astro-ph.GA]}
  \BibitemShut {NoStop}%
\bibitem [{\citenamefont {{Shen}}\ \emph {et~al.}(2022)\citenamefont {{Shen}},
  \citenamefont {{Hopkins}}, \citenamefont {{Necib}}, \citenamefont {{Jiang}},
  \citenamefont {{Boylan-Kolchin}},\ and\ \citenamefont {{Wetzel}}}]{Shen++22}%
  \BibitemOpen
  \bibfield  {author} {\bibinfo {author} {\bibfnamefont {X.}~\bibnamefont
  {{Shen}}}, \bibinfo {author} {\bibfnamefont {P.~F.}\ \bibnamefont
  {{Hopkins}}}, \bibinfo {author} {\bibfnamefont {L.}~\bibnamefont {{Necib}}},
  \bibinfo {author} {\bibfnamefont {F.}~\bibnamefont {{Jiang}}}, \bibinfo
  {author} {\bibfnamefont {M.}~\bibnamefont {{Boylan-Kolchin}}},\ and\ \bibinfo
  {author} {\bibfnamefont {A.}~\bibnamefont {{Wetzel}}},\ }\bibfield  {title}
  {\bibinfo {title} {{Dissipative Dark Matter on FIRE: II. Observational
  signatures and constraints from local dwarf galaxies}},\ }\href@noop {}
  {\bibfield  {journal} {\bibinfo  {journal} {arXiv e-prints}\ ,\ \bibinfo
  {eid} {arXiv:2206.05327}} (\bibinfo {year} {2022})},\ \Eprint
  {https://arxiv.org/abs/2206.05327} {arXiv:2206.05327 [astro-ph.GA]}
  \BibitemShut {NoStop}%
\bibitem [{\citenamefont {{Vegetti}}\ \emph {et~al.}(2012)\citenamefont
  {{Vegetti}}, \citenamefont {{Lagattuta}}, \citenamefont {{McKean}},
  \citenamefont {{Auger}}, \citenamefont {{Fassnacht}},\ and\ \citenamefont
  {{Koopmans}}}]{Vegetti++12}%
  \BibitemOpen
  \bibfield  {author} {\bibinfo {author} {\bibfnamefont {S.}~\bibnamefont
  {{Vegetti}}}, \bibinfo {author} {\bibfnamefont {D.~J.}\ \bibnamefont
  {{Lagattuta}}}, \bibinfo {author} {\bibfnamefont {J.~P.}\ \bibnamefont
  {{McKean}}}, \bibinfo {author} {\bibfnamefont {M.~W.}\ \bibnamefont
  {{Auger}}}, \bibinfo {author} {\bibfnamefont {C.~D.}\ \bibnamefont
  {{Fassnacht}}},\ and\ \bibinfo {author} {\bibfnamefont {L.~V.~E.}\
  \bibnamefont {{Koopmans}}},\ }\bibfield  {title} {\bibinfo {title}
  {{Gravitational detection of a low-mass dark satellite galaxy at cosmological
  distance}},\ }\href {https://doi.org/10.1038/nature10669} {\bibfield
  {journal} {\bibinfo  {journal} {\nat}\ }\textbf {\bibinfo {volume} {481}},\
  \bibinfo {pages} {341} (\bibinfo {year} {2012})},\ \Eprint
  {https://arxiv.org/abs/1201.3643} {arXiv:1201.3643 [astro-ph.CO]}
  \BibitemShut {NoStop}%
\bibitem [{\citenamefont {{Vegetti}}\ \emph {et~al.}(2014)\citenamefont
  {{Vegetti}}, \citenamefont {{Koopmans}}, \citenamefont {{Auger}},
  \citenamefont {{Treu}},\ and\ \citenamefont {{Bolton}}}]{Vegetti++14}%
  \BibitemOpen
  \bibfield  {author} {\bibinfo {author} {\bibfnamefont {S.}~\bibnamefont
  {{Vegetti}}}, \bibinfo {author} {\bibfnamefont {L.~V.~E.}\ \bibnamefont
  {{Koopmans}}}, \bibinfo {author} {\bibfnamefont {M.~W.}\ \bibnamefont
  {{Auger}}}, \bibinfo {author} {\bibfnamefont {T.}~\bibnamefont {{Treu}}},\
  and\ \bibinfo {author} {\bibfnamefont {A.~S.}\ \bibnamefont {{Bolton}}},\
  }\bibfield  {title} {\bibinfo {title} {{Inference of the cold dark matter
  substructure mass function at z = 0.2 using strong gravitational lenses}},\
  }\href {https://doi.org/10.1093/mnras/stu943} {\bibfield  {journal} {\bibinfo
   {journal} {\mnras}\ }\textbf {\bibinfo {volume} {442}},\ \bibinfo {pages}
  {2017} (\bibinfo {year} {2014})},\ \Eprint {https://arxiv.org/abs/1405.3666}
  {arXiv:1405.3666 [astro-ph.GA]} \BibitemShut {NoStop}%
\bibitem [{\citenamefont {{Hezaveh}}\ \emph {et~al.}(2016)\citenamefont
  {{Hezaveh}}, \citenamefont {{Dalal}}, \citenamefont {{Marrone}},
  \citenamefont {{Mao}}, \citenamefont {{Morningstar}}, \citenamefont {{Wen}},
  \citenamefont {{Blandford}}, \citenamefont {{Carlstrom}}, \citenamefont
  {{Fassnacht}}, \citenamefont {{Holder}}, \citenamefont {{Kemball}},
  \citenamefont {{Marshall}}, \citenamefont {{Murray}}, \citenamefont
  {{Perreault Levasseur}}, \citenamefont {{Vieira}},\ and\ \citenamefont
  {{Wechsler}}}]{Hezaveh++16}%
  \BibitemOpen
  \bibfield  {author} {\bibinfo {author} {\bibfnamefont {Y.~D.}\ \bibnamefont
  {{Hezaveh}}}, \bibinfo {author} {\bibfnamefont {N.}~\bibnamefont {{Dalal}}},
  \bibinfo {author} {\bibfnamefont {D.~P.}\ \bibnamefont {{Marrone}}}, \bibinfo
  {author} {\bibfnamefont {Y.-Y.}\ \bibnamefont {{Mao}}}, \bibinfo {author}
  {\bibfnamefont {W.}~\bibnamefont {{Morningstar}}}, \bibinfo {author}
  {\bibfnamefont {D.}~\bibnamefont {{Wen}}}, \bibinfo {author} {\bibfnamefont
  {R.~D.}\ \bibnamefont {{Blandford}}}, \bibinfo {author} {\bibfnamefont
  {J.~E.}\ \bibnamefont {{Carlstrom}}}, \bibinfo {author} {\bibfnamefont
  {C.~D.}\ \bibnamefont {{Fassnacht}}}, \bibinfo {author} {\bibfnamefont
  {G.~P.}\ \bibnamefont {{Holder}}}, \bibinfo {author} {\bibfnamefont
  {A.}~\bibnamefont {{Kemball}}}, \bibinfo {author} {\bibfnamefont {P.~J.}\
  \bibnamefont {{Marshall}}}, \bibinfo {author} {\bibfnamefont
  {N.}~\bibnamefont {{Murray}}}, \bibinfo {author} {\bibfnamefont
  {L.}~\bibnamefont {{Perreault Levasseur}}}, \bibinfo {author} {\bibfnamefont
  {J.~D.}\ \bibnamefont {{Vieira}}},\ and\ \bibinfo {author} {\bibfnamefont
  {R.~H.}\ \bibnamefont {{Wechsler}}},\ }\bibfield  {title} {\bibinfo {title}
  {{Detection of Lensing Substructure Using ALMA Observations of the Dusty
  Galaxy SDP.81}},\ }\href {https://doi.org/10.3847/0004-637X/823/1/37}
  {\bibfield  {journal} {\bibinfo  {journal} {\apj}\ }\textbf {\bibinfo
  {volume} {823}},\ \bibinfo {eid} {37} (\bibinfo {year} {2016})},\ \Eprint
  {https://arxiv.org/abs/1601.01388} {arXiv:1601.01388 [astro-ph.CO]}
  \BibitemShut {NoStop}%
\bibitem [{\citenamefont {{Birrer}}\ \emph {et~al.}(2017)\citenamefont
  {{Birrer}}, \citenamefont {{Amara}},\ and\ \citenamefont
  {{Refregier}}}]{Birrer++17}%
  \BibitemOpen
  \bibfield  {author} {\bibinfo {author} {\bibfnamefont {S.}~\bibnamefont
  {{Birrer}}}, \bibinfo {author} {\bibfnamefont {A.}~\bibnamefont {{Amara}}},\
  and\ \bibinfo {author} {\bibfnamefont {A.}~\bibnamefont {{Refregier}}},\
  }\bibfield  {title} {\bibinfo {title} {{Lensing substructure quantification
  in RXJ1131-1231: a 2 keV lower bound on dark matter thermal relic mass}},\
  }\href {https://doi.org/10.1088/1475-7516/2017/05/037} {\bibfield  {journal}
  {\bibinfo  {journal} {\jcap}\ }\textbf {\bibinfo {volume} {2017}},\ \bibinfo
  {eid} {037} (\bibinfo {year} {2017})},\ \Eprint
  {https://arxiv.org/abs/1702.00009} {arXiv:1702.00009 [astro-ph.CO]}
  \BibitemShut {NoStop}%
\bibitem [{\citenamefont {{{\c{C}}a{\u{g}}an {\c{S}}eng{\"u}l}}\ \emph
  {et~al.}(2021)\citenamefont {{{\c{C}}a{\u{g}}an {\c{S}}eng{\"u}l}},
  \citenamefont {{Dvorkin}}, \citenamefont {{Ostdiek}},\ and\ \citenamefont
  {{Tsang}}}]{CaganSengul++21}%
  \BibitemOpen
  \bibfield  {author} {\bibinfo {author} {\bibfnamefont {A.}~\bibnamefont
  {{{\c{C}}a{\u{g}}an {\c{S}}eng{\"u}l}}}, \bibinfo {author} {\bibfnamefont
  {C.}~\bibnamefont {{Dvorkin}}}, \bibinfo {author} {\bibfnamefont
  {B.}~\bibnamefont {{Ostdiek}}},\ and\ \bibinfo {author} {\bibfnamefont
  {A.}~\bibnamefont {{Tsang}}},\ }\bibfield  {title} {\bibinfo {title}
  {{Substructure Detection Reanalyzed: Dark Perturber shown to be a
  Line-of-Sight Halo}},\ }\href@noop {} {\bibfield  {journal} {\bibinfo
  {journal} {arXiv e-prints}\ ,\ \bibinfo {eid} {arXiv:2112.00749}} (\bibinfo
  {year} {2021})},\ \Eprint {https://arxiv.org/abs/2112.00749}
  {arXiv:2112.00749 [astro-ph.CO]} \BibitemShut {NoStop}%
\bibitem [{\citenamefont {{Minor}}\ \emph {et~al.}(2021)\citenamefont
  {{Minor}}, \citenamefont {{Gad-Nasr}}, \citenamefont {{Kaplinghat}},\ and\
  \citenamefont {{Vegetti}}}]{Minor++21}%
  \BibitemOpen
  \bibfield  {author} {\bibinfo {author} {\bibfnamefont {Q.}~\bibnamefont
  {{Minor}}}, \bibinfo {author} {\bibfnamefont {S.}~\bibnamefont {{Gad-Nasr}}},
  \bibinfo {author} {\bibfnamefont {M.}~\bibnamefont {{Kaplinghat}}},\ and\
  \bibinfo {author} {\bibfnamefont {S.}~\bibnamefont {{Vegetti}}},\ }\bibfield
  {title} {\bibinfo {title} {{An unexpected high concentration for the dark
  substructure in the gravitational lens SDSSJ0946+1006}},\ }\href
  {https://doi.org/10.1093/mnras/stab2247} {\bibfield  {journal} {\bibinfo
  {journal} {\mnras}\ }\textbf {\bibinfo {volume} {507}},\ \bibinfo {pages}
  {1662} (\bibinfo {year} {2021})},\ \Eprint {https://arxiv.org/abs/2011.10627}
  {arXiv:2011.10627 [astro-ph.GA]} \BibitemShut {NoStop}%
\bibitem [{\citenamefont {{He}}\ \emph {et~al.}(2022)\citenamefont {{He}},
  \citenamefont {{Robertson}}, \citenamefont {{Nightingale}}, \citenamefont
  {{Cole}}, \citenamefont {{Frenk}}, \citenamefont {{Massey}}, \citenamefont
  {{Amvrosiadis}}, \citenamefont {{Li}}, \citenamefont {{Cao}},\ and\
  \citenamefont {{Etherington}}}]{He++22}%
  \BibitemOpen
  \bibfield  {author} {\bibinfo {author} {\bibfnamefont {Q.}~\bibnamefont
  {{He}}}, \bibinfo {author} {\bibfnamefont {A.}~\bibnamefont {{Robertson}}},
  \bibinfo {author} {\bibfnamefont {J.}~\bibnamefont {{Nightingale}}}, \bibinfo
  {author} {\bibfnamefont {S.}~\bibnamefont {{Cole}}}, \bibinfo {author}
  {\bibfnamefont {C.~S.}\ \bibnamefont {{Frenk}}}, \bibinfo {author}
  {\bibfnamefont {R.}~\bibnamefont {{Massey}}}, \bibinfo {author}
  {\bibfnamefont {A.}~\bibnamefont {{Amvrosiadis}}}, \bibinfo {author}
  {\bibfnamefont {R.}~\bibnamefont {{Li}}}, \bibinfo {author} {\bibfnamefont
  {X.}~\bibnamefont {{Cao}}},\ and\ \bibinfo {author} {\bibfnamefont
  {A.}~\bibnamefont {{Etherington}}},\ }\bibfield  {title} {\bibinfo {title}
  {{A forward-modelling method to infer the dark matter particle mass from
  strong gravitational lenses}},\ }\href
  {https://doi.org/10.1093/mnras/stac191} {\bibfield  {journal} {\bibinfo
  {journal} {\mnras}\ }\textbf {\bibinfo {volume} {511}},\ \bibinfo {pages}
  {3046} (\bibinfo {year} {2022})},\ \Eprint {https://arxiv.org/abs/2010.13221}
  {arXiv:2010.13221 [astro-ph.CO]} \BibitemShut {NoStop}%
\bibitem [{\citenamefont {{Despali}}\ \emph {et~al.}(2022)\citenamefont
  {{Despali}}, \citenamefont {{Vegetti}}, \citenamefont {{White}},
  \citenamefont {{Powell}}, \citenamefont {{Stacey}}, \citenamefont
  {{Fassnacht}}, \citenamefont {{Rizzo}},\ and\ \citenamefont
  {{Enzi}}}]{Despali++22}%
  \BibitemOpen
  \bibfield  {author} {\bibinfo {author} {\bibfnamefont {G.}~\bibnamefont
  {{Despali}}}, \bibinfo {author} {\bibfnamefont {S.}~\bibnamefont
  {{Vegetti}}}, \bibinfo {author} {\bibfnamefont {S.~D.~M.}\ \bibnamefont
  {{White}}}, \bibinfo {author} {\bibfnamefont {D.~M.}\ \bibnamefont
  {{Powell}}}, \bibinfo {author} {\bibfnamefont {H.~R.}\ \bibnamefont
  {{Stacey}}}, \bibinfo {author} {\bibfnamefont {C.~D.}\ \bibnamefont
  {{Fassnacht}}}, \bibinfo {author} {\bibfnamefont {F.}~\bibnamefont
  {{Rizzo}}},\ and\ \bibinfo {author} {\bibfnamefont {W.}~\bibnamefont
  {{Enzi}}},\ }\bibfield  {title} {\bibinfo {title} {{Detecting low-mass haloes
  with strong gravitational lensing I: the effect of data quality and lensing
  configuration}},\ }\href {https://doi.org/10.1093/mnras/stab3537} {\bibfield
  {journal} {\bibinfo  {journal} {\mnras}\ }\textbf {\bibinfo {volume} {510}},\
  \bibinfo {pages} {2480} (\bibinfo {year} {2022})},\ \Eprint
  {https://arxiv.org/abs/2111.08718} {arXiv:2111.08718 [astro-ph.GA]}
  \BibitemShut {NoStop}%
\bibitem [{\citenamefont {{Amorisco}}\ \emph {et~al.}(2022)\citenamefont
  {{Amorisco}}, \citenamefont {{Nightingale}}, \citenamefont {{He}},
  \citenamefont {{Amvrosiadis}}, \citenamefont {{Cao}}, \citenamefont {{Cole}},
  \citenamefont {{Etherington}}, \citenamefont {{Frenk}}, \citenamefont {{Li}},
  \citenamefont {{Massey}},\ and\ \citenamefont {{Robertson}}}]{Amorisco++22}%
  \BibitemOpen
  \bibfield  {author} {\bibinfo {author} {\bibfnamefont {N.~C.}\ \bibnamefont
  {{Amorisco}}}, \bibinfo {author} {\bibfnamefont {J.}~\bibnamefont
  {{Nightingale}}}, \bibinfo {author} {\bibfnamefont {Q.}~\bibnamefont {{He}}},
  \bibinfo {author} {\bibfnamefont {A.}~\bibnamefont {{Amvrosiadis}}}, \bibinfo
  {author} {\bibfnamefont {X.}~\bibnamefont {{Cao}}}, \bibinfo {author}
  {\bibfnamefont {S.}~\bibnamefont {{Cole}}}, \bibinfo {author} {\bibfnamefont
  {A.}~\bibnamefont {{Etherington}}}, \bibinfo {author} {\bibfnamefont {C.~S.}\
  \bibnamefont {{Frenk}}}, \bibinfo {author} {\bibfnamefont {R.}~\bibnamefont
  {{Li}}}, \bibinfo {author} {\bibfnamefont {R.}~\bibnamefont {{Massey}}},\
  and\ \bibinfo {author} {\bibfnamefont {A.}~\bibnamefont {{Robertson}}},\
  }\bibfield  {title} {\bibinfo {title} {{Halo concentration strengthens dark
  matter constraints in galaxy-galaxy strong lensing analyses}},\ }\href
  {https://doi.org/10.1093/mnras/stab3527} {\bibfield  {journal} {\bibinfo
  {journal} {\mnras}\ }\textbf {\bibinfo {volume} {510}},\ \bibinfo {pages}
  {2464} (\bibinfo {year} {2022})},\ \Eprint {https://arxiv.org/abs/2109.00018}
  {arXiv:2109.00018 [astro-ph.CO]} \BibitemShut {NoStop}%
\bibitem [{\citenamefont {{Casadio}}\ \emph {et~al.}(2021)\citenamefont
  {{Casadio}}, \citenamefont {{Blinov}}, \citenamefont {{Readhead}},
  \citenamefont {{Browne}}, \citenamefont {{Wilkinson}}, \citenamefont
  {{Hovatta}}, \citenamefont {{Mandarakas}}, \citenamefont {{Pavlidou}},
  \citenamefont {{Tassis}}, \citenamefont {{Vedantham}}, \citenamefont
  {{Zensus}}, \citenamefont {{Diamantopoulos}}, \citenamefont {{Dolapsaki}},
  \citenamefont {{Gkimisi}}, \citenamefont {{Kalaitzidakis}}, \citenamefont
  {{Mastorakis}}, \citenamefont {{Nikolaou}}, \citenamefont {{Ntormousi}},
  \citenamefont {{Pelgrims}},\ and\ \citenamefont {{Psarras}}}]{Casadio++21}%
  \BibitemOpen
  \bibfield  {author} {\bibinfo {author} {\bibfnamefont {C.}~\bibnamefont
  {{Casadio}}}, \bibinfo {author} {\bibfnamefont {D.}~\bibnamefont {{Blinov}}},
  \bibinfo {author} {\bibfnamefont {A.~C.~S.}\ \bibnamefont {{Readhead}}},
  \bibinfo {author} {\bibfnamefont {I.~W.~A.}\ \bibnamefont {{Browne}}},
  \bibinfo {author} {\bibfnamefont {P.~N.}\ \bibnamefont {{Wilkinson}}},
  \bibinfo {author} {\bibfnamefont {T.}~\bibnamefont {{Hovatta}}}, \bibinfo
  {author} {\bibfnamefont {N.}~\bibnamefont {{Mandarakas}}}, \bibinfo {author}
  {\bibfnamefont {V.}~\bibnamefont {{Pavlidou}}}, \bibinfo {author}
  {\bibfnamefont {K.}~\bibnamefont {{Tassis}}}, \bibinfo {author}
  {\bibfnamefont {H.~K.}\ \bibnamefont {{Vedantham}}}, \bibinfo {author}
  {\bibfnamefont {J.~A.}\ \bibnamefont {{Zensus}}}, \bibinfo {author}
  {\bibfnamefont {V.}~\bibnamefont {{Diamantopoulos}}}, \bibinfo {author}
  {\bibfnamefont {K.~E.}\ \bibnamefont {{Dolapsaki}}}, \bibinfo {author}
  {\bibfnamefont {K.}~\bibnamefont {{Gkimisi}}}, \bibinfo {author}
  {\bibfnamefont {G.}~\bibnamefont {{Kalaitzidakis}}}, \bibinfo {author}
  {\bibfnamefont {M.}~\bibnamefont {{Mastorakis}}}, \bibinfo {author}
  {\bibfnamefont {K.}~\bibnamefont {{Nikolaou}}}, \bibinfo {author}
  {\bibfnamefont {E.}~\bibnamefont {{Ntormousi}}}, \bibinfo {author}
  {\bibfnamefont {V.}~\bibnamefont {{Pelgrims}}},\ and\ \bibinfo {author}
  {\bibfnamefont {K.}~\bibnamefont {{Psarras}}},\ }\bibfield  {title} {\bibinfo
  {title} {{SMILE: Search for MIlli-LEnses}},\ }\href
  {https://doi.org/10.1093/mnrasl/slab082} {\bibfield  {journal} {\bibinfo
  {journal} {\mnras}\ }\textbf {\bibinfo {volume} {507}},\ \bibinfo {pages}
  {L6} (\bibinfo {year} {2021})},\ \Eprint {https://arxiv.org/abs/2107.06896}
  {arXiv:2107.06896 [astro-ph.CO]} \BibitemShut {NoStop}%
\bibitem [{\citenamefont {{Loudas}}\ \emph {et~al.}(2022)\citenamefont
  {{Loudas}}, \citenamefont {{Pavlidou}}, \citenamefont {{Casadio}},\ and\
  \citenamefont {{Tassis}}}]{Loudas++22}%
  \BibitemOpen
  \bibfield  {author} {\bibinfo {author} {\bibfnamefont {N.}~\bibnamefont
  {{Loudas}}}, \bibinfo {author} {\bibfnamefont {V.}~\bibnamefont
  {{Pavlidou}}}, \bibinfo {author} {\bibfnamefont {C.}~\bibnamefont
  {{Casadio}}},\ and\ \bibinfo {author} {\bibfnamefont {K.}~\bibnamefont
  {{Tassis}}},\ }\bibfield  {title} {\bibinfo {title} {{Discriminating power of
  milli-lensing observations for dark matter models}},\ }\href
  {https://doi.org/10.1051/0004-6361/202244978} {\bibfield  {journal} {\bibinfo
   {journal} {\aap}\ }\textbf {\bibinfo {volume} {668}},\ \bibinfo {eid} {A166}
  (\bibinfo {year} {2022})},\ \Eprint {https://arxiv.org/abs/2209.13393}
  {arXiv:2209.13393 [astro-ph.CO]} \BibitemShut {NoStop}%
\bibitem [{\citenamefont {{Nierenberg}}\ \emph {et~al.}(2021)\citenamefont
  {{Nierenberg}}, \citenamefont {{Bennert}}, \citenamefont {{Benson}},
  \citenamefont {{Birrer}}, \citenamefont {{Djorgovski}}, \citenamefont {{Du}},
  \citenamefont {{Fassnacht}}, \citenamefont {{Gilman}}, \citenamefont
  {{Hoenig}}, \citenamefont {{Kusenko}}, \citenamefont {{Malkan}},
  \citenamefont {{Motta}}, \citenamefont {{Moustakas}}, \citenamefont
  {{Sluse}}, \citenamefont {{Stern}},\ and\ \citenamefont
  {{Treu}}}]{NierenbergJWST}%
  \BibitemOpen
  \bibfield  {author} {\bibinfo {author} {\bibfnamefont {A.}~\bibnamefont
  {{Nierenberg}}}, \bibinfo {author} {\bibfnamefont {V.~N.}\ \bibnamefont
  {{Bennert}}}, \bibinfo {author} {\bibfnamefont {A.}~\bibnamefont {{Benson}}},
  \bibinfo {author} {\bibfnamefont {S.}~\bibnamefont {{Birrer}}}, \bibinfo
  {author} {\bibfnamefont {S.~G.}\ \bibnamefont {{Djorgovski}}}, \bibinfo
  {author} {\bibfnamefont {X.}~\bibnamefont {{Du}}}, \bibinfo {author}
  {\bibfnamefont {C.}~\bibnamefont {{Fassnacht}}}, \bibinfo {author}
  {\bibfnamefont {D.}~\bibnamefont {{Gilman}}}, \bibinfo {author}
  {\bibfnamefont {S.~F.}\ \bibnamefont {{Hoenig}}}, \bibinfo {author}
  {\bibfnamefont {A.}~\bibnamefont {{Kusenko}}}, \bibinfo {author}
  {\bibfnamefont {M.~A.}\ \bibnamefont {{Malkan}}}, \bibinfo {author}
  {\bibfnamefont {V.}~\bibnamefont {{Motta}}}, \bibinfo {author} {\bibfnamefont
  {L.~A.}\ \bibnamefont {{Moustakas}}}, \bibinfo {author} {\bibfnamefont
  {D.}~\bibnamefont {{Sluse}}}, \bibinfo {author} {\bibfnamefont {D.~K.}\
  \bibnamefont {{Stern}}},\ and\ \bibinfo {author} {\bibfnamefont {T.~L.}\
  \bibnamefont {{Treu}}},\ }\href@noop {} {\bibinfo {title} {{A definitive test
  of the dark matter paradigm on small scales}}},\ \bibinfo {howpublished}
  {JWST Proposal. Cycle 1, ID. \#2046} (\bibinfo {year} {2021})\BibitemShut
  {NoStop}%
\bibitem [{\citenamefont {{Bhattacharyya}}\ \emph {et~al.}(2022)\citenamefont
  {{Bhattacharyya}}, \citenamefont {{Adhikari}}, \citenamefont {{Banerjee}},
  \citenamefont {{More}}, \citenamefont {{Kumar}}, \citenamefont {{Nadler}},\
  and\ \citenamefont {{Chatterjee}}}]{Bhattacharyya++22}%
  \BibitemOpen
  \bibfield  {author} {\bibinfo {author} {\bibfnamefont {S.}~\bibnamefont
  {{Bhattacharyya}}}, \bibinfo {author} {\bibfnamefont {S.}~\bibnamefont
  {{Adhikari}}}, \bibinfo {author} {\bibfnamefont {A.}~\bibnamefont
  {{Banerjee}}}, \bibinfo {author} {\bibfnamefont {S.}~\bibnamefont {{More}}},
  \bibinfo {author} {\bibfnamefont {A.}~\bibnamefont {{Kumar}}}, \bibinfo
  {author} {\bibfnamefont {E.~O.}\ \bibnamefont {{Nadler}}},\ and\ \bibinfo
  {author} {\bibfnamefont {S.}~\bibnamefont {{Chatterjee}}},\ }\bibfield
  {title} {\bibinfo {title} {{The Signatures of Self-interacting Dark Matter
  and Subhalo Disruption on Cluster Substructure}},\ }\href
  {https://doi.org/10.3847/1538-4357/ac68e9} {\bibfield  {journal} {\bibinfo
  {journal} {\apj}\ }\textbf {\bibinfo {volume} {932}},\ \bibinfo {eid} {30}
  (\bibinfo {year} {2022})},\ \Eprint {https://arxiv.org/abs/2106.08292}
  {arXiv:2106.08292 [astro-ph.CO]} \BibitemShut {NoStop}%
\bibitem [{\citenamefont {{Loken}}\ \emph {et~al.}(2010)\citenamefont
  {{Loken}}, \citenamefont {{Gruner}}, \citenamefont {{Groer}}, \citenamefont
  {{Peltier}}, \citenamefont {{Bunn}}, \citenamefont {{Craig}}, \citenamefont
  {{Henriques}}, \citenamefont {{Dempsey}}, \citenamefont {{Yu}}, \citenamefont
  {{Chen}}, \citenamefont {{Dursi}}, \citenamefont {{Chong}}, \citenamefont
  {{Northrup}}, \citenamefont {{Pinto}}, \citenamefont {{Knecht}},\ and\
  \citenamefont {{Van Zon}}}]{Loken2010}%
  \BibitemOpen
  \bibfield  {author} {\bibinfo {author} {\bibfnamefont {C.}~\bibnamefont
  {{Loken}}}, \bibinfo {author} {\bibfnamefont {D.}~\bibnamefont {{Gruner}}},
  \bibinfo {author} {\bibfnamefont {L.}~\bibnamefont {{Groer}}}, \bibinfo
  {author} {\bibfnamefont {R.}~\bibnamefont {{Peltier}}}, \bibinfo {author}
  {\bibfnamefont {N.}~\bibnamefont {{Bunn}}}, \bibinfo {author} {\bibfnamefont
  {M.}~\bibnamefont {{Craig}}}, \bibinfo {author} {\bibfnamefont
  {T.}~\bibnamefont {{Henriques}}}, \bibinfo {author} {\bibfnamefont
  {J.}~\bibnamefont {{Dempsey}}}, \bibinfo {author} {\bibfnamefont {C.-H.}\
  \bibnamefont {{Yu}}}, \bibinfo {author} {\bibfnamefont {J.}~\bibnamefont
  {{Chen}}}, \bibinfo {author} {\bibfnamefont {L.~J.}\ \bibnamefont {{Dursi}}},
  \bibinfo {author} {\bibfnamefont {J.}~\bibnamefont {{Chong}}}, \bibinfo
  {author} {\bibfnamefont {S.}~\bibnamefont {{Northrup}}}, \bibinfo {author}
  {\bibfnamefont {J.}~\bibnamefont {{Pinto}}}, \bibinfo {author} {\bibfnamefont
  {N.}~\bibnamefont {{Knecht}}},\ and\ \bibinfo {author} {\bibfnamefont
  {R.}~\bibnamefont {{Van Zon}}},\ }\bibfield  {title} {\bibinfo {title}
  {{SciNet: Lessons Learned from Building a Power-efficient Top-20 System and
  Data Centre}},\ }in\ \href {https://doi.org/10.1088/1742-6596/256/1/012026}
  {\emph {\bibinfo {booktitle} {Journal of Physics Conference Series}}},\
  \bibinfo {series} {Journal of Physics Conference Series}, Vol.\ \bibinfo
  {volume} {256}\ (\bibinfo {year} {2010})\ p.\ \bibinfo {pages}
  {012026}\BibitemShut {NoStop}%
\bibitem [{\citenamefont {{Ponce}}\ \emph {et~al.}(2019)\citenamefont
  {{Ponce}}, \citenamefont {{van Zon}}, \citenamefont {{Northrup}},
  \citenamefont {{Gruner}}, \citenamefont {{Chen}}, \citenamefont {{Ertinaz}},
  \citenamefont {{Fedoseev}}, \citenamefont {{Groer}}, \citenamefont {{Mao}},
  \citenamefont {{Mundim}}, \citenamefont {{Nolta}}, \citenamefont {{Pinto}},
  \citenamefont {{Saldarriaga}}, \citenamefont {{Slavnic}}, \citenamefont
  {{Spence}}, \citenamefont {{Yu}},\ and\ \citenamefont
  {{Peltier}}}]{Ponce2019}%
  \BibitemOpen
  \bibfield  {author} {\bibinfo {author} {\bibfnamefont {M.}~\bibnamefont
  {{Ponce}}}, \bibinfo {author} {\bibfnamefont {R.}~\bibnamefont {{van Zon}}},
  \bibinfo {author} {\bibfnamefont {S.}~\bibnamefont {{Northrup}}}, \bibinfo
  {author} {\bibfnamefont {D.}~\bibnamefont {{Gruner}}}, \bibinfo {author}
  {\bibfnamefont {J.}~\bibnamefont {{Chen}}}, \bibinfo {author} {\bibfnamefont
  {F.}~\bibnamefont {{Ertinaz}}}, \bibinfo {author} {\bibfnamefont
  {A.}~\bibnamefont {{Fedoseev}}}, \bibinfo {author} {\bibfnamefont
  {L.}~\bibnamefont {{Groer}}}, \bibinfo {author} {\bibfnamefont
  {F.}~\bibnamefont {{Mao}}}, \bibinfo {author} {\bibfnamefont {B.~C.}\
  \bibnamefont {{Mundim}}}, \bibinfo {author} {\bibfnamefont {M.}~\bibnamefont
  {{Nolta}}}, \bibinfo {author} {\bibfnamefont {J.}~\bibnamefont {{Pinto}}},
  \bibinfo {author} {\bibfnamefont {M.}~\bibnamefont {{Saldarriaga}}}, \bibinfo
  {author} {\bibfnamefont {V.}~\bibnamefont {{Slavnic}}}, \bibinfo {author}
  {\bibfnamefont {E.}~\bibnamefont {{Spence}}}, \bibinfo {author}
  {\bibfnamefont {C.-H.}\ \bibnamefont {{Yu}}},\ and\ \bibinfo {author}
  {\bibfnamefont {W.~R.}\ \bibnamefont {{Peltier}}},\ }\bibfield  {title}
  {\bibinfo {title} {{Deploying a Top-100 Supercomputer for Large Parallel
  Workloads: the Niagara Supercomputer}},\ }\href@noop {} {\bibfield  {journal}
  {\bibinfo  {journal} {arXiv e-prints}\ ,\ \bibinfo {eid} {arXiv:1907.13600}}
  (\bibinfo {year} {2019})},\ \Eprint {https://arxiv.org/abs/1907.13600}
  {arXiv:1907.13600 [cs.DC]} \BibitemShut {NoStop}%
\bibitem [{\citenamefont {{Birrer}}\ and\ \citenamefont
  {{Amara}}(2018)}]{BirrerAmara++18}%
  \BibitemOpen
  \bibfield  {author} {\bibinfo {author} {\bibfnamefont {S.}~\bibnamefont
  {{Birrer}}}\ and\ \bibinfo {author} {\bibfnamefont {A.}~\bibnamefont
  {{Amara}}},\ }\bibfield  {title} {\bibinfo {title} {{lenstronomy:
  Multi-purpose gravitational lens modelling software package}},\ }\href
  {https://doi.org/10.1016/j.dark.2018.11.002} {\bibfield  {journal} {\bibinfo
  {journal} {Physics of the Dark Universe}\ }\textbf {\bibinfo {volume} {22}},\
  \bibinfo {pages} {189} (\bibinfo {year} {2018})},\ \Eprint
  {https://arxiv.org/abs/1803.09746} {arXiv:1803.09746 [astro-ph.CO]}
  \BibitemShut {NoStop}%
\bibitem [{\citenamefont {{Birrer}}\ \emph {et~al.}(2021)\citenamefont
  {{Birrer}}, \citenamefont {{Shajib}}, \citenamefont {{Gilman}}, \citenamefont
  {{Galan}}, \citenamefont {{Aalbers}}, \citenamefont {{Millon}}, \citenamefont
  {{Morgan}}, \citenamefont {{Pagano}}, \citenamefont {{Park}}, \citenamefont
  {{Teodori}}, \citenamefont {{Tessore}}, \citenamefont {{Ueland}},
  \citenamefont {{Van de Vyvere}}, \citenamefont {{Wagner-Carena}},
  \citenamefont {{Wempe}}, \citenamefont {{Yang}}, \citenamefont {{Ding}},
  \citenamefont {{Schmidt}}, \citenamefont {{Sluse}}, \citenamefont {{Zhang}},\
  and\ \citenamefont {{Amara}}}]{Birrer++21}%
  \BibitemOpen
  \bibfield  {author} {\bibinfo {author} {\bibfnamefont {S.}~\bibnamefont
  {{Birrer}}}, \bibinfo {author} {\bibfnamefont {A.}~\bibnamefont {{Shajib}}},
  \bibinfo {author} {\bibfnamefont {D.}~\bibnamefont {{Gilman}}}, \bibinfo
  {author} {\bibfnamefont {A.}~\bibnamefont {{Galan}}}, \bibinfo {author}
  {\bibfnamefont {J.}~\bibnamefont {{Aalbers}}}, \bibinfo {author}
  {\bibfnamefont {M.}~\bibnamefont {{Millon}}}, \bibinfo {author}
  {\bibfnamefont {R.}~\bibnamefont {{Morgan}}}, \bibinfo {author}
  {\bibfnamefont {G.}~\bibnamefont {{Pagano}}}, \bibinfo {author}
  {\bibfnamefont {J.}~\bibnamefont {{Park}}}, \bibinfo {author} {\bibfnamefont
  {L.}~\bibnamefont {{Teodori}}}, \bibinfo {author} {\bibfnamefont
  {N.}~\bibnamefont {{Tessore}}}, \bibinfo {author} {\bibfnamefont
  {M.}~\bibnamefont {{Ueland}}}, \bibinfo {author} {\bibfnamefont
  {L.}~\bibnamefont {{Van de Vyvere}}}, \bibinfo {author} {\bibfnamefont
  {S.}~\bibnamefont {{Wagner-Carena}}}, \bibinfo {author} {\bibfnamefont
  {E.}~\bibnamefont {{Wempe}}}, \bibinfo {author} {\bibfnamefont
  {L.}~\bibnamefont {{Yang}}}, \bibinfo {author} {\bibfnamefont
  {X.}~\bibnamefont {{Ding}}}, \bibinfo {author} {\bibfnamefont
  {T.}~\bibnamefont {{Schmidt}}}, \bibinfo {author} {\bibfnamefont
  {D.}~\bibnamefont {{Sluse}}}, \bibinfo {author} {\bibfnamefont
  {M.}~\bibnamefont {{Zhang}}},\ and\ \bibinfo {author} {\bibfnamefont
  {A.}~\bibnamefont {{Amara}}},\ }\bibfield  {title} {\bibinfo {title}
  {{lenstronomy II: A gravitational lensing software ecosystem}},\ }\href
  {https://doi.org/10.21105/joss.03283} {\bibfield  {journal} {\bibinfo
  {journal} {The Journal of Open Source Software}\ }\textbf {\bibinfo {volume}
  {6}},\ \bibinfo {eid} {3283} (\bibinfo {year} {2021})},\ \Eprint
  {https://arxiv.org/abs/2106.05976} {arXiv:2106.05976 [astro-ph.CO]}
  \BibitemShut {NoStop}%
\bibitem [{\citenamefont {{Diemer}}(2018)}]{Diemer18}%
  \BibitemOpen
  \bibfield  {author} {\bibinfo {author} {\bibfnamefont {B.}~\bibnamefont
  {{Diemer}}},\ }\bibfield  {title} {\bibinfo {title} {{COLOSSUS: A Python
  Toolkit for Cosmology, Large-scale Structure, and Dark Matter Halos}},\
  }\href {https://doi.org/10.3847/1538-4365/aaee8c} {\bibfield  {journal}
  {\bibinfo  {journal} {\apjs}\ }\textbf {\bibinfo {volume} {239}},\ \bibinfo
  {eid} {35} (\bibinfo {year} {2018})},\ \Eprint
  {https://arxiv.org/abs/1712.04512} {arXiv:1712.04512 [astro-ph.CO]}
  \BibitemShut {NoStop}%
\bibitem [{\citenamefont {{Blandford}}\ and\ \citenamefont
  {{Narayan}}(1986)}]{Blandford86}%
  \BibitemOpen
  \bibfield  {author} {\bibinfo {author} {\bibfnamefont {R.}~\bibnamefont
  {{Blandford}}}\ and\ \bibinfo {author} {\bibfnamefont {R.}~\bibnamefont
  {{Narayan}}},\ }\bibfield  {title} {\bibinfo {title} {{Fermat's Principle,
  Caustics, and the Classification of Gravitational Lens Images}},\ }\href
  {https://doi.org/10.1086/164709} {\bibfield  {journal} {\bibinfo  {journal}
  {\apj}\ }\textbf {\bibinfo {volume} {310}},\ \bibinfo {pages} {568} (\bibinfo
  {year} {1986})}\BibitemShut {NoStop}%
\bibitem [{\citenamefont {{Garrison-Kimmel}}\ \emph {et~al.}(2017)\citenamefont
  {{Garrison-Kimmel}}, \citenamefont {{Wetzel}}, \citenamefont {{Bullock}},
  \citenamefont {{Hopkins}}, \citenamefont {{Boylan-Kolchin}}, \citenamefont
  {{Faucher-Gigu{\`e}re}}, \citenamefont {{Kere{\v{s}}}}, \citenamefont
  {{Quataert}}, \citenamefont {{Sanderson}}, \citenamefont {{Graus}},\ and\
  \citenamefont {{Kelley}}}]{GarrisonKimmel}%
  \BibitemOpen
  \bibfield  {author} {\bibinfo {author} {\bibfnamefont {S.}~\bibnamefont
  {{Garrison-Kimmel}}}, \bibinfo {author} {\bibfnamefont {A.}~\bibnamefont
  {{Wetzel}}}, \bibinfo {author} {\bibfnamefont {J.~S.}\ \bibnamefont
  {{Bullock}}}, \bibinfo {author} {\bibfnamefont {P.~F.}\ \bibnamefont
  {{Hopkins}}}, \bibinfo {author} {\bibfnamefont {M.}~\bibnamefont
  {{Boylan-Kolchin}}}, \bibinfo {author} {\bibfnamefont {C.-A.}\ \bibnamefont
  {{Faucher-Gigu{\`e}re}}}, \bibinfo {author} {\bibfnamefont {D.}~\bibnamefont
  {{Kere{\v{s}}}}}, \bibinfo {author} {\bibfnamefont {E.}~\bibnamefont
  {{Quataert}}}, \bibinfo {author} {\bibfnamefont {R.~E.}\ \bibnamefont
  {{Sanderson}}}, \bibinfo {author} {\bibfnamefont {A.~S.}\ \bibnamefont
  {{Graus}}},\ and\ \bibinfo {author} {\bibfnamefont {T.}~\bibnamefont
  {{Kelley}}},\ }\bibfield  {title} {\bibinfo {title} {{Not so lumpy after all:
  modelling the depletion of dark matter subhaloes by Milky Way-like
  galaxies}},\ }\href {https://doi.org/10.1093/mnras/stx1710} {\bibfield
  {journal} {\bibinfo  {journal} {\mnras}\ }\textbf {\bibinfo {volume} {471}},\
  \bibinfo {pages} {1709} (\bibinfo {year} {2017})},\ \Eprint
  {https://arxiv.org/abs/1701.03792} {arXiv:1701.03792 [astro-ph.GA]}
  \BibitemShut {NoStop}%
\bibitem [{\citenamefont {{Webb}}\ and\ \citenamefont
  {{Bovy}}(2020)}]{Webb++20}%
  \BibitemOpen
  \bibfield  {author} {\bibinfo {author} {\bibfnamefont {J.~J.}\ \bibnamefont
  {{Webb}}}\ and\ \bibinfo {author} {\bibfnamefont {J.}~\bibnamefont
  {{Bovy}}},\ }\bibfield  {title} {\bibinfo {title} {{High-resolution
  simulations of dark matter subhalo disruption in a Milky-Way-like tidal
  field}},\ }\href {https://doi.org/10.1093/mnras/staa2852} {\bibfield
  {journal} {\bibinfo  {journal} {\mnras}\ }\textbf {\bibinfo {volume} {499}},\
  \bibinfo {pages} {116} (\bibinfo {year} {2020})},\ \Eprint
  {https://arxiv.org/abs/2006.06695} {arXiv:2006.06695 [astro-ph.GA]}
  \BibitemShut {NoStop}%
\bibitem [{\citenamefont {{Sheth}}\ \emph {et~al.}(2001)\citenamefont
  {{Sheth}}, \citenamefont {{Mo}},\ and\ \citenamefont
  {{Tormen}}}]{ShethTormen}%
  \BibitemOpen
  \bibfield  {author} {\bibinfo {author} {\bibfnamefont {R.~K.}\ \bibnamefont
  {{Sheth}}}, \bibinfo {author} {\bibfnamefont {H.~J.}\ \bibnamefont {{Mo}}},\
  and\ \bibinfo {author} {\bibfnamefont {G.}~\bibnamefont {{Tormen}}},\
  }\bibfield  {title} {\bibinfo {title} {{Ellipsoidal collapse and an improved
  model for the number and spatial distribution of dark matter haloes}},\
  }\href {https://doi.org/10.1046/j.1365-8711.2001.04006.x} {\bibfield
  {journal} {\bibinfo  {journal} {\mnras}\ }\textbf {\bibinfo {volume} {323}},\
  \bibinfo {pages} {1} (\bibinfo {year} {2001})},\ \Eprint
  {https://arxiv.org/abs/astro-ph/9907024} {arXiv:astro-ph/9907024 [astro-ph]}
  \BibitemShut {NoStop}%
\bibitem [{\citenamefont {{Planck Collaboration}}\ \emph
  {et~al.}(2020)\citenamefont {{Planck Collaboration}}, \citenamefont
  {{Aghanim}}, \citenamefont {{Akrami}}, \citenamefont {{Ashdown}},
  \citenamefont {{Aumont}}, \citenamefont {{Baccigalupi}}, \citenamefont
  {{Ballardini}}, \citenamefont {{Banday}}, \citenamefont {{Barreiro}},
  \citenamefont {{Bartolo}}, \citenamefont {{Basak}}, \citenamefont {{Battye}},
  \citenamefont {{Benabed}}, \citenamefont {{Bernard}}, \citenamefont
  {{Bersanelli}}, \citenamefont {{Bielewicz}}, \citenamefont {{Bock}},
  \citenamefont {{Bond}}, \citenamefont {{Borrill}}, \citenamefont {{Bouchet}},
  \citenamefont {{Boulanger}}, \citenamefont {{Bucher}}, \citenamefont
  {{Burigana}}, \citenamefont {{Butler}}, \citenamefont {{Calabrese}},
  \citenamefont {{Cardoso}}, \citenamefont {{Carron}}, \citenamefont
  {{Challinor}}, \citenamefont {{Chiang}}, \citenamefont {{Chluba}},
  \citenamefont {{Colombo}}, \citenamefont {{Combet}}, \citenamefont
  {{Contreras}}, \citenamefont {{Crill}}, \citenamefont {{Cuttaia}},
  \citenamefont {{de Bernardis}}, \citenamefont {{de Zotti}}, \citenamefont
  {{Delabrouille}}, \citenamefont {{Delouis}}, \citenamefont {{Di Valentino}},
  \citenamefont {{Diego}}, \citenamefont {{Dor{\'e}}}, \citenamefont
  {{Douspis}}, \citenamefont {{Ducout}}, \citenamefont {{Dupac}}, \citenamefont
  {{Dusini}}, \citenamefont {{Efstathiou}}, \citenamefont {{Elsner}},
  \citenamefont {{En{\ss}lin}}, \citenamefont {{Eriksen}}, \citenamefont
  {{Fantaye}}, \citenamefont {{Farhang}}, \citenamefont {{Fergusson}},
  \citenamefont {{Fernandez-Cobos}}, \citenamefont {{Finelli}}, \citenamefont
  {{Forastieri}}, \citenamefont {{Frailis}}, \citenamefont {{Fraisse}},
  \citenamefont {{Franceschi}}, \citenamefont {{Frolov}}, \citenamefont
  {{Galeotta}}, \citenamefont {{Galli}}, \citenamefont {{Ganga}}, \citenamefont
  {{G{\'e}nova-Santos}}, \citenamefont {{Gerbino}}, \citenamefont {{Ghosh}},
  \citenamefont {{Gonz{\'a}lez-Nuevo}}, \citenamefont {{G{\'o}rski}},
  \citenamefont {{Gratton}}, \citenamefont {{Gruppuso}}, \citenamefont
  {{Gudmundsson}}, \citenamefont {{Hamann}}, \citenamefont {{Handley}},
  \citenamefont {{Hansen}}, \citenamefont {{Herranz}}, \citenamefont
  {{Hildebrandt}}, \citenamefont {{Hivon}}, \citenamefont {{Huang}},
  \citenamefont {{Jaffe}}, \citenamefont {{Jones}}, \citenamefont {{Karakci}},
  \citenamefont {{Keih{\"a}nen}}, \citenamefont {{Keskitalo}}, \citenamefont
  {{Kiiveri}}, \citenamefont {{Kim}}, \citenamefont {{Kisner}}, \citenamefont
  {{Knox}}, \citenamefont {{Krachmalnicoff}}, \citenamefont {{Kunz}},
  \citenamefont {{Kurki-Suonio}}, \citenamefont {{Lagache}}, \citenamefont
  {{Lamarre}}, \citenamefont {{Lasenby}}, \citenamefont {{Lattanzi}},
  \citenamefont {{Lawrence}}, \citenamefont {{Le Jeune}}, \citenamefont
  {{Lemos}}, \citenamefont {{Lesgourgues}}, \citenamefont {{Levrier}},
  \citenamefont {{Lewis}}, \citenamefont {{Liguori}}, \citenamefont {{Lilje}},
  \citenamefont {{Lilley}}, \citenamefont {{Lindholm}}, \citenamefont
  {{L{\'o}pez-Caniego}}, \citenamefont {{Lubin}}, \citenamefont {{Ma}},
  \citenamefont {{Mac{\'\i}as-P{\'e}rez}}, \citenamefont {{Maggio}},
  \citenamefont {{Maino}}, \citenamefont {{Mandolesi}}, \citenamefont
  {{Mangilli}}, \citenamefont {{Marcos-Caballero}}, \citenamefont {{Maris}},
  \citenamefont {{Martin}}, \citenamefont {{Martinelli}}, \citenamefont
  {{Mart{\'\i}nez-Gonz{\'a}lez}}, \citenamefont {{Matarrese}}, \citenamefont
  {{Mauri}}, \citenamefont {{McEwen}}, \citenamefont {{Meinhold}},
  \citenamefont {{Melchiorri}}, \citenamefont {{Mennella}}, \citenamefont
  {{Migliaccio}}, \citenamefont {{Millea}}, \citenamefont {{Mitra}},
  \citenamefont {{Miville-Desch{\^e}nes}}, \citenamefont {{Molinari}},
  \citenamefont {{Montier}}, \citenamefont {{Morgante}}, \citenamefont
  {{Moss}}, \citenamefont {{Natoli}}, \citenamefont {{N{\o}rgaard-Nielsen}},
  \citenamefont {{Pagano}}, \citenamefont {{Paoletti}}, \citenamefont
  {{Partridge}}, \citenamefont {{Patanchon}}, \citenamefont {{Peiris}},
  \citenamefont {{Perrotta}}, \citenamefont {{Pettorino}}, \citenamefont
  {{Piacentini}}, \citenamefont {{Polastri}}, \citenamefont {{Polenta}},
  \citenamefont {{Puget}}, \citenamefont {{Rachen}}, \citenamefont
  {{Reinecke}}, \citenamefont {{Remazeilles}}, \citenamefont {{Renzi}},
  \citenamefont {{Rocha}}, \citenamefont {{Rosset}}, \citenamefont {{Roudier}},
  \citenamefont {{Rubi{\~n}o-Mart{\'\i}n}}, \citenamefont {{Ruiz-Granados}},
  \citenamefont {{Salvati}}, \citenamefont {{Sandri}}, \citenamefont
  {{Savelainen}}, \citenamefont {{Scott}}, \citenamefont {{Shellard}},
  \citenamefont {{Sirignano}}, \citenamefont {{Sirri}}, \citenamefont
  {{Spencer}}, \citenamefont {{Sunyaev}}, \citenamefont {{Suur-Uski}},
  \citenamefont {{Tauber}}, \citenamefont {{Tavagnacco}}, \citenamefont
  {{Tenti}}, \citenamefont {{Toffolatti}}, \citenamefont {{Tomasi}},
  \citenamefont {{Trombetti}}, \citenamefont {{Valenziano}}, \citenamefont
  {{Valiviita}}, \citenamefont {{Van Tent}}, \citenamefont {{Vibert}},
  \citenamefont {{Vielva}}, \citenamefont {{Villa}}, \citenamefont
  {{Vittorio}}, \citenamefont {{Wandelt}}, \citenamefont {{Wehus}},
  \citenamefont {{White}}, \citenamefont {{White}}, \citenamefont {{Zacchei}},\
  and\ \citenamefont {{Zonca}}}]{Planck2020}%
  \BibitemOpen
  \bibfield  {author} {\bibinfo {author} {\bibnamefont {{Planck
  Collaboration}}}, \bibinfo {author} {\bibfnamefont {N.}~\bibnamefont
  {{Aghanim}}}, \bibinfo {author} {\bibfnamefont {Y.}~\bibnamefont {{Akrami}}},
  \bibinfo {author} {\bibfnamefont {M.}~\bibnamefont {{Ashdown}}}, \bibinfo
  {author} {\bibfnamefont {J.}~\bibnamefont {{Aumont}}}, \bibinfo {author}
  {\bibfnamefont {C.}~\bibnamefont {{Baccigalupi}}}, \bibinfo {author}
  {\bibfnamefont {M.}~\bibnamefont {{Ballardini}}}, \bibinfo {author}
  {\bibfnamefont {A.~J.}\ \bibnamefont {{Banday}}}, \bibinfo {author}
  {\bibfnamefont {R.~B.}\ \bibnamefont {{Barreiro}}}, \bibinfo {author}
  {\bibfnamefont {N.}~\bibnamefont {{Bartolo}}}, \bibinfo {author}
  {\bibfnamefont {S.}~\bibnamefont {{Basak}}}, \bibinfo {author} {\bibfnamefont
  {R.}~\bibnamefont {{Battye}}}, \bibinfo {author} {\bibfnamefont
  {K.}~\bibnamefont {{Benabed}}}, \bibinfo {author} {\bibfnamefont {J.~P.}\
  \bibnamefont {{Bernard}}}, \bibinfo {author} {\bibfnamefont {M.}~\bibnamefont
  {{Bersanelli}}}, \bibinfo {author} {\bibfnamefont {P.}~\bibnamefont
  {{Bielewicz}}}, \bibinfo {author} {\bibfnamefont {J.~J.}\ \bibnamefont
  {{Bock}}}, \bibinfo {author} {\bibfnamefont {J.~R.}\ \bibnamefont {{Bond}}},
  \bibinfo {author} {\bibfnamefont {J.}~\bibnamefont {{Borrill}}}, \bibinfo
  {author} {\bibfnamefont {F.~R.}\ \bibnamefont {{Bouchet}}}, \bibinfo {author}
  {\bibfnamefont {F.}~\bibnamefont {{Boulanger}}}, \bibinfo {author}
  {\bibfnamefont {M.}~\bibnamefont {{Bucher}}}, \bibinfo {author}
  {\bibfnamefont {C.}~\bibnamefont {{Burigana}}}, \bibinfo {author}
  {\bibfnamefont {R.~C.}\ \bibnamefont {{Butler}}}, \bibinfo {author}
  {\bibfnamefont {E.}~\bibnamefont {{Calabrese}}}, \bibinfo {author}
  {\bibfnamefont {J.~F.}\ \bibnamefont {{Cardoso}}}, \bibinfo {author}
  {\bibfnamefont {J.}~\bibnamefont {{Carron}}}, \bibinfo {author}
  {\bibfnamefont {A.}~\bibnamefont {{Challinor}}}, \bibinfo {author}
  {\bibfnamefont {H.~C.}\ \bibnamefont {{Chiang}}}, \bibinfo {author}
  {\bibfnamefont {J.}~\bibnamefont {{Chluba}}}, \bibinfo {author}
  {\bibfnamefont {L.~P.~L.}\ \bibnamefont {{Colombo}}}, \bibinfo {author}
  {\bibfnamefont {C.}~\bibnamefont {{Combet}}}, \bibinfo {author}
  {\bibfnamefont {D.}~\bibnamefont {{Contreras}}}, \bibinfo {author}
  {\bibfnamefont {B.~P.}\ \bibnamefont {{Crill}}}, \bibinfo {author}
  {\bibfnamefont {F.}~\bibnamefont {{Cuttaia}}}, \bibinfo {author}
  {\bibfnamefont {P.}~\bibnamefont {{de Bernardis}}}, \bibinfo {author}
  {\bibfnamefont {G.}~\bibnamefont {{de Zotti}}}, \bibinfo {author}
  {\bibfnamefont {J.}~\bibnamefont {{Delabrouille}}}, \bibinfo {author}
  {\bibfnamefont {J.~M.}\ \bibnamefont {{Delouis}}}, \bibinfo {author}
  {\bibfnamefont {E.}~\bibnamefont {{Di Valentino}}}, \bibinfo {author}
  {\bibfnamefont {J.~M.}\ \bibnamefont {{Diego}}}, \bibinfo {author}
  {\bibfnamefont {O.}~\bibnamefont {{Dor{\'e}}}}, \bibinfo {author}
  {\bibfnamefont {M.}~\bibnamefont {{Douspis}}}, \bibinfo {author}
  {\bibfnamefont {A.}~\bibnamefont {{Ducout}}}, \bibinfo {author}
  {\bibfnamefont {X.}~\bibnamefont {{Dupac}}}, \bibinfo {author} {\bibfnamefont
  {S.}~\bibnamefont {{Dusini}}}, \bibinfo {author} {\bibfnamefont
  {G.}~\bibnamefont {{Efstathiou}}}, \bibinfo {author} {\bibfnamefont
  {F.}~\bibnamefont {{Elsner}}}, \bibinfo {author} {\bibfnamefont {T.~A.}\
  \bibnamefont {{En{\ss}lin}}}, \bibinfo {author} {\bibfnamefont {H.~K.}\
  \bibnamefont {{Eriksen}}}, \bibinfo {author} {\bibfnamefont {Y.}~\bibnamefont
  {{Fantaye}}}, \bibinfo {author} {\bibfnamefont {M.}~\bibnamefont
  {{Farhang}}}, \bibinfo {author} {\bibfnamefont {J.}~\bibnamefont
  {{Fergusson}}}, \bibinfo {author} {\bibfnamefont {R.}~\bibnamefont
  {{Fernandez-Cobos}}}, \bibinfo {author} {\bibfnamefont {F.}~\bibnamefont
  {{Finelli}}}, \bibinfo {author} {\bibfnamefont {F.}~\bibnamefont
  {{Forastieri}}}, \bibinfo {author} {\bibfnamefont {M.}~\bibnamefont
  {{Frailis}}}, \bibinfo {author} {\bibfnamefont {A.~A.}\ \bibnamefont
  {{Fraisse}}}, \bibinfo {author} {\bibfnamefont {E.}~\bibnamefont
  {{Franceschi}}}, \bibinfo {author} {\bibfnamefont {A.}~\bibnamefont
  {{Frolov}}}, \bibinfo {author} {\bibfnamefont {S.}~\bibnamefont
  {{Galeotta}}}, \bibinfo {author} {\bibfnamefont {S.}~\bibnamefont {{Galli}}},
  \bibinfo {author} {\bibfnamefont {K.}~\bibnamefont {{Ganga}}}, \bibinfo
  {author} {\bibfnamefont {R.~T.}\ \bibnamefont {{G{\'e}nova-Santos}}},
  \bibinfo {author} {\bibfnamefont {M.}~\bibnamefont {{Gerbino}}}, \bibinfo
  {author} {\bibfnamefont {T.}~\bibnamefont {{Ghosh}}}, \bibinfo {author}
  {\bibfnamefont {J.}~\bibnamefont {{Gonz{\'a}lez-Nuevo}}}, \bibinfo {author}
  {\bibfnamefont {K.~M.}\ \bibnamefont {{G{\'o}rski}}}, \bibinfo {author}
  {\bibfnamefont {S.}~\bibnamefont {{Gratton}}}, \bibinfo {author}
  {\bibfnamefont {A.}~\bibnamefont {{Gruppuso}}}, \bibinfo {author}
  {\bibfnamefont {J.~E.}\ \bibnamefont {{Gudmundsson}}}, \bibinfo {author}
  {\bibfnamefont {J.}~\bibnamefont {{Hamann}}}, \bibinfo {author}
  {\bibfnamefont {W.}~\bibnamefont {{Handley}}}, \bibinfo {author}
  {\bibfnamefont {F.~K.}\ \bibnamefont {{Hansen}}}, \bibinfo {author}
  {\bibfnamefont {D.}~\bibnamefont {{Herranz}}}, \bibinfo {author}
  {\bibfnamefont {S.~R.}\ \bibnamefont {{Hildebrandt}}}, \bibinfo {author}
  {\bibfnamefont {E.}~\bibnamefont {{Hivon}}}, \bibinfo {author} {\bibfnamefont
  {Z.}~\bibnamefont {{Huang}}}, \bibinfo {author} {\bibfnamefont {A.~H.}\
  \bibnamefont {{Jaffe}}}, \bibinfo {author} {\bibfnamefont {W.~C.}\
  \bibnamefont {{Jones}}}, \bibinfo {author} {\bibfnamefont {A.}~\bibnamefont
  {{Karakci}}}, \bibinfo {author} {\bibfnamefont {E.}~\bibnamefont
  {{Keih{\"a}nen}}}, \bibinfo {author} {\bibfnamefont {R.}~\bibnamefont
  {{Keskitalo}}}, \bibinfo {author} {\bibfnamefont {K.}~\bibnamefont
  {{Kiiveri}}}, \bibinfo {author} {\bibfnamefont {J.}~\bibnamefont {{Kim}}},
  \bibinfo {author} {\bibfnamefont {T.~S.}\ \bibnamefont {{Kisner}}}, \bibinfo
  {author} {\bibfnamefont {L.}~\bibnamefont {{Knox}}}, \bibinfo {author}
  {\bibfnamefont {N.}~\bibnamefont {{Krachmalnicoff}}}, \bibinfo {author}
  {\bibfnamefont {M.}~\bibnamefont {{Kunz}}}, \bibinfo {author} {\bibfnamefont
  {H.}~\bibnamefont {{Kurki-Suonio}}}, \bibinfo {author} {\bibfnamefont
  {G.}~\bibnamefont {{Lagache}}}, \bibinfo {author} {\bibfnamefont {J.~M.}\
  \bibnamefont {{Lamarre}}}, \bibinfo {author} {\bibfnamefont {A.}~\bibnamefont
  {{Lasenby}}}, \bibinfo {author} {\bibfnamefont {M.}~\bibnamefont
  {{Lattanzi}}}, \bibinfo {author} {\bibfnamefont {C.~R.}\ \bibnamefont
  {{Lawrence}}}, \bibinfo {author} {\bibfnamefont {M.}~\bibnamefont {{Le
  Jeune}}}, \bibinfo {author} {\bibfnamefont {P.}~\bibnamefont {{Lemos}}},
  \bibinfo {author} {\bibfnamefont {J.}~\bibnamefont {{Lesgourgues}}}, \bibinfo
  {author} {\bibfnamefont {F.}~\bibnamefont {{Levrier}}}, \bibinfo {author}
  {\bibfnamefont {A.}~\bibnamefont {{Lewis}}}, \bibinfo {author} {\bibfnamefont
  {M.}~\bibnamefont {{Liguori}}}, \bibinfo {author} {\bibfnamefont {P.~B.}\
  \bibnamefont {{Lilje}}}, \bibinfo {author} {\bibfnamefont {M.}~\bibnamefont
  {{Lilley}}}, \bibinfo {author} {\bibfnamefont {V.}~\bibnamefont
  {{Lindholm}}}, \bibinfo {author} {\bibfnamefont {M.}~\bibnamefont
  {{L{\'o}pez-Caniego}}}, \bibinfo {author} {\bibfnamefont {P.~M.}\
  \bibnamefont {{Lubin}}}, \bibinfo {author} {\bibfnamefont {Y.~Z.}\
  \bibnamefont {{Ma}}}, \bibinfo {author} {\bibfnamefont {J.~F.}\ \bibnamefont
  {{Mac{\'\i}as-P{\'e}rez}}}, \bibinfo {author} {\bibfnamefont
  {G.}~\bibnamefont {{Maggio}}}, \bibinfo {author} {\bibfnamefont
  {D.}~\bibnamefont {{Maino}}}, \bibinfo {author} {\bibfnamefont
  {N.}~\bibnamefont {{Mandolesi}}}, \bibinfo {author} {\bibfnamefont
  {A.}~\bibnamefont {{Mangilli}}}, \bibinfo {author} {\bibfnamefont
  {A.}~\bibnamefont {{Marcos-Caballero}}}, \bibinfo {author} {\bibfnamefont
  {M.}~\bibnamefont {{Maris}}}, \bibinfo {author} {\bibfnamefont {P.~G.}\
  \bibnamefont {{Martin}}}, \bibinfo {author} {\bibfnamefont {M.}~\bibnamefont
  {{Martinelli}}}, \bibinfo {author} {\bibfnamefont {E.}~\bibnamefont
  {{Mart{\'\i}nez-Gonz{\'a}lez}}}, \bibinfo {author} {\bibfnamefont
  {S.}~\bibnamefont {{Matarrese}}}, \bibinfo {author} {\bibfnamefont
  {N.}~\bibnamefont {{Mauri}}}, \bibinfo {author} {\bibfnamefont {J.~D.}\
  \bibnamefont {{McEwen}}}, \bibinfo {author} {\bibfnamefont {P.~R.}\
  \bibnamefont {{Meinhold}}}, \bibinfo {author} {\bibfnamefont
  {A.}~\bibnamefont {{Melchiorri}}}, \bibinfo {author} {\bibfnamefont
  {A.}~\bibnamefont {{Mennella}}}, \bibinfo {author} {\bibfnamefont
  {M.}~\bibnamefont {{Migliaccio}}}, \bibinfo {author} {\bibfnamefont
  {M.}~\bibnamefont {{Millea}}}, \bibinfo {author} {\bibfnamefont
  {S.}~\bibnamefont {{Mitra}}}, \bibinfo {author} {\bibfnamefont {M.~A.}\
  \bibnamefont {{Miville-Desch{\^e}nes}}}, \bibinfo {author} {\bibfnamefont
  {D.}~\bibnamefont {{Molinari}}}, \bibinfo {author} {\bibfnamefont
  {L.}~\bibnamefont {{Montier}}}, \bibinfo {author} {\bibfnamefont
  {G.}~\bibnamefont {{Morgante}}}, \bibinfo {author} {\bibfnamefont
  {A.}~\bibnamefont {{Moss}}}, \bibinfo {author} {\bibfnamefont
  {P.}~\bibnamefont {{Natoli}}}, \bibinfo {author} {\bibfnamefont {H.~U.}\
  \bibnamefont {{N{\o}rgaard-Nielsen}}}, \bibinfo {author} {\bibfnamefont
  {L.}~\bibnamefont {{Pagano}}}, \bibinfo {author} {\bibfnamefont
  {D.}~\bibnamefont {{Paoletti}}}, \bibinfo {author} {\bibfnamefont
  {B.}~\bibnamefont {{Partridge}}}, \bibinfo {author} {\bibfnamefont
  {G.}~\bibnamefont {{Patanchon}}}, \bibinfo {author} {\bibfnamefont {H.~V.}\
  \bibnamefont {{Peiris}}}, \bibinfo {author} {\bibfnamefont {F.}~\bibnamefont
  {{Perrotta}}}, \bibinfo {author} {\bibfnamefont {V.}~\bibnamefont
  {{Pettorino}}}, \bibinfo {author} {\bibfnamefont {F.}~\bibnamefont
  {{Piacentini}}}, \bibinfo {author} {\bibfnamefont {L.}~\bibnamefont
  {{Polastri}}}, \bibinfo {author} {\bibfnamefont {G.}~\bibnamefont
  {{Polenta}}}, \bibinfo {author} {\bibfnamefont {J.~L.}\ \bibnamefont
  {{Puget}}}, \bibinfo {author} {\bibfnamefont {J.~P.}\ \bibnamefont
  {{Rachen}}}, \bibinfo {author} {\bibfnamefont {M.}~\bibnamefont
  {{Reinecke}}}, \bibinfo {author} {\bibfnamefont {M.}~\bibnamefont
  {{Remazeilles}}}, \bibinfo {author} {\bibfnamefont {A.}~\bibnamefont
  {{Renzi}}}, \bibinfo {author} {\bibfnamefont {G.}~\bibnamefont {{Rocha}}},
  \bibinfo {author} {\bibfnamefont {C.}~\bibnamefont {{Rosset}}}, \bibinfo
  {author} {\bibfnamefont {G.}~\bibnamefont {{Roudier}}}, \bibinfo {author}
  {\bibfnamefont {J.~A.}\ \bibnamefont {{Rubi{\~n}o-Mart{\'\i}n}}}, \bibinfo
  {author} {\bibfnamefont {B.}~\bibnamefont {{Ruiz-Granados}}}, \bibinfo
  {author} {\bibfnamefont {L.}~\bibnamefont {{Salvati}}}, \bibinfo {author}
  {\bibfnamefont {M.}~\bibnamefont {{Sandri}}}, \bibinfo {author}
  {\bibfnamefont {M.}~\bibnamefont {{Savelainen}}}, \bibinfo {author}
  {\bibfnamefont {D.}~\bibnamefont {{Scott}}}, \bibinfo {author} {\bibfnamefont
  {E.~P.~S.}\ \bibnamefont {{Shellard}}}, \bibinfo {author} {\bibfnamefont
  {C.}~\bibnamefont {{Sirignano}}}, \bibinfo {author} {\bibfnamefont
  {G.}~\bibnamefont {{Sirri}}}, \bibinfo {author} {\bibfnamefont {L.~D.}\
  \bibnamefont {{Spencer}}}, \bibinfo {author} {\bibfnamefont {R.}~\bibnamefont
  {{Sunyaev}}}, \bibinfo {author} {\bibfnamefont {A.~S.}\ \bibnamefont
  {{Suur-Uski}}}, \bibinfo {author} {\bibfnamefont {J.~A.}\ \bibnamefont
  {{Tauber}}}, \bibinfo {author} {\bibfnamefont {D.}~\bibnamefont
  {{Tavagnacco}}}, \bibinfo {author} {\bibfnamefont {M.}~\bibnamefont
  {{Tenti}}}, \bibinfo {author} {\bibfnamefont {L.}~\bibnamefont
  {{Toffolatti}}}, \bibinfo {author} {\bibfnamefont {M.}~\bibnamefont
  {{Tomasi}}}, \bibinfo {author} {\bibfnamefont {T.}~\bibnamefont
  {{Trombetti}}}, \bibinfo {author} {\bibfnamefont {L.}~\bibnamefont
  {{Valenziano}}}, \bibinfo {author} {\bibfnamefont {J.}~\bibnamefont
  {{Valiviita}}}, \bibinfo {author} {\bibfnamefont {B.}~\bibnamefont {{Van
  Tent}}}, \bibinfo {author} {\bibfnamefont {L.}~\bibnamefont {{Vibert}}},
  \bibinfo {author} {\bibfnamefont {P.}~\bibnamefont {{Vielva}}}, \bibinfo
  {author} {\bibfnamefont {F.}~\bibnamefont {{Villa}}}, \bibinfo {author}
  {\bibfnamefont {N.}~\bibnamefont {{Vittorio}}}, \bibinfo {author}
  {\bibfnamefont {B.~D.}\ \bibnamefont {{Wandelt}}}, \bibinfo {author}
  {\bibfnamefont {I.~K.}\ \bibnamefont {{Wehus}}}, \bibinfo {author}
  {\bibfnamefont {M.}~\bibnamefont {{White}}}, \bibinfo {author} {\bibfnamefont
  {S.~D.~M.}\ \bibnamefont {{White}}}, \bibinfo {author} {\bibfnamefont
  {A.}~\bibnamefont {{Zacchei}}},\ and\ \bibinfo {author} {\bibfnamefont
  {A.}~\bibnamefont {{Zonca}}},\ }\bibfield  {title} {\bibinfo {title} {{Planck
  2018 results. VI. Cosmological parameters}},\ }\href
  {https://doi.org/10.1051/0004-6361/201833910} {\bibfield  {journal} {\bibinfo
   {journal} {\aap}\ }\textbf {\bibinfo {volume} {641}},\ \bibinfo {eid} {A6}
  (\bibinfo {year} {2020})},\ \Eprint {https://arxiv.org/abs/1807.06209}
  {arXiv:1807.06209 [astro-ph.CO]} \BibitemShut {NoStop}%
\bibitem [{\citenamefont {{Cyr-Racine}}\ \emph
  {et~al.}(2016{\natexlab{b}})\citenamefont {{Cyr-Racine}}, \citenamefont
  {{Moustakas}}, \citenamefont {{Keeton}}, \citenamefont {{Sigurdson}},\ and\
  \citenamefont {{Gilman}}}]{Cyr-Racine++16b}%
  \BibitemOpen
  \bibfield  {author} {\bibinfo {author} {\bibfnamefont {F.-Y.}\ \bibnamefont
  {{Cyr-Racine}}}, \bibinfo {author} {\bibfnamefont {L.~A.}\ \bibnamefont
  {{Moustakas}}}, \bibinfo {author} {\bibfnamefont {C.~R.}\ \bibnamefont
  {{Keeton}}}, \bibinfo {author} {\bibfnamefont {K.}~\bibnamefont
  {{Sigurdson}}},\ and\ \bibinfo {author} {\bibfnamefont {D.~A.}\ \bibnamefont
  {{Gilman}}},\ }\bibfield  {title} {\bibinfo {title} {{Dark census:
  Statistically detecting the satellite populations of distant galaxies}},\
  }\href {https://doi.org/10.1103/PhysRevD.94.043505} {\bibfield  {journal}
  {\bibinfo  {journal} {\prd}\ }\textbf {\bibinfo {volume} {94}},\ \bibinfo
  {eid} {043505} (\bibinfo {year} {2016}{\natexlab{b}})},\ \Eprint
  {https://arxiv.org/abs/1506.01724} {arXiv:1506.01724 [astro-ph.CO]}
  \BibitemShut {NoStop}%
\bibitem [{\citenamefont {{Diemer}}(2017)}]{Diemer17}%
  \BibitemOpen
  \bibfield  {author} {\bibinfo {author} {\bibfnamefont {B.}~\bibnamefont
  {{Diemer}}},\ }\bibfield  {title} {\bibinfo {title} {{The Splashback Radius
  of Halos from Particle Dynamics. I. The SPARTA Algorithm}},\ }\href
  {https://doi.org/10.3847/1538-4365/aa799c} {\bibfield  {journal} {\bibinfo
  {journal} {\apjs}\ }\textbf {\bibinfo {volume} {231}},\ \bibinfo {eid} {5}
  (\bibinfo {year} {2017})},\ \Eprint {https://arxiv.org/abs/1703.09712}
  {arXiv:1703.09712 [astro-ph.CO]} \BibitemShut {NoStop}%
\bibitem [{\citenamefont {{Wang}}\ \emph {et~al.}(2020)\citenamefont {{Wang}},
  \citenamefont {{Mao}}, \citenamefont {{Zentner}}, \citenamefont {{Lange}},
  \citenamefont {{van den Bosch}},\ and\ \citenamefont
  {{Wechsler}}}]{Wang++20}%
  \BibitemOpen
  \bibfield  {author} {\bibinfo {author} {\bibfnamefont {K.}~\bibnamefont
  {{Wang}}}, \bibinfo {author} {\bibfnamefont {Y.-Y.}\ \bibnamefont {{Mao}}},
  \bibinfo {author} {\bibfnamefont {A.~R.}\ \bibnamefont {{Zentner}}}, \bibinfo
  {author} {\bibfnamefont {J.~U.}\ \bibnamefont {{Lange}}}, \bibinfo {author}
  {\bibfnamefont {F.~C.}\ \bibnamefont {{van den Bosch}}},\ and\ \bibinfo
  {author} {\bibfnamefont {R.~H.}\ \bibnamefont {{Wechsler}}},\ }\bibfield
  {title} {\bibinfo {title} {{Concentrations of dark haloes emerge from their
  merger histories}},\ }\href {https://doi.org/10.1093/mnras/staa2733}
  {\bibfield  {journal} {\bibinfo  {journal} {\mnras}\ }\textbf {\bibinfo
  {volume} {498}},\ \bibinfo {pages} {4450} (\bibinfo {year} {2020})},\ \Eprint
  {https://arxiv.org/abs/2004.13732} {arXiv:2004.13732 [astro-ph.GA]}
  \BibitemShut {NoStop}%
\bibitem [{\citenamefont {{M{\"u}ller-S{\'a}nchez}}\ \emph
  {et~al.}(2011)\citenamefont {{M{\"u}ller-S{\'a}nchez}}, \citenamefont
  {{Prieto}}, \citenamefont {{Hicks}}, \citenamefont {{Vives-Arias}},
  \citenamefont {{Davies}}, \citenamefont {{Malkan}}, \citenamefont
  {{Tacconi}},\ and\ \citenamefont {{Genzel}}}]{MullerSanchez++11}%
  \BibitemOpen
  \bibfield  {author} {\bibinfo {author} {\bibfnamefont {F.}~\bibnamefont
  {{M{\"u}ller-S{\'a}nchez}}}, \bibinfo {author} {\bibfnamefont {M.~A.}\
  \bibnamefont {{Prieto}}}, \bibinfo {author} {\bibfnamefont {E.~K.~S.}\
  \bibnamefont {{Hicks}}}, \bibinfo {author} {\bibfnamefont {H.}~\bibnamefont
  {{Vives-Arias}}}, \bibinfo {author} {\bibfnamefont {R.~I.}\ \bibnamefont
  {{Davies}}}, \bibinfo {author} {\bibfnamefont {M.}~\bibnamefont {{Malkan}}},
  \bibinfo {author} {\bibfnamefont {L.~J.}\ \bibnamefont {{Tacconi}}},\ and\
  \bibinfo {author} {\bibfnamefont {R.}~\bibnamefont {{Genzel}}},\ }\bibfield
  {title} {\bibinfo {title} {{Outflows from Active Galactic Nuclei: Kinematics
  of the Narrow-line and Coronal-line Regions in Seyfert Galaxies}},\ }\href
  {https://doi.org/10.1088/0004-637X/739/2/69} {\bibfield  {journal} {\bibinfo
  {journal} {\apj}\ }\textbf {\bibinfo {volume} {739}},\ \bibinfo {eid} {69}
  (\bibinfo {year} {2011})},\ \Eprint {https://arxiv.org/abs/1107.3140}
  {arXiv:1107.3140 [astro-ph.CO]} \BibitemShut {NoStop}%
\bibitem [{\citenamefont {{Auger}}\ \emph {et~al.}(2010)\citenamefont
  {{Auger}}, \citenamefont {{Treu}}, \citenamefont {{Bolton}}, \citenamefont
  {{Gavazzi}}, \citenamefont {{Koopmans}}, \citenamefont {{Marshall}},
  \citenamefont {{Moustakas}},\ and\ \citenamefont {{Burles}}}]{Auger++10}%
  \BibitemOpen
  \bibfield  {author} {\bibinfo {author} {\bibfnamefont {M.~W.}\ \bibnamefont
  {{Auger}}}, \bibinfo {author} {\bibfnamefont {T.}~\bibnamefont {{Treu}}},
  \bibinfo {author} {\bibfnamefont {A.~S.}\ \bibnamefont {{Bolton}}}, \bibinfo
  {author} {\bibfnamefont {R.}~\bibnamefont {{Gavazzi}}}, \bibinfo {author}
  {\bibfnamefont {L.~V.~E.}\ \bibnamefont {{Koopmans}}}, \bibinfo {author}
  {\bibfnamefont {P.~J.}\ \bibnamefont {{Marshall}}}, \bibinfo {author}
  {\bibfnamefont {L.~A.}\ \bibnamefont {{Moustakas}}},\ and\ \bibinfo {author}
  {\bibfnamefont {S.}~\bibnamefont {{Burles}}},\ }\bibfield  {title} {\bibinfo
  {title} {{The Sloan Lens ACS Survey. X. Stellar, Dynamical, and Total Mass
  Correlations of Massive Early-type Galaxies}},\ }\href
  {https://doi.org/10.1088/0004-637X/724/1/511} {\bibfield  {journal} {\bibinfo
   {journal} {\apj}\ }\textbf {\bibinfo {volume} {724}},\ \bibinfo {pages}
  {511} (\bibinfo {year} {2010})},\ \Eprint {https://arxiv.org/abs/1007.2880}
  {arXiv:1007.2880 [astro-ph.CO]} \BibitemShut {NoStop}%
\bibitem [{\citenamefont {{Hsueh}}\ \emph {et~al.}(2018)\citenamefont
  {{Hsueh}}, \citenamefont {{Despali}}, \citenamefont {{Vegetti}},
  \citenamefont {{Xu}}, \citenamefont {{Fassnacht}},\ and\ \citenamefont
  {{Metcalf}}}]{Hsueh++18}%
  \BibitemOpen
  \bibfield  {author} {\bibinfo {author} {\bibfnamefont {J.-W.}\ \bibnamefont
  {{Hsueh}}}, \bibinfo {author} {\bibfnamefont {G.}~\bibnamefont {{Despali}}},
  \bibinfo {author} {\bibfnamefont {S.}~\bibnamefont {{Vegetti}}}, \bibinfo
  {author} {\bibfnamefont {D.}~\bibnamefont {{Xu}}}, \bibinfo {author}
  {\bibfnamefont {C.~D.}\ \bibnamefont {{Fassnacht}}},\ and\ \bibinfo {author}
  {\bibfnamefont {R.~B.}\ \bibnamefont {{Metcalf}}},\ }\bibfield  {title}
  {\bibinfo {title} {{Flux-ratio anomalies from discs and other baryonic
  structures in the Illustris simulation}},\ }\href
  {https://doi.org/10.1093/mnras/stx3320} {\bibfield  {journal} {\bibinfo
  {journal} {\mnras}\ }\textbf {\bibinfo {volume} {475}},\ \bibinfo {pages}
  {2438} (\bibinfo {year} {2018})},\ \Eprint {https://arxiv.org/abs/1707.07680}
  {arXiv:1707.07680 [astro-ph.GA]} \BibitemShut {NoStop}%
\bibitem [{\citenamefont {{Powell}}\ \emph {et~al.}(2022)\citenamefont
  {{Powell}}, \citenamefont {{Vegetti}}, \citenamefont {{McKean}},
  \citenamefont {{Spingola}}, \citenamefont {{Stacey}},\ and\ \citenamefont
  {{Fassnacht}}}]{Powell++22}%
  \BibitemOpen
  \bibfield  {author} {\bibinfo {author} {\bibfnamefont {D.~M.}\ \bibnamefont
  {{Powell}}}, \bibinfo {author} {\bibfnamefont {S.}~\bibnamefont {{Vegetti}}},
  \bibinfo {author} {\bibfnamefont {J.~P.}\ \bibnamefont {{McKean}}}, \bibinfo
  {author} {\bibfnamefont {C.}~\bibnamefont {{Spingola}}}, \bibinfo {author}
  {\bibfnamefont {H.~R.}\ \bibnamefont {{Stacey}}},\ and\ \bibinfo {author}
  {\bibfnamefont {C.~D.}\ \bibnamefont {{Fassnacht}}},\ }\bibfield  {title}
  {\bibinfo {title} {{A lensed radio jet at milli-arcsecond resolution I:
  Bayesian comparison of parametric lens models}},\ }\href@noop {} {\bibfield
  {journal} {\bibinfo  {journal} {arXiv e-prints}\ ,\ \bibinfo {eid}
  {arXiv:2207.03375}} (\bibinfo {year} {2022})},\ \Eprint
  {https://arxiv.org/abs/2207.03375} {arXiv:2207.03375 [astro-ph.GA]}
  \BibitemShut {NoStop}%
\bibitem [{\citenamefont {{Kochanek}}\ and\ \citenamefont
  {{Dalal}}(2004)}]{Kochanek++04}%
  \BibitemOpen
  \bibfield  {author} {\bibinfo {author} {\bibfnamefont {C.~S.}\ \bibnamefont
  {{Kochanek}}}\ and\ \bibinfo {author} {\bibfnamefont {N.}~\bibnamefont
  {{Dalal}}},\ }\bibfield  {title} {\bibinfo {title} {{Tests for Substructure
  in Gravitational Lenses}},\ }\href {https://doi.org/10.1086/421436}
  {\bibfield  {journal} {\bibinfo  {journal} {\apj}\ }\textbf {\bibinfo
  {volume} {610}},\ \bibinfo {pages} {69} (\bibinfo {year} {2004})},\ \Eprint
  {https://arxiv.org/abs/astro-ph/0302036} {arXiv:astro-ph/0302036 [astro-ph]}
  \BibitemShut {NoStop}%
\bibitem [{\citenamefont {{Congdon}}\ and\ \citenamefont
  {{Keeton}}(2005)}]{Congdon++05}%
  \BibitemOpen
  \bibfield  {author} {\bibinfo {author} {\bibfnamefont {A.~B.}\ \bibnamefont
  {{Congdon}}}\ and\ \bibinfo {author} {\bibfnamefont {C.~R.}\ \bibnamefont
  {{Keeton}}},\ }\bibfield  {title} {\bibinfo {title} {{Multipole models of
  four-image gravitational lenses with anomalous flux ratios}},\ }\href
  {https://doi.org/10.1111/j.1365-2966.2005.09699.x} {\bibfield  {journal}
  {\bibinfo  {journal} {\mnras}\ }\textbf {\bibinfo {volume} {364}},\ \bibinfo
  {pages} {1459} (\bibinfo {year} {2005})},\ \Eprint
  {https://arxiv.org/abs/astro-ph/0510232} {arXiv:astro-ph/0510232 [astro-ph]}
  \BibitemShut {NoStop}%
\bibitem [{\citenamefont {{Bender}}\ \emph {et~al.}(1989)\citenamefont
  {{Bender}}, \citenamefont {{Surma}}, \citenamefont {{Doebereiner}},
  \citenamefont {{Moellenhoff}},\ and\ \citenamefont
  {{Madejsky}}}]{Bender1989}%
  \BibitemOpen
  \bibfield  {author} {\bibinfo {author} {\bibfnamefont {R.}~\bibnamefont
  {{Bender}}}, \bibinfo {author} {\bibfnamefont {P.}~\bibnamefont {{Surma}}},
  \bibinfo {author} {\bibfnamefont {S.}~\bibnamefont {{Doebereiner}}}, \bibinfo
  {author} {\bibfnamefont {C.}~\bibnamefont {{Moellenhoff}}},\ and\ \bibinfo
  {author} {\bibfnamefont {R.}~\bibnamefont {{Madejsky}}},\ }\bibfield  {title}
  {\bibinfo {title} {{Isophote shapes of elliptical galaxies. II. Correlations
  with global optical, radio and X-ray properties.}},\ }\href@noop {}
  {\bibfield  {journal} {\bibinfo  {journal} {\aap}\ }\textbf {\bibinfo
  {volume} {217}},\ \bibinfo {pages} {35} (\bibinfo {year} {1989})}\BibitemShut
  {NoStop}%
\bibitem [{\citenamefont {{Hao}}\ \emph {et~al.}(2006)\citenamefont {{Hao}},
  \citenamefont {{Mao}}, \citenamefont {{Deng}}, \citenamefont {{Xia}},\ and\
  \citenamefont {{Wu}}}]{Hao++06}%
  \BibitemOpen
  \bibfield  {author} {\bibinfo {author} {\bibfnamefont {C.~N.}\ \bibnamefont
  {{Hao}}}, \bibinfo {author} {\bibfnamefont {S.}~\bibnamefont {{Mao}}},
  \bibinfo {author} {\bibfnamefont {Z.~G.}\ \bibnamefont {{Deng}}}, \bibinfo
  {author} {\bibfnamefont {X.~Y.}\ \bibnamefont {{Xia}}},\ and\ \bibinfo
  {author} {\bibfnamefont {H.}~\bibnamefont {{Wu}}},\ }\bibfield  {title}
  {\bibinfo {title} {{Isophotal shapes of elliptical/lenticular galaxies from
  the Sloan Digital Sky Survey}},\ }\href
  {https://doi.org/10.1111/j.1365-2966.2006.10545.x} {\bibfield  {journal}
  {\bibinfo  {journal} {\mnras}\ }\textbf {\bibinfo {volume} {370}},\ \bibinfo
  {pages} {1339} (\bibinfo {year} {2006})},\ \Eprint
  {https://arxiv.org/abs/astro-ph/0605319} {arXiv:astro-ph/0605319 [astro-ph]}
  \BibitemShut {NoStop}%
\bibitem [{\citenamefont {{Mu{\~n}oz}}\ \emph {et~al.}(2001)\citenamefont
  {{Mu{\~n}oz}}, \citenamefont {{Kochanek}},\ and\ \citenamefont
  {{Keeton}}}]{Munoz++01}%
  \BibitemOpen
  \bibfield  {author} {\bibinfo {author} {\bibfnamefont {J.~A.}\ \bibnamefont
  {{Mu{\~n}oz}}}, \bibinfo {author} {\bibfnamefont {C.~S.}\ \bibnamefont
  {{Kochanek}}},\ and\ \bibinfo {author} {\bibfnamefont {C.~R.}\ \bibnamefont
  {{Keeton}}},\ }\bibfield  {title} {\bibinfo {title} {{Cusped Mass Models of
  Gravitational Lenses}},\ }\href {https://doi.org/10.1086/322314} {\bibfield
  {journal} {\bibinfo  {journal} {\apj}\ }\textbf {\bibinfo {volume} {558}},\
  \bibinfo {pages} {657} (\bibinfo {year} {2001})},\ \Eprint
  {https://arxiv.org/abs/astro-ph/0103009} {arXiv:astro-ph/0103009 [astro-ph]}
  \BibitemShut {NoStop}%
\end{thebibliography}%
	\newpage 
	\appendix
	
	These appendices provide technical details relevant for the analysis presented in the main article, and additional discussion. First, in Section \ref{sec:supinference} we provide additional details regarding the inference methodology used to analyze the data. In particular, Sections \ref{ssec:supcollapsescatter} and \ref{ssec:supSHMF} discuss how our results depend on the prior assigned to the subhalo mass function amplitude, and the scatter in the core collapse times. Section \ref{sec:supmodels} describes the models we implement for the subhalo and field halo mass functions, the background source, and the main deflector mass profile. In  Section \ref{sec:supcollapsedprofile}, we investigate our sensitivity to the inner structure of collapsed profiles. 
	
	\section{Inference method}
	\label{sec:supinference}
	This section discusses how we compute likelihood functions and posterior distributions. We begin in Section \ref{ssec:generalinf} by explaining the methodology in general terms. The methodology outlined in Section \ref{ssec:generalinf} is discussed at length by previous analyses that utilize the method \citep{Gilman++20,Gilman++20b,Gilman++21,Gilman++22,Laroche++22}. For completeness, however, we have included an explanation of the full methodology in this work as well. For details pertaining to the likelihood function that we use specifically within the context of the analysis presented in the main article, one can skip ahead to Section \ref{ssec:inf}. Sections \ref{ssec:supcollapsescatter} and \ref{ssec:supSHMF} discuss how our results depend our knowledge of the scatter in the core collapse timescales, and the amplitude of the subhalo mass function. 
	
	\subsection{Bayesian inference in substructure lensing}
	\label{ssec:generalinf}
	
	Our goal is to compute a posterior distribution $p\left({\bf{q}}|{\bf{D}}\right)$, where ${\bf{q}}$ represents a set of parameters of interest and ${\bf{D}}$ represents the full set of data from a sample of lenses. As each lens contributes statistically independent information, we can compute this distribution as a prior probability density $\pi\left({\bf{q}}\right)$ times the product of likelihood functions computed for individual lenses
	\begin{equation}
		\label{eqn:posterior}
		p\left({\bf{q}}|{\bf{D}}\right) \propto \pi\left({\bf{q}}\right) \prod_n \mathcal{L}\left({\bf{d}_n} | \bf{q} \right).
	\end{equation}
	The quantity ${\bf{d_n}}$ represents the data for the $n$-th lens, which specifies the four image positions and three magnification ratios (flux ratios) of a lensed background quasar. The parameters ${\bf{q}}$ are, in most cases, hyper-parameters that define properties of dark matter halos, such as the slope and amplitude of the subhalo and field halo mass functions, or the number of core-collapsed halos. 
	
	Individual realizations of dark matter structure, ${\bf{m}}$, generated from the model specified by ${\bf{q}}$, mediate the connection between ${\bf{q}}$ and the data. A single realization specifies the masses, positions, and density profiles for dark matter halos and subhalos distributed between the observer and the source. In addition to a realization ${\bf{m}}$, the size of the lensed background source, the density profile of the main deflector, and measurement uncertainties can also affect the interpretation of the data. Collecting these nuisance parameters into a vector ${\bf{x}}$, the likelihood function is given by
	\begin{equation}
		\label{eqn:likelihoodfunction}
		\mathcal{L}\left(\bf{d}_n | {\bf{q}}\right) = \int p\left(\bf{d}_n | \bf{m}_{\rm{sub}}, \bf{x}\right) p\left(\bf{m}, \bf{x} | \bf{q}_{\rm{sub}}\right) d {\bf{m}} d {\bf{x}}.
	\end{equation}
	This integral is computationally intractable due the high dimension of the parameter space. In particular, the overwhelming majority of random draws of a lens mass profile for the main deflector will not produce a lens system with the same image positions as in the observed data. 
	
	With the framework developed by \citep{Gilman++18,Gilman++19}, we can bypass the direct evaluation of Equation \ref{eqn:likelihoodfunction} with a forward modeling approach that, by construction, computes flux ratios of simulated lensed images at the same coordinates as the observed image positions. First, we generate a realization of dark matter structure ${\bf{m}}$ from the model ${\bf{q}}$. Next, using the recursive form of the multi-plane lens equation \cite{Blandford86}
	\begin{equation}
		\label{eqn:lenseqn}
		{\boldsymbol{\theta_K}}= \boldsymbol{\theta} - \frac{1}{D_{\rm{s}}} \sum_{k=1}^{K-1} D_{\rm{ks}}{\boldsymbol{\alpha_{\rm{k}}}} \left(D_{\rm{k}} \boldsymbol{\theta_{\rm{k}}}\right)
	\end{equation} 
	we solve for a set of parameters that describe the mass profile of the main deflector (hereafter the `macromodel') that map the four observed image positions in the data to a common source position. The quantity $D_{\rm ks}$ represents an angular diameter distance between the $k$th lens plane and the source plane, and $D_{\rm{s}}$ ($D_{\rm{k}}$) represents the angular diameter distance to the source plane (the $k$th lens plane). The vector $\bf{\theta}$ represents an angle on the sky, and $\bf{\alpha_{\rm{k}}}$ represents the deflection field from all halos at the $k$th lens plane. For each realization, we first run a randomly-initialized particle swarm optimization, followed by a downhill simplex routine, to obtain a precise solution for the macromodel parameters. 
	
	When performing the optimization, we sample parameters that describe the main deflector mass profile from their respective priors. For our analysis, we model the main deflector as a power-law ellipsoid embedded in external shear, with the addition of an octopole mass moment that adds boxyness or diskyness (see Section \ref{sec:supmodels}) to the main deflector mass profile. The optimization is performed with all halos included in the lens model, such that configurations of the macromodel returned by the optimization routine express any potential covariance with the hyper-parameters ${\bf{q}}$. The Einstein radius, mass centroid, ellipticity, ellipticity position angle, and external shear position angle are left free to vary during the optimization, while the logarithmic profile slope, external shear strength, and amplitude of the octopole mass moment are sampled from a prior, and held fixed during the optimization. We account for astrometric uncertainties in the measured image positions by adding random perturbations to the image positions in the simulated lens before performing the optimization.  
	
	At this stage, we have a lens system that includes a realization of dark matter structure with the same image positions as the observed data. We now proceed to compute the flux ratios. First, we sample a source size from a prior, and then ray-trace through the lens system to compute the image magnification with the extended source. After adding measurement uncertainties to the resulting image magnifications, we compute a summary statistic using the observed flux ratios $\bf{f}_{\rm{data}}$ and the model-predicted flux ratios $\bf{f}_{\rm{model}}$
	\begin{equation}
		\label{eqn:supsummarystat}
		S = \sqrt{\sum_{i=1}^{3}\left(f_{\rm{data}(i)} - f_{\rm{model(i)}}\right)^2},
	\end{equation}
	where the sum runs over the three flux ratios. We accept a proposals of model parameters ${\bf{q}}$ if the corresponding statistic satisfies $S < \epsilon$, where $\epsilon$ is a tolerance threshold. In practice, we generate hundreds of thousands to millions of realizations per lens, and retain the several thousand proposals corresponding to the smallest values of $S$, resulting in a tolerance threshold $\epsilon \sim 0.05$. 
	
	As $\epsilon \rightarrow 0$, the ratio of the number of accepted samples between two models will approach the relative likelihood of the two models. Thus, our method returns the relative likelihood of different points throughout the prior volume $\pi\left({\bf{q}}\right)$, which is enough information to determine the likelihood function for an individual lens $\mathcal{L}\left(\bf{d_n}|{\bf{q}}\right)$ in Equation \ref{eqn:likelihoodfunction}, and the posterior probability density in Equation \ref{eqn:posterior}, up to an irrelevant (for our purposes) numerical prefactor. 
	
	\subsection{The likelihood function for core collapse in SIDM}
	\label{ssec:inf}
	
	We now discuss the likelihood function and posterior distribution used to obtain the key results presented in the main article. We compute the likelihood in terms of the fraction of collapsed halos as a function of halo mass. Due to the finite computational resources available to us, we must approximate the fraction of collapsed halos, which is a continuous function of halo mass, as a sequence of discrete mass bins. We write the fraction of collapsed subhalos in a mass range $10^a - 10^b M_{\odot}$ as $f_{a/b}$, and compute the fraction of collapsed field halos in the same mass range as a function of redshift as $r_{a/b}\times f_{a/b} \times \left[T\left(z\right) / T\left(z_d\right)\right]$. The parameter $r_{a/b}$ sets the fraction of collapsed field halos near the main deflector redshift, $z_d$, relative to the fraction of collapsed subhalos, and we multiply by the ratio of halo ages $ T\left(z\right) / T\left(z_d\right)$ such that field halos at $z > z_d$ ($z < z_d$) are less likely (more likely) to core collapse, with a probability that scales linearly with the elapsed time since formation. This approximation works to better than $10 \%$ for $s_{\rm{field}} < 1.5 \ \rm{Gyr}$.  It breaks down for larger values of $s_{\rm{field}}$ halos begin collapsing early, and continue to collapse with approximately equal probability until the present time (as opposed to the collapse probability scaling directly with halo age). 
	
	The timescale for core collapse also has a strong concentration on concentration, $c$. For a velocity independent cross section, $t_0\propto c^{-7/2}$ \citep{Essig++19}. At first order, scatter in the concentration mass relation changes the overall number of collapsed halos. This effect can be absorbed into the collapse timescales $\lambda_{\rm{sub}}$ and $\lambda_{\rm{field}}$. At second order, the probability that halo has collapsed becomes correlated with the shape of halo profile (as determined by the concentration) outside of the collapsed central region. We do not account for this second-order effect in our model, but it could be included in the future. 
	
	Collecting terms, the vector of hyper-parameters associated with SIDM physics is ${\bf{q}} \equiv \left(f_{6/7.5}, f_{7.5/8.5}, f_{8.5/10}, r_{6/7.5},r_{7.5/8.5},r_{8.5/10}\right)$. Working in terms of the fraction of collapsed halos allows us analyze virtually any form of the interaction cross section with any model for structure formation. Computing the likelihood in terms of the fraction of collapsed halos also has utility from a computational perspective. Resonances in the cross section cause a highly stochastic response of the cross section strength to small changes in $\alpha_{\chi}$ and $m_{\chi}/m_{\phi}$, causing the resulting likelihood function to fluctuate unpredictably over a range of scales. This poses significant numerical challenges, because it is prohibitively difficult to obtain an accurate approximation of the target probability density by applying a kernel density estimator to the likelihood function or posterior distribution. On the other hand, the likelihood function is a smoothly-varying function of the collapse fractions, allowing us to apply a kernel density estimator to obtain a continuous approximation of the likelihood. 
	
	For each lens, we use the forward modeling approach discussed in Section \ref{ssec:generalinf} to compute the likelihood of the data in terms of the fraction of collapsed halos, the amplitude of the subhalo mass function $\Sigma_{\rm{sub}}$, the logarithmic slope of the subhalo mass function $\alpha$, and a term that rescales the amplitude of the line-of-sight halo mass function, $\delta_{\rm{LOS}}$. For each lens, we write this likelihood $\mathcal{L}\left({\bf{d_n}} | {\bf{q}}, {\bf{v}}\right)$, where we have introduced the shorthand notation ${\bf{v}} \equiv \left(\Sigma_{\rm{sub}},\delta_{\rm{LOS}}, \alpha\right)$. We include terms that alter the amplitude of the subhalo and field halo mass functions in the computation of the likelihood function because the constraining power over halo density profiles scales with the overall number of halos. 
	
\begin{figure}
		\includegraphics[trim=0.25cm 1.cm 0.25cm
		0.5cm,width=0.45\textwidth]{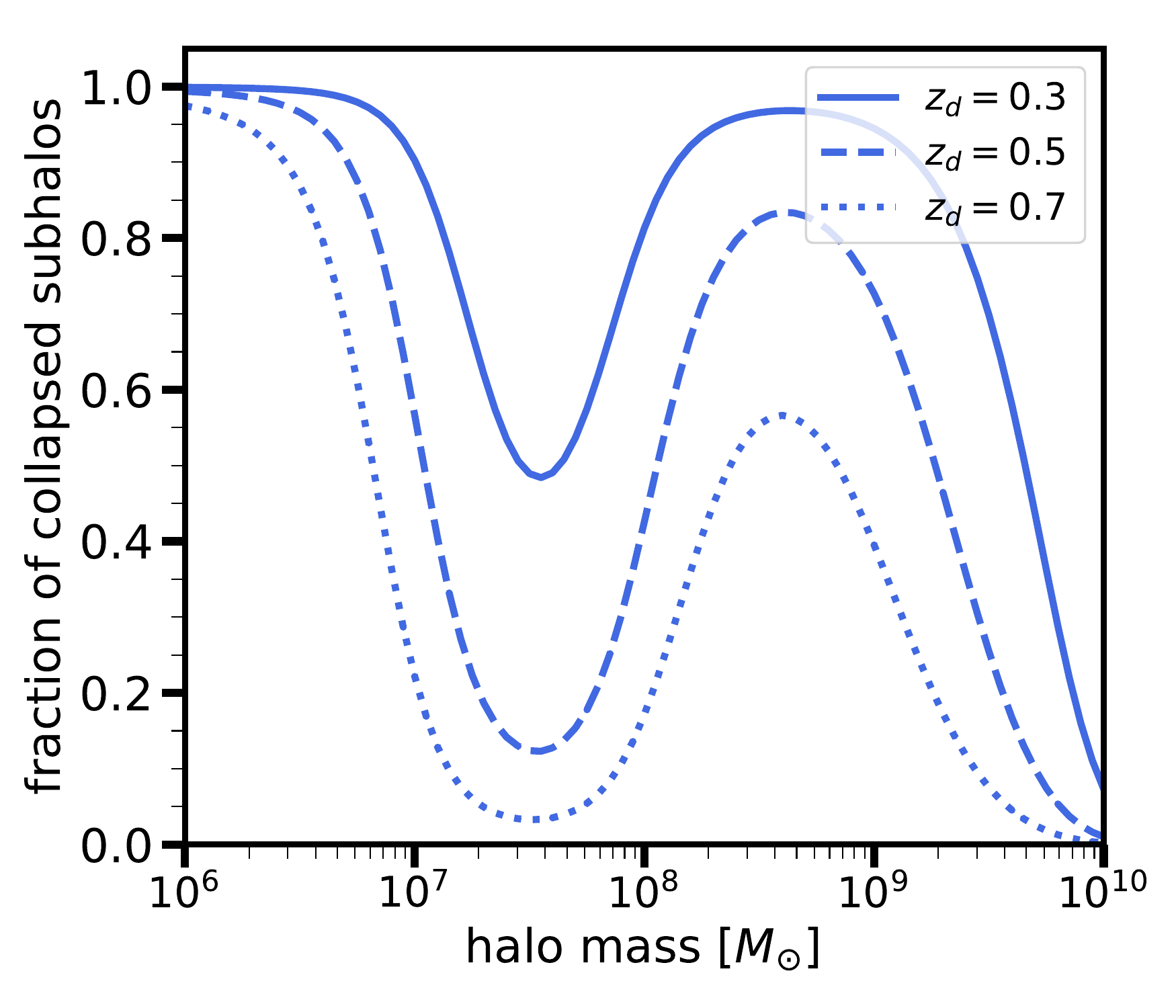}
		\caption{\label{fig:ssubz} The fraction of collapsed subhalos for Model 4 as a function of the main deflector redshift $z_d$, using the same values of $\lambda_{\rm{sub}}$, $\lambda_{\rm{field}}$, and $s_{\rm{sub}}$ as in Figure \ref{fig:collapsefrac}.}
	\end{figure}

	When computing the likelihood function in terms of the fraction of collapsed halos in each mass bin, we evaluate the collapse fraction $f_{a/b}$ at $m_{c} + \delta m_c$, where $\log_{10} m_c = 0.5\left(a + b\right)$ represents the (logarithmic) center of the mass bin, and $\delta m_c$ is a random step in $\log_{10} m$ drawn from a uniform distribution centered at zero with a width equal to the bin size. When computing the likelihood function, we evaluate the structure formation model at the main deflector redshift, $z_d$, as the population of subhalos and field halos affecting each lens reflect the properties of SIDM halos at different times. To illustrate this evolution, Figure \ref{fig:ssubz} shows the fraction of collapsed subhalos for Model 4 as a function of the main deflector redshift. Multiplying the likelihood functions for different lenses computed in terms of the fraction of core collapsed objects, i.e. multiplying likelihoods $\mathcal{L}\left({\bf{d}_n} | {\bf{q}}, {\bf{v}}, \sigma_V \right)$ instead of $\mathcal{L}\left({\bf{d}_n} | {\bf{p}}, {\bf{v}}, \sigma_V \right)$, would discard potentially useful information associated with the temporal evolution of SIDM halos, and possibly give a misleading result. 
	
	\subsection{The effect of scatter in the collapse timescales}
	\label{ssec:supcollapsescatter}
	
\begin{figure}
		\includegraphics[trim=0.25cm 1.cm 0.25cm
		0.5cm,width=0.45\textwidth]{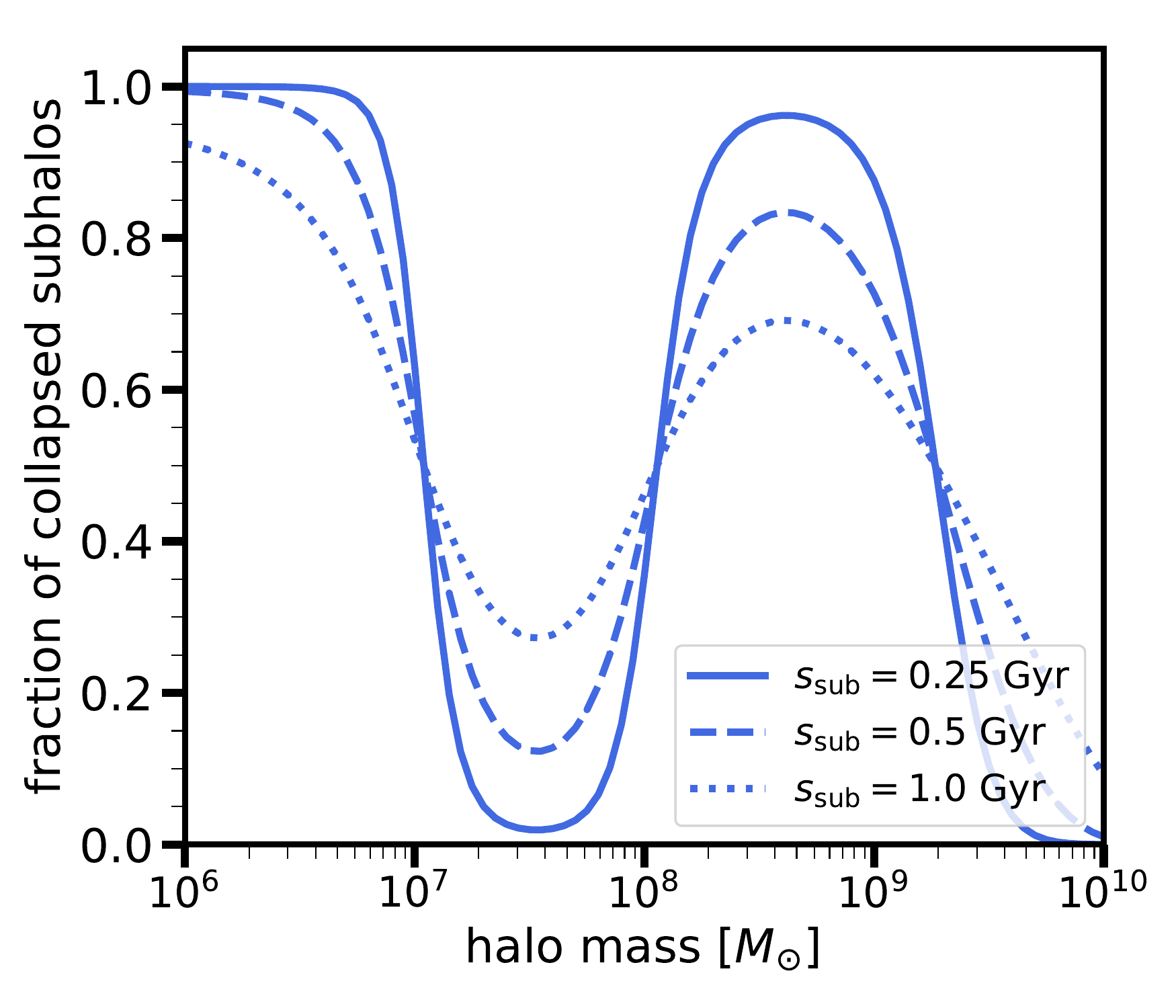}
		\caption{\label{fig:ssubscatter} The fraction of collapsed subhalos for Model 3 for different values of the scatter in the collapse time, $s_{\rm{sub}}$, evaluated at $z_d = 0.5$, with the same values of $\lambda_{\rm{sub}}$ and $\lambda_{\rm{field}}$ as shown in Figure \ref{fig:collapsefrac}.}
	\end{figure}

	The function used to determine the core collapse probability (Equation \ref{eqn:collapseprob}) is the cumulative distribution function for the logistic distribution 
	\begin{equation}
		p\left(x\right) = \frac{\exp\left(-x\right)}{\left(1+\exp\left(-x\right)\right)^2}
	\end{equation}
	with $x=\left(T\left(z\right) - t_{\rm{sub}}\left(m,z,\sigma_V\right)\right)/s_{\rm{sub}}$ for subhalos, and a similar expression for field halos. This probability density results in a standard deviation of collapse times given by $\left(\pi/\sqrt{3}\right) s_{\rm{sub}} \sim 1.8 s_{\rm{sub}}$. Figure \ref{fig:ssubscatter} shows how increasing the value of $s_{\rm{sub}}$ from $0.25 \rm{Gyr}$ to $1.0 \ \rm{Gyr}$ affects the fraction of collapsed subhalos for Model 4. The effect of increasing the scatter is qualitatively similar among each of the five benchmark models. 
	
	Figure \ref{fig:posteriorssigmasub} shows how our constraints on Model 2 change assuming a different prior for the collapse timescales and the amplitude of the subhalo mass function (see next section). We focus on Model 2 because we obtain the strongest constraints on this model, and the effects of the various prior choices become most apparent. The top left panel shows the constraints presented in the main article,  which assume a uniform prior on both $s_{\rm{field}}$ and $s_{\rm{field}}$ between $0.25 - 1.0 \ \rm{Gyr}$. The top right panel shows the constraints if we assume a model in which the scatter in collapse times in field halos is roughly three times less than the scatter in collapse times for subhalos. This could occur, for example, if tidal stripping, heating, and evaporation differentially impact subhalo evolution with a strong dependence on orbital pericenter, while field halos have similar evolutionary histories, and collapse within $1 \rm{Gyr}$ of each other. Folding in this more informative prior, the constraints on the collapse timescales strengthen by $\sim 15\%$ for the combinations of $\lambda_{\rm{sub}}$ and $\lambda_{\rm{field}}$ listed in Table \ref{tab:models}. Thus, theoretical predictions for these parameters will likely important increasingly important as the data quality improves  and the sample size of lenses grows.
	
	\subsection{The subhalo mass function in SIDM}
	\label{ssec:supSHMF}
	The amplitude of the subhalo mass function (SHMF), or $\Sigma_{\rm{sub}}$ in our notation, is subject to significant theoretical uncertainty associated with tidal stripping. While one generally expects some fraction of subhalos to become tidally disrupted and eventually destroyed as they orbit the central galaxy, the extent to which this occurs around massive elliptical galaxies remains somewhat unexplored. This is mainly due to the computational expense required to perform cosmological simulations of a host halo with mass $\sim 10^{13} M_{\odot}$ while resolving subhalos down to $10^7 M_{\odot}$. At the present time, we are only aware of one study that examines this problem in detail \citep{Fiacconi++16}
	
	Recently, \citet{Nadler++21} and \citet{Banik++21} inferred the amplitude of the SHMF in the Milky Way from dwarf galaxy counts and stellar streams, respectively. The inferences on the subhalo mass function amplitude between the two probes are consistent with each other, and when extrapolated up to the halo mass scales relevant for strong lensing, correspond to subhalo mass function amplitudes $\Sigma_{\rm{sub}} \sim 0.025 \times q$, where $q$ represents the differential tidal stripping efficiency between the Milky Way and massive ellipticals. If subhalos are disrupted by the Milky Way's disk twice as efficiency as by an ellpitical galaxy, we would have $q=2$ and expect $\Sigma_{\rm{sub}} = 0.05 \rm{kpc^{-2}}$ \citep{Nadler++21}. A value $q>1$ is likely correct, as stellar disks prove particularly adept at tidally disrupting subhalos in N-body simulations \citep{GarrisonKimmel,Webb++20}. 
	
	Self-interactions add an additional layer of complexity to predicting the SHMF amplitude. As shown by \citet{Nadler++20}, self-interactions between subhalo particles and host halo particles remove mass from subhalos through ram pressure stripping. For the cross sections considered by \citet{Nadler++20}, the ram pressure stripping from self-interactions suppressed the number of surviving subhalos by $20\%-50\%$. The efficiency of this process depends on the amplitude of the cross section at the velocity scale set by the host halo velocity dispersion. While \citet{Nadler++20} studied this effect for Milky Way-like systems, the effect likely persists in early-type galaxy halos, although the amount of suppression may differ due to the somewhat higher velocity dispersion of typical lens galaxy host halos ($\sim 10^{13} M_{\odot}$ instead of $\sim 10^{12} M_{\odot}$). 
	
	\begin{figure*}
		\includegraphics[trim=0.25cm 1.cm 0.25cm
		0.5cm,width=0.45\textwidth]{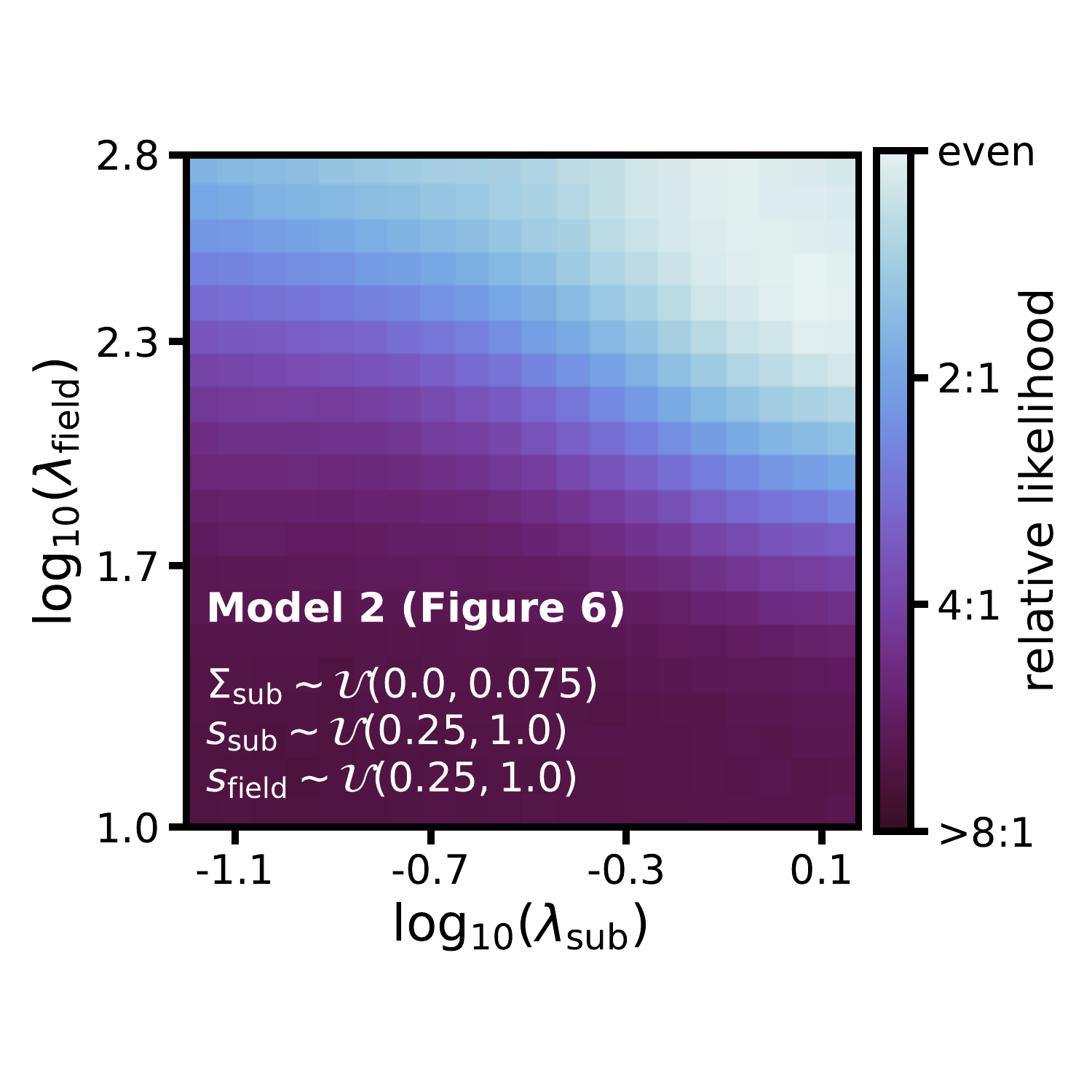}
		\includegraphics[trim=0.25cm 1.cm 0.25cm
		0.5cm,width=0.45\textwidth]{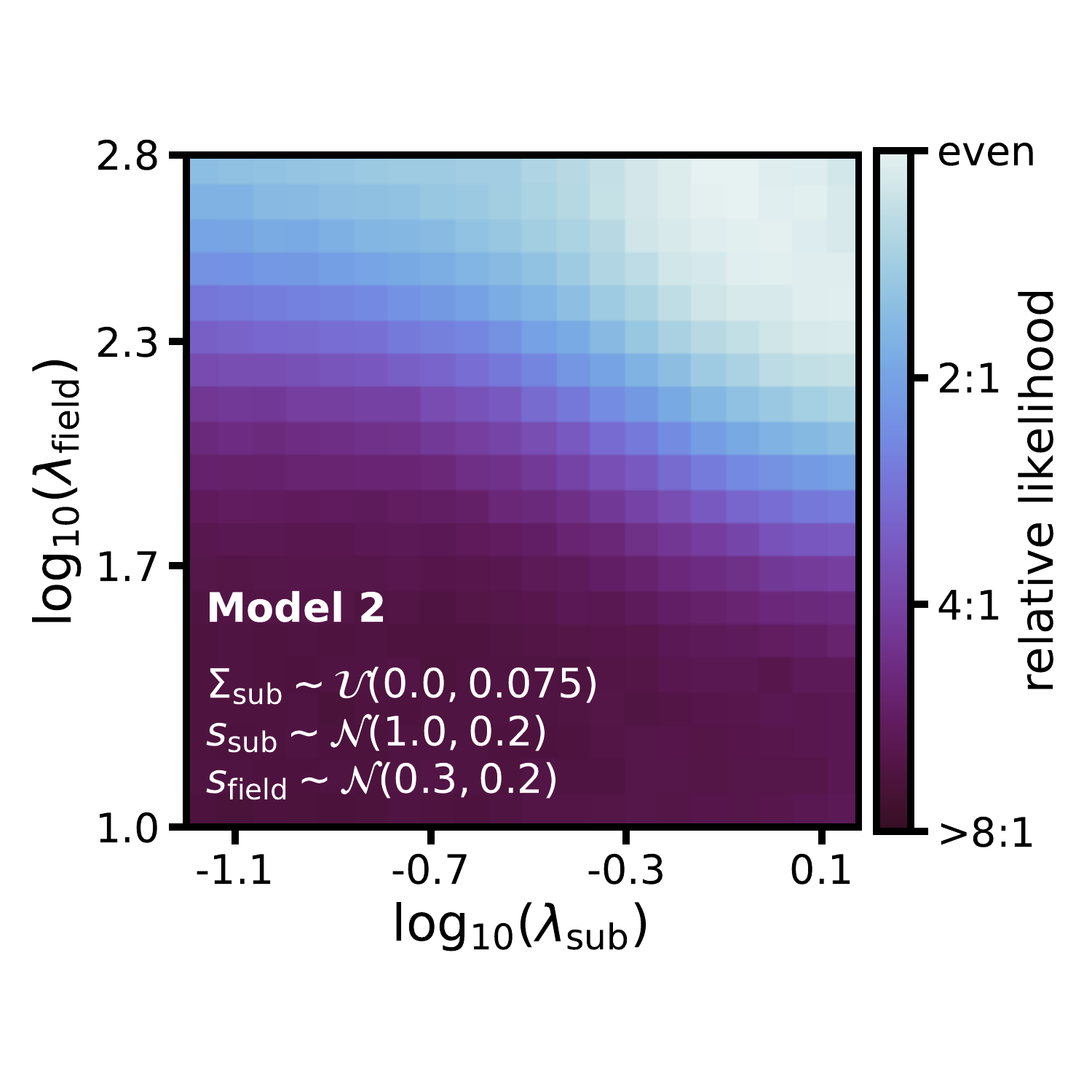}
		\includegraphics[trim=0.25cm 1.cm 0.25cm
		0.5cm,width=0.45\textwidth]{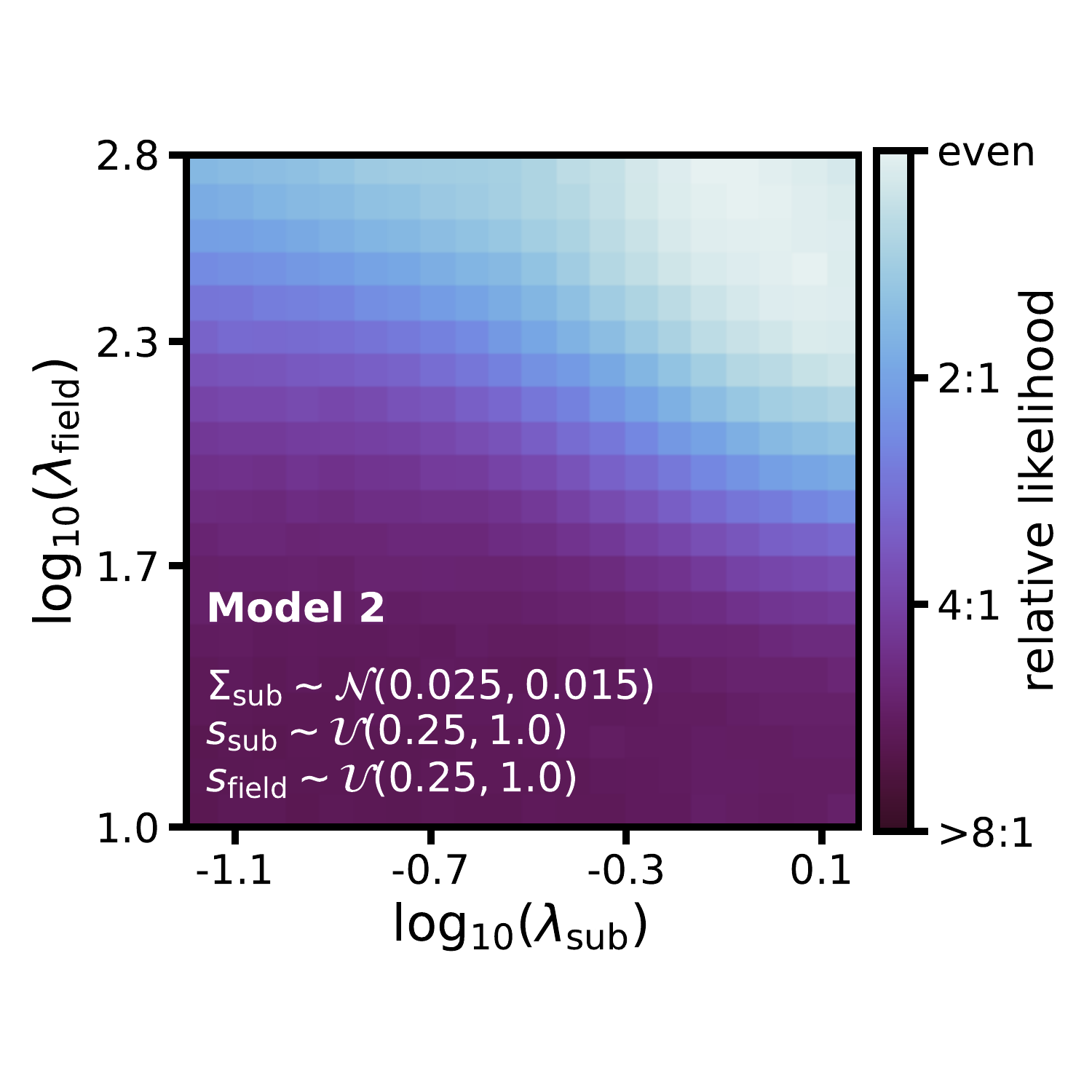}
		\includegraphics[trim=0.25cm 1.cm 0.25cm
		0.5cm,width=0.45\textwidth]{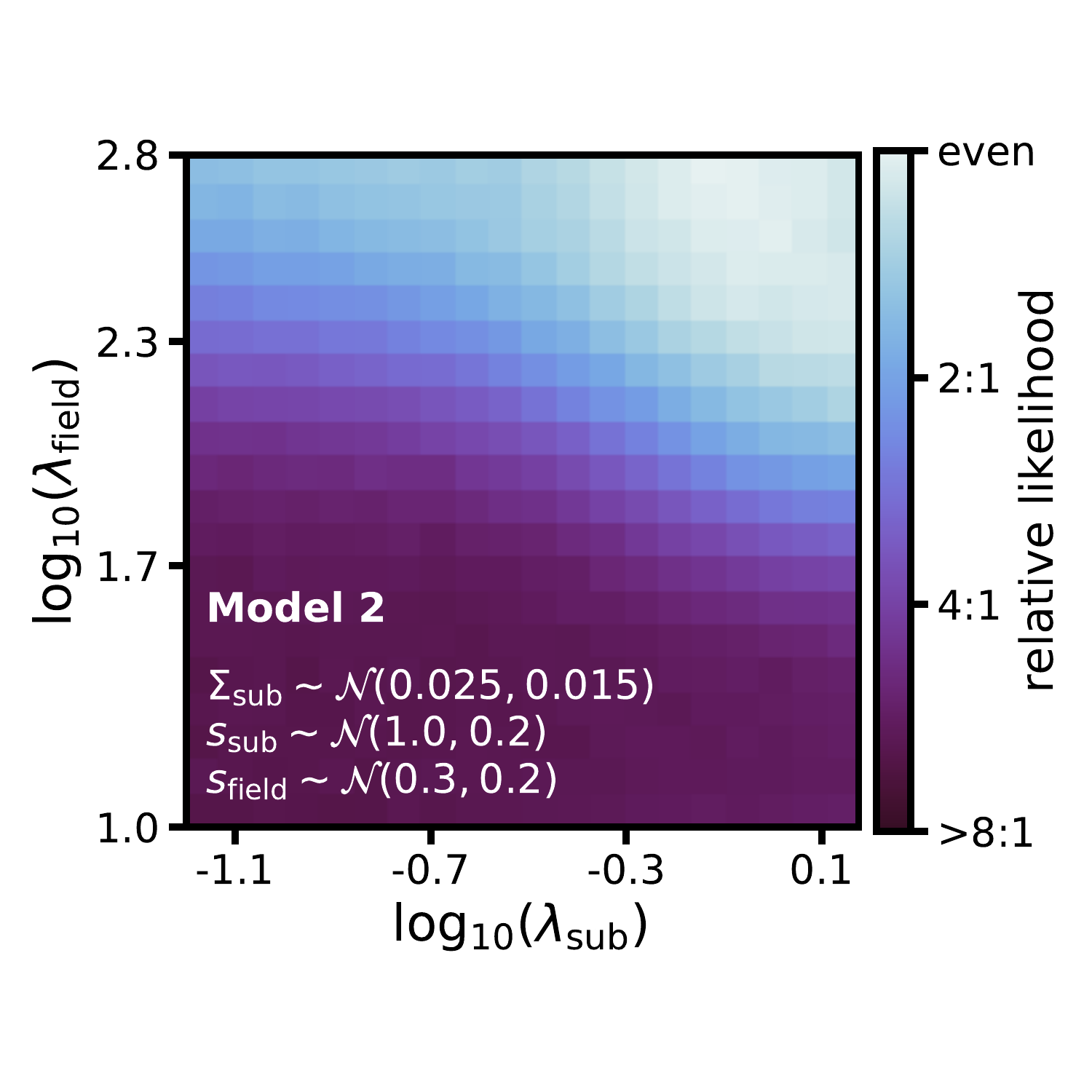}
		\caption{\label{fig:posteriorssigmasub} Constraints on the collapse timescales $\lambda_{\rm{sub}}$ and $\lambda_{\rm{field}}$ for Model 2 after marginalizing over different priors for the amplitude of the subhalo mass function $\Sigma_{\rm{sub}}$, and the scatter in collapse times for subhalos $s_{\rm{sub}}$ and field halos $s_{\rm{field}}$. The top left panel shows the constraints presented in Figure \ref{fig:posteriors}, for comparison. For discussion, see Sections \ref{ssec:supcollapsescatter} and \ref{ssec:supSHMF}).}
	\end{figure*}
	
	In our framework, we can account for this effect by folding in a prior on the subhalo mass function amplitude. For the main analysis, we do not impose strong assumptions on the amplitude of the subhalo mass function and marginalize over a uniform prior on $\Sigma_{\rm{sub}}$ between $0.0 -0.075 \rm{kpc^{-2}}$. For reference, if tidal stripping by stellar disks is twice as efficient as tidal stripping by an elliptical central galaxy and ram pressure stripping from self-interactions is negligible, we would expect $\Sigma_{\rm{sub}} \sim 0.05 \ \rm{kpc^{-2}}$\citep{Nadler++21}, while if ram-pressure stripping suppresses the amplitude of the SHMF by $50 \%$ with doubly-efficient tidal stripping, we would expect $\Sigma_{\rm{sub}}\sim 0.025 \rm{kpc^{-2}}$. 
	
	Figures \ref{fig:posteriorssigmasub} shows how a prior on the SHMF amplitude affects the inference on $\lambda_{\rm{sub}}$ and $\lambda_{\rm{field}}$ for Model 2. With lower subhalo mass function amplitudes, we obtain somewhat weaker constraints on timescales by $\sim 20\%$. The effect of the prior on the subhalo mass function amplitude is comparable to the effect of changing the prior on $s_{\rm{sub}}$ and $s_{\rm{field}}$, comparing the bottom panels of Figure \ref{fig:posteriorssigmasub} with the top right panel. The effect of ram pressure stripping in SIDM on the amplitude of the SHMF further underscores the need for sound theoretical predictions for the tidal evolution of dark subhalos around early-type galaxies for velocity-dependent cross sections. 
	
	\section{Parameterization of the mass function, background source, and main deflector mass profile}
	\label{sec:supmodels}
	In this section, we provide technical details regarding the models we implement for the subhalo and field halo mass functions, the background source, and the main deflector mass profile. Many of the modeling choices outlined in the following sections are the same as used in previous analyses \citep{Gilman++20,Gilman++21}; we refer to these works for further details and discussion. 
	
	\subsection{Subhalo and field halo mass functions}
	We consider models of self-interacting dark matter in which the linear matter power spectrum is unaffected on scales probed by our data. We note that some SIDM theories predict a suppression in the matter power spectrum from significant interactions between dark matter and dark radiation \cite[e.g.][]{Vogelsberger++16,Cyr-Racine++16}, but this property is not inherent to all SIDM frameworks. We could easily extend our analysis to such models, however, given a prediction for the halo mass function and the concentrataion-mass relation.
	
	We generate populations of halos in field, along the line of sight from observer to source, using the mass function model presented by \citet{ShethTormen}
	\begin{equation}
		\label{eqn:mfunclos}
		\frac{dN_{\rm{CDM}}}{dm dV} = \delta_{\rm{LOS}} \times \xi\left(M_{\rm{host}},z\right) \frac{dN}{dm dV} \Big \vert_{\rm{Sheth\,Tormen}}.
	\end{equation}
	assuming a flat $\Lambda$CDM cosmology with parameters from Planck \citep{Planck2020}. $M_{\rm{host}} \sim 10^{13} M_{\odot}$ is the host halo mass. The term $\delta_{\rm{LOS}}$ rescales the amplitude of the mass function everywhere, and absorbs uncertainties associated with cosmological parameters, and the definition of halo mass. We marginalize over values of $\delta_{\rm{LOS}}$ between 0.8 and 1.2 in the forward model to account for these uncertainties. The term $\xi\left(M_{\rm{host}},z\right)$ adds additional halos near the main lens plane, due to the local enhancement of the dark matter density associated with the host dark matter halo. We refer to \citet{Gilman++19} and \citet{Gilman++20} for additional details regarding this term.  
	
	We model the subhalo mass function as
	\begin{equation}
		\frac{dN_{\rm{CDM}}}{dm dA} = \frac{\Sigma_{\rm{sub}}}{m_0} \left(\frac{m}{m_0}\right)^{-\alpha} \mathcal{F}\left(M_{\rm{host}},z\right),
	\end{equation}
	where $\Sigma_{\rm{sub}}$ sets the normalization, and $\alpha$ is the logarithmic slope pivoting around $m_0 = 10^8 M_{\odot}$. The term $\mathcal{F}\left(M_{\rm{host}},z\right)$ accounts for the evolution of the projected mass in substructure with host halo mass and redshift \citep{Gilman++20}. By factoring the evolution with redshift and host halo mass out of the definition of $\Sigma_{\rm{sub}}$, we can combine inferences of $\Sigma_{\rm{sub}}$ from multiple lenses (in other words, we can multiply together likelihoods computed for different lenses conditioned on $\Sigma_{\rm{sub}}$). In the specific context of SIDM $\Sigma_{\rm{sub}}$ absorbs effects from ram pressure stripping, which can suppress the number of subhalos \cite{Nadler++20}. 
	
	We model the density profiles of halos generated from the mass functions as tidally truncated NFW profiles 
	\begin{equation}
		\rho\left(r,r_s,r_t\right) = \rho_s \left(\frac{r}{r_s}\right)^{-1}\left(1+\frac{r}{r_s}\right)^{-2}\left(\frac{r_t^2}{r^2+r_t^2}\right).
	\end{equation}
	We assign a truncation radius $r_t$ to subhalos based on their mass and three dimensional position inside the host halo \citep{Cyr-Racine++16b}. We truncate field halos at $r_{50}$, corresponding to $r_t / r_s \sim 25$ for most objects. Field halos in CDM become effectively truncated at the splash-back radius \citep{Diemer17}, which is the same order as $r_{50}$. We use the concentration-mass relation presented by \cite{Diemer++19} with a scatter of 0.2 dex \cite{Wang++20}. 
	
	\subsection{Background source}
	We model the extended structure of the background source as a circular Gaussian with a size (defined as the full-width at half-maximum) determined by the emission feature used to measure the relative image brightness. For lenses with fluxes measured from nuclear narrow-line emission, we sample source sizes in the forward model between 25-60 pc \citep{MullerSanchez++11}. To model the mid-IR source, we assume a size of $0.5-10$ pc \citep{Chiba++05}, and for the CO 11-10 emission we assume a size of 5-15 pc \citep{Stacey++18}. 
	
	\subsection{Main deflector mass profile}
	The models we use for the main deflector mass profile are motivated by observations of the early-type galaxies that typically act as strong lenses \citep{Auger++10}. These systems typically have approximately isothermal mass density profiles, with approximately elliptical iso-density contours. We therefore model the mass profile of the main deflector as an isothermal ellipsoid, plus external shear from nearby structure. We marginalize over the logarithmic profile slope $\gamma_{\rm{macro}}$ between $-1.9$ and $-2.2$, and over the external shear strength in the main lens plane. 
	
	Higher-order angular structure in lens galaxies can impact image flux ratios \citep{Gilman++17,Hsueh++18,Powell++22}, although in order to explain the data completely the multipoles require unreasonably large amplitudes \citep{Kochanek++04,Congdon++05}. To account for deviations from ellipsoidal symmetry, we add additional angular structure to the main deflector mass profile through a multipole term with amplitude $a_4$
	\begin{equation}
		\kappa_{\rm{oct}}\left(r,a_4,\theta\right) = \frac{a_4}{r} \cos\left(4\left(\theta - \theta_{\epsilon}\right)\right)
	\end{equation}
	where $\theta_{\epsilon}$ is the position angle of the main deflector's elliptical mass profile. The sign of $a_4$ determines whether this additional component results in iso-density contours that are boxy or disky. We marginalize over a prior on $a_4$ with mean zero and variance 0.01, where the variance is determined from observations of isophotes of elliptical galaxies, assuming light traces mass \citep{Bender1989,Hao++06}. This choice overestimates the magnitude of the ellipsoidal symmetry-breaking mass component, because the shape of the light profile will be more boxy or disky that the shape of the projected mass after accounting for the additional projected mass from the host dark matter halo. In this sense, we implement a conservative model for the multipole terms. These multipole terms were also included in recent lensing analyses \citep{Gilman++21,Gilman++22,Laroche++22}.
	
	Finally, if the observed lens system has a nearby satellite galaxy, we include it explicitly in the lens model as an isothermal sphere, with the position and Einstein radius, and uncertainties on these quantities, determined by the observations that report the discovery of the satellite. For additional details, see \citet{Gilman++20}. 
	
	\section{The effect of the collapsed halo profile}
	\begin{figure}
		\centering
		\includegraphics[trim=0.cm 0.cm 0cm
		0.cm,width=0.45\textwidth]{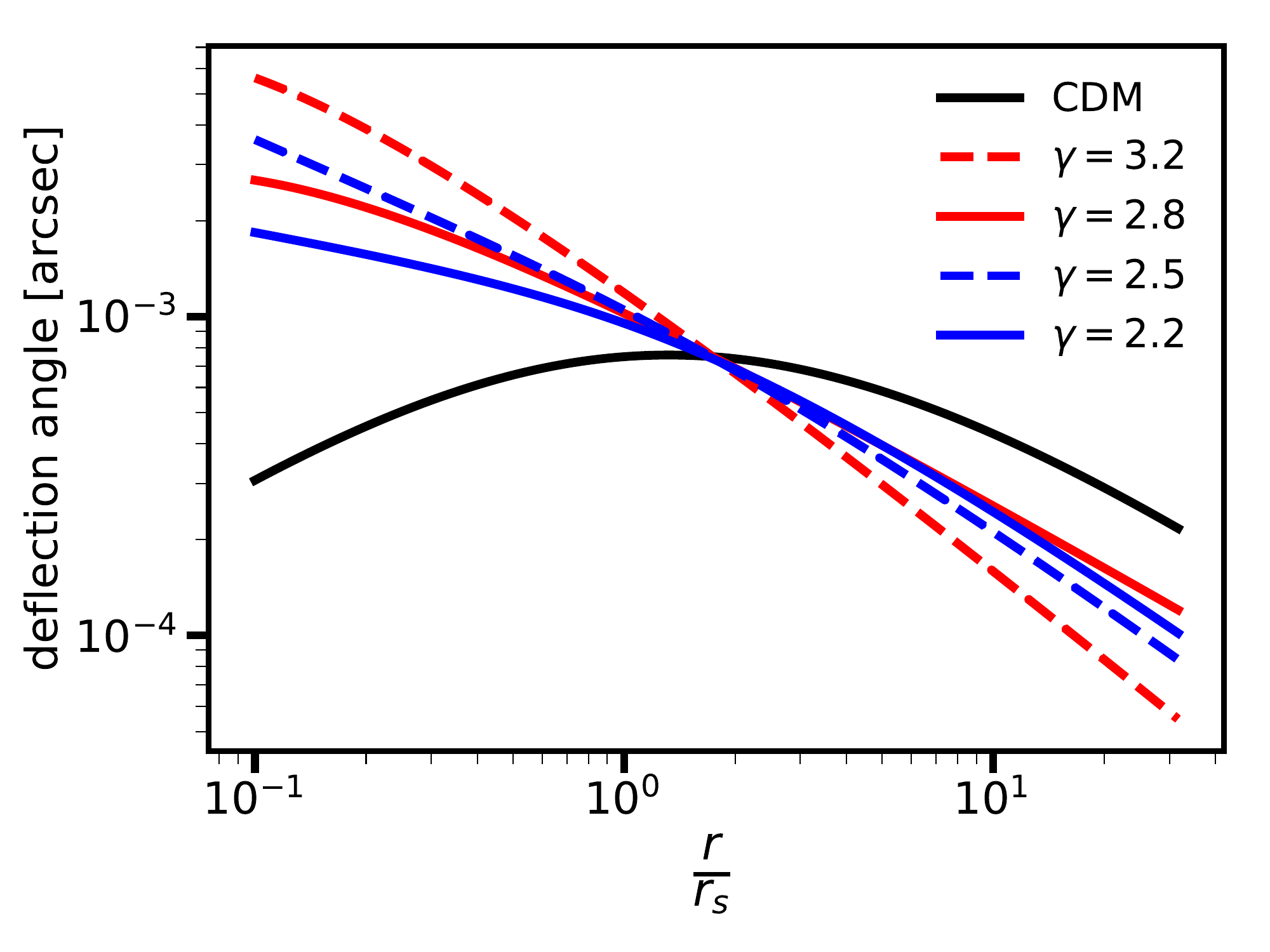}
		\includegraphics[trim=0.cm 0.cm 0cm
		0.cm,width=0.45\textwidth]{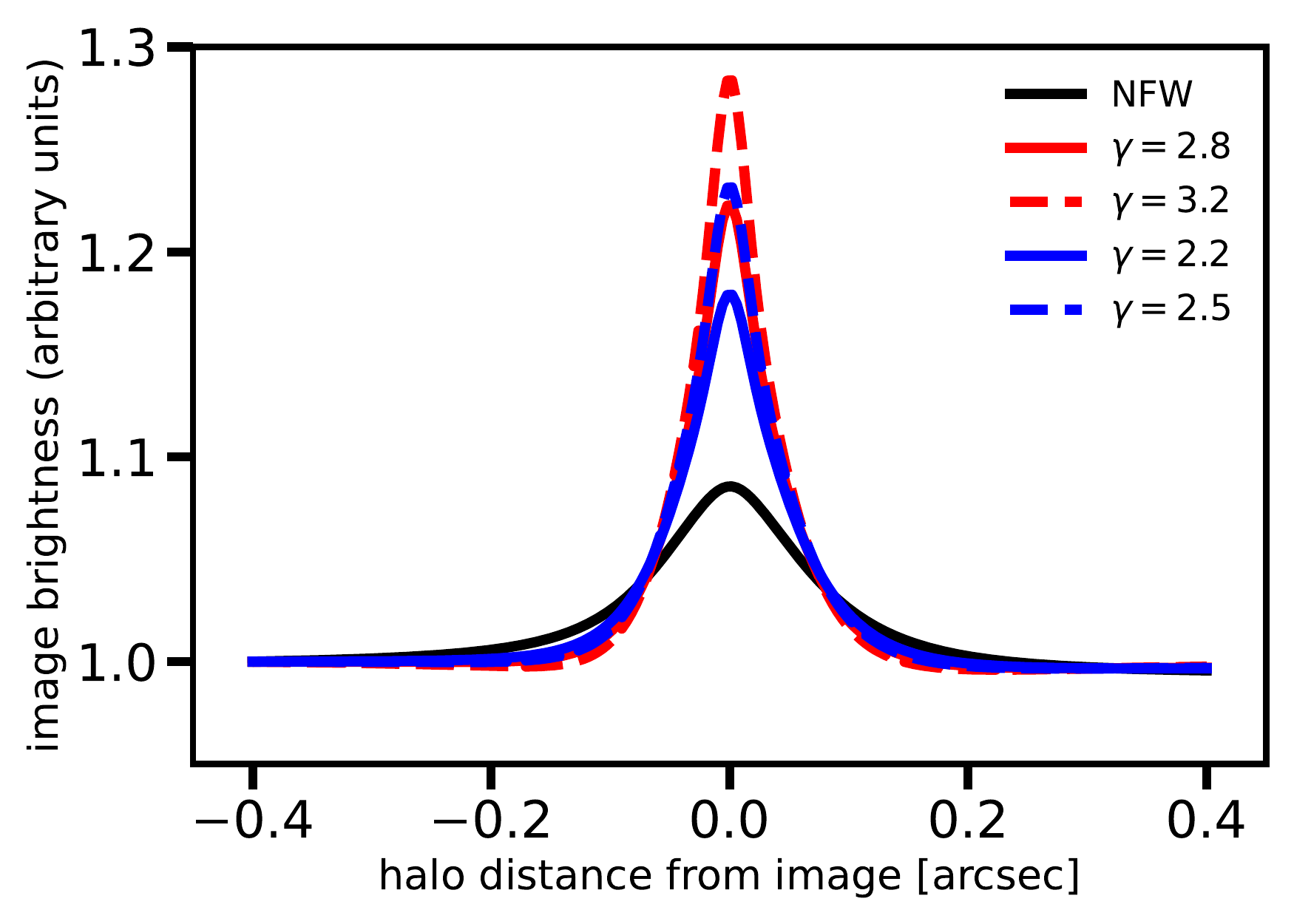}
		\caption{\label{fig:proflensing} The deflection angle (top) and magnification cross section (bottom) produced by a $10^8 M_{\odot}$ halo intersecting a lensed source with a size of 40 pc. Black curves show a CDM halo modeled as an NFW profile, red curves correspond to the core-collapsed profile given by Equation 8 in the main article with a core size $0.025 \ r_s$ and logarithmic slope $\gamma$, and blue curves show a profile (Equation \ref{eqn:supcollapseprof2}) with a logarithmic slope inside the scale radius $\gamma$, and a logarithmic slope beyond the scale radius of $-3$. Differences between the magnification cross section with different models of the collapsed halo density profile are much smaller than the difference between the magnification cross section of core-collapsed halos and NFW profiles.}
	\end{figure}
	\label{sec:supcollapsedprofile}
	Dark matter halo mass profiles are not observable, but the deflection angle associated with them can impact strong lensing data. Therefore, in order to obtain robust constraints on SIDM models through core collapse and strong lensing, we require an accurate model for the deflection angle
	\begin{equation}
		\label{eqn:deflectionangle}
		\alpha\left(r\right) \propto \frac{1}{r} \int_{0}^{r} r^{\prime}\kappa\left(r^{\prime}\right) dr^{\prime}.
	\end{equation}
	where $\kappa\left(r\right) = \int_{-\infty}^{\infty} \rho\left(\sqrt{r^2+z^2}\right)dz $ represents the projected mass of a halo.  
	
	Core-collapsed halos produced in simulations have logarithmic slopes interior to $r_s$ that range from approximately isothermal ($-2$) to around ${-3}$, transitioning to $-3$, like an NFW profile, beyond $r_s$. The profile we use to model collapsed objects
	\begin{equation}
		\label{eqn:supcollapseprof}
		\rho\left(r,r_c,x_{\rm{match}}\right) = \rho_0\left(x_{\rm{match}}\right) \left(1 + r^2 / r_c^2\right)^{-\gamma/2}.
	\end{equation}
	captures the steep inner profile slope, the small central core that persists throughout the various stages of collapse, and approximately matches the slope beyond $r_s$, provided $\gamma \sim 3$. 
	
	We may ask how a different logarithmic profile slope, closer to $\gamma \sim 2$, might affect the deflection angle and the magnification cross section. Performing this analysis using the density profile in Equation \ref{eqn:supcollapseprof}, however, could give misleading results, because with $\gamma = 2$ the collapsed profile becomes substantially more massive than a CDM halo. To make comparisons between halo mass profiles with different central profiles, we compute deflection angles and magnification cross sections with a modified NFW halo profile with density
	\begin{equation}
		\label{eqn:supcollapseprof2}
		\rho\left(r\right) = \rho_0\left(x_{\rm{match}}\right)\left(\frac{r}{r_s}\right)^{-\gamma} \left(1+\frac{r^2}{r_s^2}\right)^{\left(\gamma-3\right)/2},
	\end{equation}
	for which we have analytic solutions for the projected mass and deflection angle \citep{Munoz++01}. We define the normalization in terms of the radius $r_{\rm{match}}\equiv x_{\rm{match}} r_s$ where the collapsed profile encloses the same mass as an NFW profile, the same way we normalize the density profile in Equation \ref{eqn:supcollapseprof}.  \citet{ShengqiYang++22} also proposed a model for SIDM halo mass profiles, but we were unable to obtain closed-form solutions for the projected mass for their model using {\tt{Mathematica}}. Analytic expressions for the deflection angle substantially increase the speed with we we can perform ray-tracing computations with {\tt{lenstronomy}} \citep{BirrerAmara++18,Birrer++21}. 
	
	By comparing the lensing properties of the profiles in Equations \ref{eqn:supcollapseprof} and \ref{eqn:supcollapseprof2}, we assess at what level systematic uncertainties associated with the form of the halo profile impact our results. The top panel of Figure \ref{fig:proflensing} shows the deflection angle of a $10^8 M_{\odot}$ halo for profiles with various logarithmic slopes. The bottom panel shows the magnification cross section, assuming a source of size of 40 pc. All profiles have $x_{\rm{match}}=2.5$, and the curves corresponding to the density profile in Equation \ref{eqn:supcollapseprof} have $r_c = 0.025 r_s$. While the magnification cross section varies between the different models for the core-collapsed profiles, these differences are small compared to the total increase in the lensing efficiency, relative to CDM. In this sense, we regard the details of the core-collapsed density profile as a second-order effect relative to the total number of collapsed halos associated with the timescales $\lambda_{\rm{sub}}$ and $\lambda_{\rm{field}}$. 
	\begin{figure}
		\centering
		\includegraphics[trim=0.cm 0.cm 0cm
		0.cm,width=0.48\textwidth]{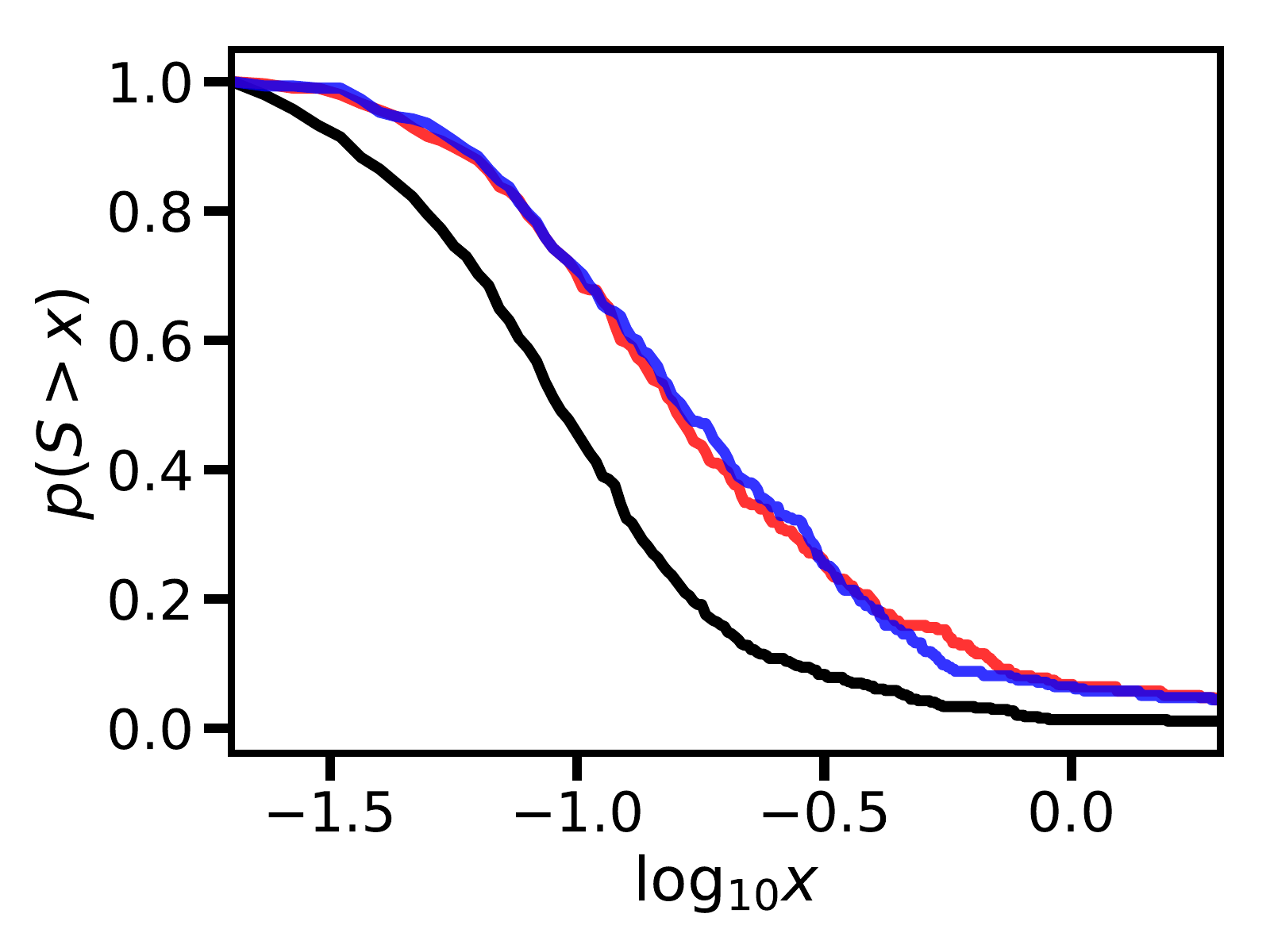}
		\caption{\label{fig:cumulativestat} The cumulative distribution of the summary statistic $S$ defined in Equation \ref{eqn:supsummarystat} for CDM (black), SIDM with collapsed profiles modeled using Equation \ref{eqn:supcollapseprof} (red), and collapsed profiles given by Equation \ref{eqn:supcollapseprof2} (blue). The x-axis shows the value of the summary statistic, and the y-axis shows the probability of obtaining a summary statistic greater than $x$. We compute the summary statistic for flux ratios perturbed by substructure with respect to flux ratios computed with a smooth lens model, such that long tails of the distribution indicate large and/or frequent perturbation to the flux ratios.}
	\end{figure}
	
	We also perform a statistical comparison of the effects of different models for collapsed halo density profiles. First, we generate a smooth lens model with no substructure, and compute a set of flux ratios from it. We then generate 200 realizations of dark matter structure in CDM, compute the flux ratios for each realization, and evaluate the summary statistic in Equation \ref{eqn:supsummarystat} using the smooth lens model flux ratio in place of ${\bf{f}}_{\rm{data}}$, and the flux ratio computed with substructure in the place of ${\bf{f}}_{\rm{model}}$. We repeat this procedure for two SIDM scenarios in which $50\%$ of all halos core collapse. The first case has collapsed halos implemented using the profile in Equation \ref{eqn:supcollapseprof}, and another uses the density profile in Equation \ref{eqn:supcollapseprof2}. 
	
	Figure \ref{fig:cumulativestat} shows the cumulative distribution of the summary statistics for each scenario. The x-axis shows a summary statistic value $x$, and the y-axis shows the probability of obtaining a summary statistic greater than $x$ for CDM (black), SIDM with the density profile in Equation \ref{eqn:supcollapseprof} (red), and SIDM with the density profile in Equation \ref{eqn:supcollapseprof2} (blue). Large values of $x$ correspond to large perturbation to an image magnification.
	
	The degree of similarity between distributions of $S$ generated under different models acts as a proxy for our ability to distinguish between the models. The small differences between the red and blue curves, relative to the difference between both of them and the black curve, show that systematic uncertainty associated with how we implement the collapsed halo profile is unlikely to affect our results at the present time. However, as other sources of statistical and systematic uncertainties become smaller with a larger sample size of lenses, more precise theoretical predictions for the core collapse timescales, and observations with JWST of more compact background sources, we may also require more precise models for the mass profile of collapsed objects.
	
\end{document}